\documentclass[12pt]{article}
\usepackage{jheppub}
\usepackage{youngtab}

\def\CB{{\cal B}}

\def\CN{{\cal N}}

\def\CO{{\cal O}}

\def\IZ {{\mathbb Z}}
\newcommand{\Tr}{{\rm Tr\,}}

\title{Duality walls and defects in 5d $\mathcal{N}=1$ theories}

\author[1]{Davide Gaiotto and Hee-Cheol Kim}

\affiliation[1]{Perimeter Institute for Theoretical Physics\\%
31 Caroline Street North, ON N2L 2Y5, Canada}

\abstract{We propose an explicit description of ``duality walls'' which encode at low energy the 
global symmetry enhancement expected in the UV completion of certain five-dimensional gauge theories.
The proposal is supported by explicit localization computations and implies that the instanton partition function of these theories 
satisfies novel and unexpected integral equations.}

\begin{document}
\maketitle

\section{Introduction}
Five-dimensional super-conformal field theories are a particularly rich subject of investigation (see \cite{Seiberg:1996bd,Morrison:1996xf,Douglas:1996xp,Intriligator:1997pq,Aharony:1997ju} for seminal work on the subject). 
The only constructions available for these theories involve  
brane constructions, in particular quarter-BPS webs of five-branes in IIB string theory. Some of the five-dimensional SCFTs admit mass deformations 
to five-dimensional gauge theories, with the inverse gauge coupling playing the role of mass deformation parameter. Several protected quantities 
in the five-dimensional SCFT are computable directly from the low-energy gauge-theory description \cite{Kim:2012gu}. 

More precisely, the space of mass deformations of the UV SCFT is usually decomposed into chambers, 
which flow in the IR to distinct-looking gauge theories, or to the same gauge theory but with different identifications of the parameters. 
With some abuse of language, these distinct IR theories may be thought of as being related by an ``UV duality'',
in the sense that protected calculations in these IR theories should match \cite{Bergman:2013aca}. 

In such a situation, one may define the notion of ``duality walls'' between the different IR theories \cite{Gaiotto:2008ak}. 
These are half-BPS interfaces which we expect to arise from RG flows starting from Janus-like configurations, where the mass deformation parameters 
vary continuously in the UV, interpolating between two chambers. Duality walls between different chambers should compose 
appropriately. 

Furthermore, if we have some BPS defect in the UV SCFT, we have in principle a distinct IR image of the defect in each 
chamber, each giving the same answer when inserted in protected quantities. 
The duality walls should intertwine, in an appropriate sense, between these images. 

In this paper we propose candidate duality walls for a large class of quiver gauge theories of unitary groups.\footnote{Duality walls of the same kind, for 5d gauge theories endowed with a six-dimensional UV completion, appeared first in \cite{Gaiotto:2015usa}.}
The UV completion of these gauge theories has a conjectural enhanced global symmetry whose Cartan generators
are the instanton number symmetries of the low-energy gauge theory. The chambers in the space of real mass deformations 
dual to these global symmetries are Weyl chambers and the duality walls generate Weyl reflections 
relating different chambers. 

The duality walls admit a Lagrangian description in the low energy gauge theory. The fusion of interfaces 
reproduces the expected relations for the Weyl group generators thanks to a beautiful collection of Seiberg dualities. 
This is the first non-trivial check of our proposal. The second set of checks involve the computation of protected quantities. 

The duality walls we propose give a direct physical interpretation to a somewhat unfamiliar object: elliptic Fourier 
transforms (See \cite{2004math.....11044S} and references within). 
These are invertible integral transformations whose kernel is built out of elliptic gamma functions. We interpret the 
integral kernel as the superconformal index of the four-dimensional degrees of freedom sitting at the duality interface 
and the integral transform as the action of the duality interface on more general boundary conditions for the five-dimensional gauge theory. 
The integral identity which encodes the invertibility of the elliptic Fourier transform 
follows from the corresponding Seiberg duality relations. 

It follows directly from the localization formulae on the $S^4\times S^1$ and the definition of a duality wall 
that the corresponding elliptic Fourier transform acting on the instanton partition function of the gauge theory 
should give back the same partition function, up to the Weyl reflection of the instanton fugacity. 
This is a surprising, counterintuitive integral relation which should be satisfied by the instanton partition function. 
Amazingly, we find that this relation is indeed satisfied to any order in the instanton expansion we cared to check. 
This is a very strong test of our proposal.

Experimentally, we find that this is the first example of an infinite series of integral identities, which 
control the duality symmetries of Wilson line operators. These relations suggest how to assemble 
naive gauge theory Wilson line operators into objects which can be expected to have an 
ancestor in the UV SCFT which is invariant under the full global symmetry group. 

We also identify a few boundary conditions and interfaces in the gauge theory which transform covariantly 
under the action of the duality interface and could thus be good candidates for 
symmetric defects in the UV SCFT. We briefly look at duality properties of defects in codimension two and three 
as well. 

Finally, we attempt to give a physical explanation to another instance of elliptic Fourier transform which we found in the literature, 
which schematically appears to represent an interface between an $Sp(N)$ and an $SU(N+1)$ gauge theories. 
We find that the AC elliptic Fourier transform maps the instanton partition function of an $Sp(N)$ gauge theory into the 
instanton partition function of an exotic version of $SU(N+1)$ gauge theory with the same number of flavors.

After this work was completed, we received \cite{Hayashi:2015fsa,Yonekura:2015ksa} which have some overlap with the 
last section of this paper. 

\section{Duality walls between $SU(N)$ gauge theories}
\subsection{Pure $\CN=1$ $SU(N)_N$ gauge theory}
Our first and key example of duality wall encodes the UV symmetries of a 
pure five-dimensional $\CN=1$ $SU(N)$ gauge theory, with 5d CS coupling $N$.

\begin{figure}[h]
    \centering
 \includegraphics[width=.75\textwidth]{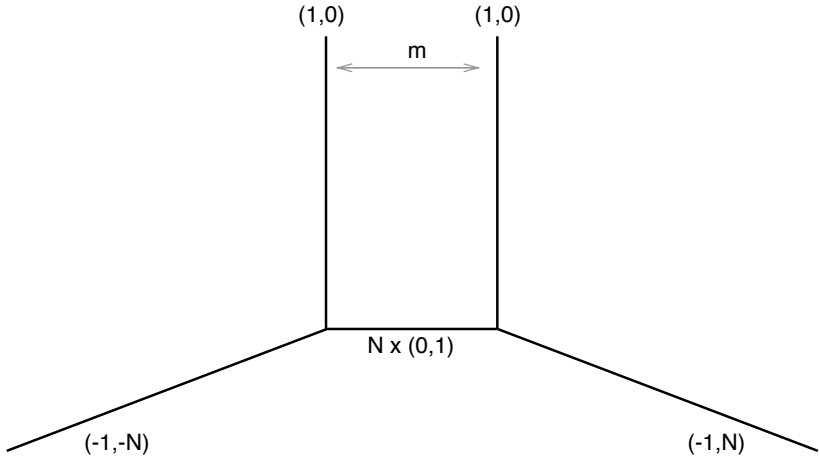}  
    \caption{The fivebrane web which engineers the UV completion of pure $SU(N)_N$ gauge theory. The gauge theory is supported on the bundle of $N$ parallel D5 branes. After removing the centre of mass, the only non-normalizable deformation is the separation $m$ between the NS5 branes}
    \label{fig:sun}
\end{figure}

This gauge theory is expected to be a low-energy description of a 5d SCFT with $SU(2)$ 
global symmetry, deformed by a real mass associated to the Cartan generator of $SU(2)$. 
In turn, the SCFT can be engineered by a BPS five-brane web involving four semi-infinite 
external legs: two parallel NS5 branes, a $(-1,N)$ and a $(-1,-N)$ fivebranes. 
The $SU(2)$ global symmetry is associated to the two parallel NS5 branes. See figure \ref{fig:sun}.

The mass deformation breaks $SU(2)$ to a $U(1)$ subgroup, which is identified with 
the instanton $U(1)_{in}$  global symmetry of the 
$SU(N)$ gauge theory, whose current is the instanton number density $J_{in}= \frac{1}{8 \pi^2} \Tr F \wedge F$.
The (absolute value of) the real mass is identified with $m= g_{YM}^{-2}$ in the IR
and with the separation between the parallel NS5 branes in the UV. 

The Weyl symmetry acts as $m \to -m$ and the corresponding duality wall should relate two copies of the 
same gauge theory, glued at the interface in such to preserve the anti-diagonal combination of 
the $U(1)_{in}$ instanton global symmetries on the two sides of the interface. 

\begin{figure}[h]
    \centering
 \includegraphics[width=.5\textwidth]{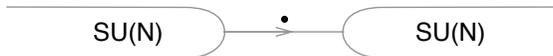}  
    \caption{Our schematic depiction of the duality wall. We denote 5d gauge groups on the two sides of an interface as open circles and the bi-fundamental matter as an arrow between them. 
    The extra baryonic coupling is denoted as a black dot over the arrow.}
    \label{fig:sunwall}
\end{figure}

We propose the following setup: a domain wall defined by Neumann b.c. for 
the $SU(N)_N$ gauge theory on the two sides of the wall, together with a set of bi-fundamental 
4d chiral multiplets $q$ living at the wall, coupled to an extra chiral multiplet $b$ by a 4d superpotential 
\begin{equation}
W = b \det q \ .
\end{equation}
See figure \ref{fig:sunwall} for a schematic depiction of the duality wall. 

This system is rife with potential gauge, mixed and global anomalies at the interface,
which originate from the 4d degrees of freedom, from the boundary conditions of the 5d gauge fields and 
and from anomaly inflow from the bulk Chern-Simons couplings. 

The cubic gauge anomaly cancels out beautifully: the bi-fundamental chiral multiplets 
behave as $N$ fundamental chiral multiplets for the gauge group on the right of the wall, 
giving $N$ units of cubic anomaly, which cancel against the anomaly inflow from the 
$N$ units of five-dimensional Chern-Simons coupling. Similarly, we get $-N$ units of cubic anomaly for the 
gauge group on the left of the wall, which also cancel against the anomaly inflow from the 
$N$ units of five-dimensional Chern-Simons coupling.

The bi-fundamental chiral multiplets also contribute to a mixed anomaly between the bulk 
gauge fields and the baryonic $U(1)_{B} $ symmetry which rotates the bi-fundamental fields with charge $1/N$
(normalized so that the baryon $B=\det q$ has charge $1$).  
The anomaly involving the left gauge fields has the same sign and magnitude as the anomaly 
involving the right gauge fields. Both are the same as the anomaly which would be associated to a single fundamental 
boundary chiral of charge $1$. 

We can make a non-anomalous $U(1)_\lambda$ global symmetry 
by combining $U(1)_B$ with $U(1)_{in}$ from both sides of the wall. Under $U(1)_\lambda$  
a boundary baryon operator will have the same charge as an instanton particle 
on the left side of the wall, or an anti-instanton particle on the right side of the wall. 
In particular, the proposed duality wall glues together $U(1)_{in}$ on the two sides of the wall 
with opposite signs and thus has a chance to implement the $Z_2$ duality symmetry. 

We can also define a non-anomalous R-symmetry by combining the Cartan generator of the bulk $SU(2)_R$ symmetry 
and a boundary symmetry which gives charge $0$ to the bifundamentals, and thus charge $2$ to $b$. 
The cancellation of the mixed gauge anomaly proceeds as follows: the bulk gauge fields with Neumann b.c. 
contribute half as much as 4d $SU(N)$ gauge fields would contribute and thus the R-symmetry assignment is the same as for a 4d SQCD with 
$N_f = N$. 

A neat check of this proposal is that two concatenated duality walls will annihilate in the IR. 
Far in the IR, a pair of consecutive duality walls looks like a single interface supporting 
four-dimensional $SU(N)$ gauge fields which arise from the compactification of the five-dimensional 
$SU(N)_N$ gauge theory on the interval. Together with the quarks associated to each duality wall, 
that gives us a four-dimensional $S(N)$ gauge theory with $N$ flavors, deformed by 
a superpotential coupling  
\begin{equation}
W = b \det q + \tilde b \det \tilde q
\end{equation}
which sets to zero the two baryon operators $\det q$ and $\det \tilde q$. 

\begin{figure}[h]
    \centering
 \includegraphics[width=.75\textwidth]{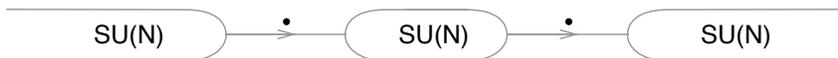}  
    \caption{A schematic depiction of the composition of two duality walls. The resulting 4d $SU(N)$ gauge theory has $N$ flavors and at low energy it glues the two 5d gauge groups together.}
    \label{fig:sunsquare}
\end{figure}

This four-dimensional theory has a well-known low-energy behaviour: it can be described as an effective non-linear sigma model 
parameterized by the mesons $M = \tilde q q$ and baryons $B = \det q$, $\tilde B = \det \tilde q$, subject to a constraint 
\begin{equation}
\det M - B \tilde B = \Lambda^{2 N} \ .
\end{equation}
Because of the $b B + \tilde b \tilde B$ superpotential couplings, we can restrict ourselves to the locus $B = \tilde B = 0$, 
where $M$ is an invertible matrix, which provides precisely the degrees of freedom required to Higgs the left and right five-dimensional 
theories back together, and thus flow in the far IR back to a trivial interface. This is the expected behaviour for $Z_2$ duality walls. 

\subsubsection{Domain wall actions.}
We should be able to use the domain walls to define a $Z_2$ duality action on 
$U(1)_{in}$-preserving half-BPS boundary conditions for the 
$SU(N)_N$ five-dimensional gauge theory. As the five-dimensional gauge theories are IR free, 
we can describe most boundary conditions in terms of their 
boundary degrees of freedom, which are in general some four-dimensional SCFTs equipped with  
an $SU(N)$ and an $U(1)_{in}$ global symmetries with specific cubic anomalies. The exceptions
are boundary conditions which (partially) break the gauge symmetry at the boundary. 

More precisely, consider a 4d $\CN=1$ theory $\CB$ with a $SU(N)$ global symmetry with $N$ units of cubic 't Hooft anomaly, 
a $U(1)_\partial$ global symmetry with a mixed 't Hooft  anomaly with the $SU(N)$ global symmetry equal to the contribution of a single fundamental chiral 
field of charge $1$ and an R-symmetry with a mixed 't Hooft  anomaly with the $SU(N)$ global symmetry equal to the contribution of $N$ quarks of R-charge $0$. Such a theory can be used to define a boundary condition for the 5d $SU(N)_N$ gauge theory which preserves a $U(1)_\lambda$ symmetry, diagonal combination of $U(1)_{in}$ and $U(1)_\partial$, and an R-symmetry. 

The action of the duality wall on this boundary condition gives a new theory $\CB'$ built from $\CB$ by adding $N$ anti-fundamental chiral multiplets $q$ of $SU(N)$, gauging the overall $SU(N)$ global symmetry and adding the $W = b \det q$ superpotential. The new theory has the same type of mixed 't Hooft  anomalies as we required for $\CB$ (involving a new choice of $U(1)_\partial$ global symmetry).

In case of boundary conditions which break the gauge group to some subgroup $H$, we can apply a similar transformation, which 
only gauges the $H$ subgroup of $SU(N)$. For example, 
the duality wall maps Dirichlet boundary conditions, which fully break the gauge group at the boundary, to Neumann boundary conditions
enriched by the set of $N$ chiral quarks $q$ and the $b$ chiral field with $W = b \det q$, and vice-versa. \footnote{It may be possible to consider a larger set of boundary conditions, 
involving singular boundary conditions for the matter and gauge fields, akin to Nahm pole boundary conditions for maximally supersymmetric gauge theories \cite{Gaiotto:2008sa,Hashimoto:2014vpa}. }

We can provide a more entertaining example: a self-dual boundary condition. 
We define the boundary condition by coupling the five-dimensional gauge fields to $N+1$ 
quarks $q'$ and a single anti-quark $\tilde q'$. For future convenience, we also add 
$N+1$ extra chiral multiplets $M$ coupled by the superpotential 
\begin{equation}
\tilde q' q' M \ .
\end{equation}
Thus the boundary condition has an extra $SU(N+1) \times U(1)_e$ global symmetries defined at the boundary. 
The $SU(N+1)$ simply rotates $q'$ as anti-fundamentals and $M$ as fundamentals. The non-anomalous R-symmetry 
assignments are akin to the ones for a 4d SQCD with $N+1$ flavors. 

The bulk instanton symmetry can be extended to a non-anomalous symmetry under which the quarks have charge $1/N$ and anti-quarks have charge $-1/N$. 
The remaining non-anomalous boundary $U(1)_e$ will act on quarks with charge $1/N$, anti-quarks with charge $-1-1/N$ and on $M$ with charge $1$.

After acting with the duality interface, we find at the boundary four-dimensional $SU(N)$ gauge theory, with $N+1$ flavors given by the quarks $q'$ and anti-quarks $q$ and $\tilde q'$. The theory has a Seiberg dual description in the IR,
involving the mesons and baryons coupled by a cubic superpotential. The $W= b \det q + \tilde q' q' M$ lift the $\tilde q' q'$ mesons and the $\det q$ anti-baryon. The remaining $q q'$ mesons give $N+1$ new fundamental chiral at the boundary, the dual version of $q'$. The remaining anti-baryons give one anti-fundamental chiral, the dual version of $\tilde q'$.  The baryons give the dual version of $M$.  

We should keep track of the Abelian global symmetries. The dual quarks have instanton charge zero and $U(1)_e$ charge $1/N$. The dual anti-quarks have instanton charge $-1$ and $U(1)_e$ charge $-1-1/N$. The dual $M$ has instanton charge $1$ and $U(1)_e$ charge $1$.

In order for the self-duality to be apparent, we should re-define our instanton symmetry to act on the quarks $q'$ with charge $1/(2N)$, anti-quarks with charge $1/2-1/(2N)$, on $M$ with charge $-1/2$. Then the action of the duality interface switches the sign of the instanton charges, but leave $U(1)_e$ unaffected. 
It is natural to conjecture that this boundary condition descends from an $SU(2)_{in}$-invariant boundary condition for the UV SCFT, equipped with an extra $SU(N+1) \times U(1)_e$ global symmetry. 

We can generalize that to a duality-covariant interface $I_{N,N'}$ between $SU(N)_N$ and $SU(N')_{N'}$, coupled to three sets of four-dimensional chiral fields: $N+N'$ fundamentals $w$ of $SU(N)$, $N+N'$ anti-fundamentals $u$ of $SU(N')$ and a set of bi-fundamentals $v$ of $SU(N')$ and $SU(N)$, coupled by a cubic superpotential $W = u v w$. 

\begin{figure}[h]
    \centering
 \includegraphics[width=.5\textwidth]{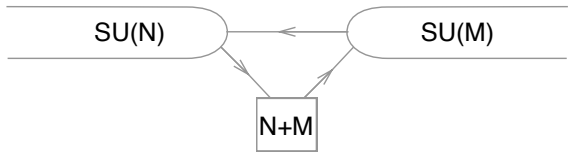}  
    \caption{A schematic depiction of the duality-covariant interface $I_{N,M}$. We include a superpotential coupling for the closed loop of three arrows.}
    \label{fig:suncov}
\end{figure}

If we act with an $SU(N)_N$ duality interface, we obtain a four-dimensional $SU(N)$ gauge theory with $N+N'$ flavors,
fundamentals $w$ and anti-fundamentals $v$ and $q$. Applying Seiberg duality, we arrive to an $SU(N')$ gauge theory with $N+N'$ flavors.
The original superpotential lifts the $u$ fields and the $v w$ mesons. The $b \det q$ superpotential maps to a similar $b \det q^\vee$
involving the Seiberg-dual quarks which transform under the five-dimensional $SU(N')_{N'}$ gauge fields.
The final result is identical as what one would obtain by acting with the $SU(N')_{N'}$ duality interface.

\begin{figure}[h]
    \centering
 \includegraphics[width=.99\textwidth]{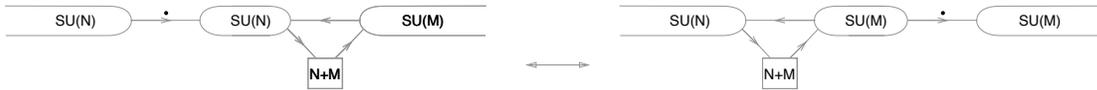}  
    \caption{The Seiberg duality transformation which implies the duality-covariance of $I_{N,M}$.}
    \label{fig:suncov2}
\end{figure}

The duality-covariant interfaces $I_{N,N'}$ have interesting properties under composition. 
Consider the composition of $I_{N,N'}$ and $I_{N',N''}$: it supports a four-dimensional $SU(N')$ gauge theory coupled to 
$N + N' + N''$ flavors, which include the $N+N'$ anti-fundamentals $u$, $N'+N''$ fundamentals $w'$, bifundamentals $v$ and $v'$. 
If we apply Seiberg duality, we find a new description of a composite interface, which is actually a modification of 
$I_{N,N''}$! Indeed, we find an $SU(N + N'')$ gauge group which is coupled to the 5d degrees of freedom just as the 
flavor group of $I_{N,N''}$, and is furthermore coupled to $N+N'$ fundamentals and $N'+N''$ anti-fundamentals
with a superpotential coupling to $(N + N')\times (N' + N'')$ mesons. This is consistent with the duality-covariance of the interface. 

The interface $I_{N,N'}$ clearly has an $SU(N + N')$ global symmetry. We can also define an $U(1)_e$ non-anomalous global symmetry, 
acting with charge $1$ on $v$, $-\frac{N'}{N+N'}$ on $w$ and $-\frac{N}{N+N'}$ on $u$. The second $U(1)_{in}$ global symmetry can be taken to act 
with charge $1$ on $w$, $-1$ on $u$ and charge $N+N'$ on instantons on the two sides. 

The $I_{N,N}$ duality-covariant interface is particularly interesting. It supports a baryon operator $\det v$ charged under $U(1)_e$ only. 
If we give it a vev, by a diagonal vev of $v$, we Higgs together the gauge fields on the two sides of the interface and the superpotential coupling 
gives a mass to $u$ and $w$. We arrive to a trivial interface. Later on in section \ref{sec:thooft} we will use $I_{N,N}$ to study the duality properties of 
of 't Hooft surface defects. 

\subsection{$SU(N)_{N-N_f/2}$ SQCD with $N_f < 2 N$ flavors}
\begin{figure}[h]
    \centering
 \includegraphics[width=.99\textwidth]{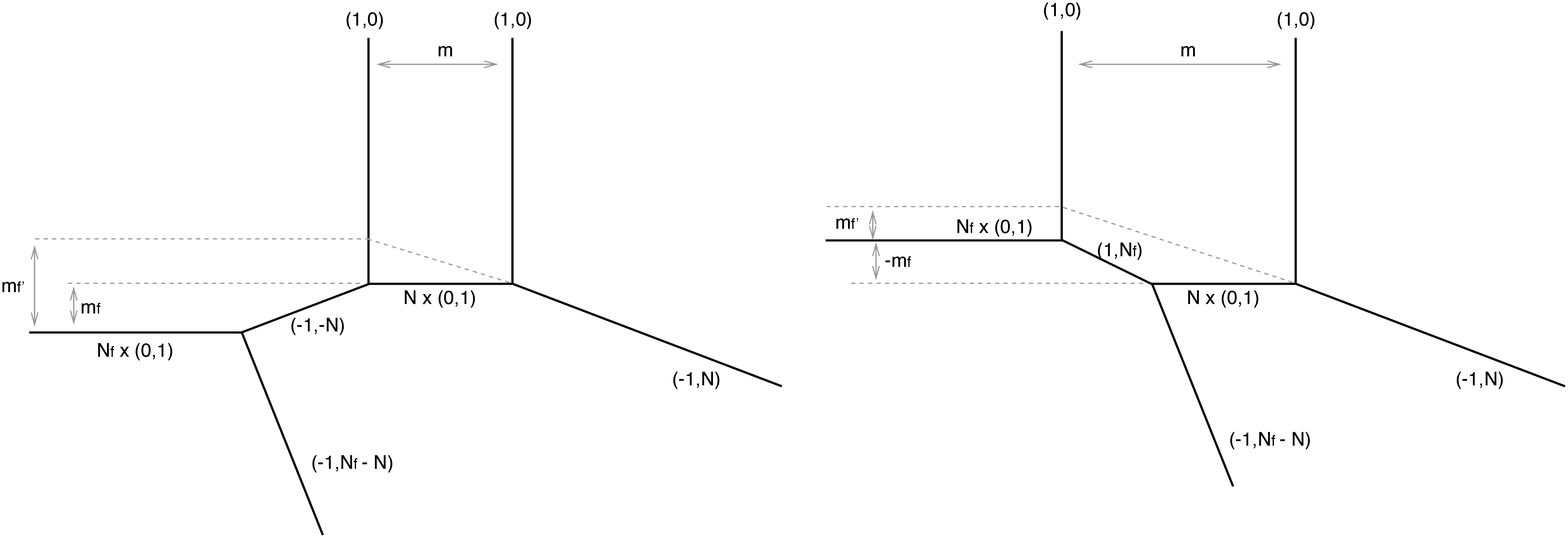}  
    \caption{The fivebrane web which engineers the UV completion of $SU(N)_{N-N_f/2}$ SQCD. The gauge theory is supported on the bundle of $N$ parallel D5 branes. After removing the centre of mass, the non-normalizable deformation are the separation $m$ between the NS5 branes and the vertical separation $m_f$ 
    between the semi-infinite D5 branes and the intersection of one of the NS5 branes and the $(-1,N)$ fivebrane. The latter parameter is the overall mass parameter for the hypermultiplets. We drew the resolved fivebrane web for positive and negative values of the overall hypermultiplet mass. The former is closely related, but not identical to the 
    gauge coupling or mass for $U(1)_{in}$. It is possible to argue that the instanton mass $m_i$ actually equals $m + \frac{N_f}{2} m_f$. The standard IR gauge theory description is valid for $m>0$ and $m + N_f m_f>0$.
    When $m$ becomes negative and we flip its sign to go to a dual parameterization, we exchange the roles of the $NS5$ branes and thus the role of $m_f$ and the auxiliary parameter $m'_f = m_f+ \frac{m}{N}$. 
    Alternatively, we can use $\frac{m_f + m_f'}{2}$ as a parameter, which remains invariant under duality }
    \label{fig:sqcd}
\end{figure}

A similar UV promotion of $U(1)_{in}$ to $SU(2)$ is expected to hold for $SU(N)_{N-N_f/2}$ 5d gauge theories with $N_f$ flavors, with $N_f < 2 N$. 
The SCFT can be engineered by a BPS five-brane web involving $N_f + 4$ semi-infinite 
external legs: two parallel NS5 branes, a $(-1,N)$ and a $(-1,N_f-N)$ fivebranes, $N_f$ D5 branes pointing to the left. 
The $SU(2)$ global symmetry is associated again to the two parallel NS5 branes, while the $N_f$ D5 branes support 
an $U(N_f)$ global symmetry. The fivebrane webs and mass parameters are depicted in figure \ref{fig:sqcd}.

As the gauge fields are IR free, we expect to be able to describe a typical half-BPS boundary condition for such gauge theories 
in terms of an $SU(N)$-preserving boundary conditions for the five-dimensional hypermultiplets, with a weak gauging of the five-dimensional $SU(N)$ symmetry. 
Of course, it is also possible to only preserve, and gauge, at the boundary some smaller subgroup $H$ 
of the five-dimensional gauge group. An extreme example would be to give Dirichlet boundary conditions 
to the gauge fields.

Half-BPS boundary conditions for five-dimensional free hypermultiplets may yet be strongly coupled. On general grounds \cite{Dimofte:2012pd}, it is always 
possible, up to D-terms, to describe such boundary conditions as deformations of simple boundary conditions which 
set a Lagrangian half of the hypermultiplet scalars (which we can denote as ``Y'') to zero at the boundary. 
The remaining hypers (which we can denote as ``X'') can be coupled to a boundary theory $\CB$ by a linear superpotential coupling 
\begin{equation}
W = X \CO
\end{equation}
involving some boundary operator $\CO$. This gives a boundary condition which we could denote as $\CB_X$. 

Conversely, if we are given some boundary condition $\CB_X$ for free hypermultiplets, we can 
produce a four-dimensional theory $\CB$ by putting the 5d hypers on a segment, with boundary conditions 
$\CB_X$ on one side and $X=0$ on the other side. Up to D-terms, this inverts the map $\CB \to \CB_X$,
with $\CO$ being the value of $Y$ at the $X=0$ boundary. 

In particular, a boundary condition $X=0$ can be engineered by a theory $\CB$ consisting of free chiral multiplets 
$\phi$ with the same quantum numbers as $Y$, and superpotential $W = X \phi$. The trivial interface can be obtained from 
a $Y=0$, $X'=0$ boundary condition by a $W=X Y'$ superpotential coupling, where the primed and un-primed fields 
live on the two sides of the interface. 

With these considerations in mind, we can evaluate the 't Hooft anomaly polynomial for a boundary condition 
$Y=0$: because of the symmetry between $X=0$ and $Y=0$, it must be exactly half of the 't Hooft anomaly polynomial
for a four-dimensional free chiral with the same quantum numbers as $X$. 

Our proposal for the duality interface generalizes the interface for pure $SU(N)_{N}$ gauge theory: we
set to zero at the boundary the fundamental half $X$ of the hypermultiplets on the right of the wall and 
anti-fundamental $Y'$ on the left of the wall, with a boundary superpotential 
\begin{equation}
W = b \det q + \Tr X' q Y \ .
\end{equation}
The combination of gauge anomalies from $q$ and the boundary condition for the hypermultiplet 
precisely matches the desired bulk Chern-Simons level $N - N_f/2$. We denote as $X$ the fields which transform as anti-fundamentals 
of $U(N_f)$. In particular, we give them charge $-1$ under the diagonal $U(1)_f$ global symmetry in $U(N_f)$. 

\begin{figure}[h]
    \centering
 \includegraphics[width=.5\textwidth]{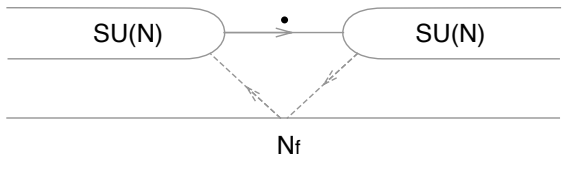}  
    \caption{Our schematic depiction of the duality wall for SQCD. We denote the 5d $SU(N_f)$ flavor group which goes through the interface as a strip. 
  The dashed arrows indicate which half of the bulk hypermultiplets survives at the wall. We include a superpotential coupling for the closed loop of three arrows.}
    \label{fig:sqcdwall}
\end{figure}

A consecutive pair of these conjectural duality walls can be analyzed just as in the pure gauge theory case, as the 
boundary conditions  prevent the five-dimensional hypers on the interval from contributing extra light four-dimensional fields. 
They can be integrated away to give a $\Tr X'' \tilde q q Y$ coupling. As the meson $q \tilde q$ is identifies with the identity 
operator in the IR, the interface flows to a trivial interface for both the gauge fields and the hypermultiplets, 
up to D-terms.  Thus the interface is a reasonable candidate for a duality wall.

\begin{figure}[h]
    \centering
 \includegraphics[width=.75\textwidth]{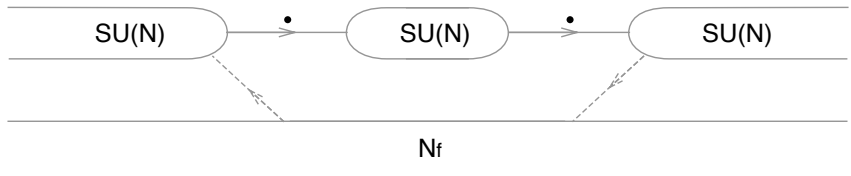}  
    \caption{A schematic depiction of the composition of two duality wall for SQCD. The resulting 4d $SU(N)$ gauge theory has $N$ flavors and at low energy it glues the two 5d gauge groups together.
    The theory includes a quartic superpotential coupling which arises from integrating away the hypermultiplets in the segment. In the IR, it glues together the hypermultiplets on the two sides of the interface}
    \label{fig:sqcdsquare}
\end{figure}

Next, we can look carefully at the anomaly cancellation conditions. It is useful to express the anomaly cancellation in terms of fugacities. 
If we ignore for a moment the R-charge and say that $q$ has fugacity $\lambda^{1/N}$, $X$ has fugacity $x$ and $X'$ has fugacity $x'$, 
the superpotential imposes $x = \lambda^{1/N} x'$, anomaly cancellation for the left gauge group sets the instanton fugacities on the right to 
$i_r = \lambda x^{- N_f/2}$ and $i_\ell = \lambda^{-1} (x')^{-N_f/2}$. 

We can re-cast the relation as a statement about one combination of bulk fugacity being inverted by the interface, 
$\lambda = i_r x^{N_f/2}$ and $\lambda^{-1} = i_\ell (x')^{N_f/2}$, and one being not inverted $i_r x^{N_f/2 - 2 N} = i_\ell (x')^{N_f/2 - 2 N}$. 

Although these relations may look unfamiliar, they can be understood in a straightforward way in therms of the $(p,q)$ fivebrane construction 
of $SU(N)_{N-N_f/2}$. Indeed, $\lambda$ is the fugacity which is associated to the mass parameter $m$ and 
$x^{-1}$ to $m_f$, $(x')^{-1}$ to $m_f'$. 

As far as R-symmetry is concerned, the bulk R-symmetry only acts on the scalar fields in the hypermultiplets, with charge $1$.
Thus we expect that assigning R-symmetry $0$ to $q$ and $2$ to $b$ will both satisfy anomaly cancellation and be compatible with the superpotential couplings. 

\begin{figure}[h]
    \centering
 \includegraphics[width=.5\textwidth]{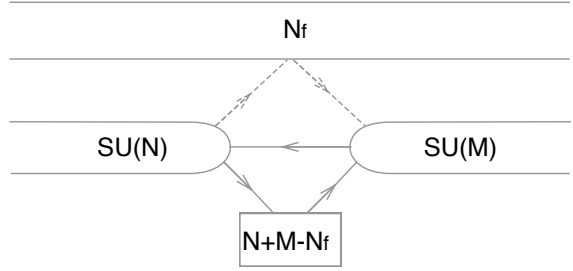}  
    \caption{A schematic depiction of the duality-covariant interface $I_{N,M}$. We include a superpotential coupling for the closed loops of three arrows.}
    \label{fig:sqcdcov}
\end{figure}

\begin{figure}[h]
    \centering
 \includegraphics[width=.99\textwidth]{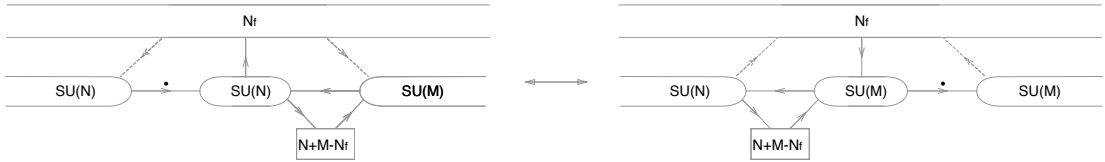}  
    \caption{The Seiberg duality transformation which implies the duality-covariance of $I_{N,M}$.}
    \label{fig:sqcdcov2}
\end{figure}

It is straightforward to extend to SQCD the duality-covariant boundary conditions and interfaces proposed for 
pure $SU(N)$ gauge theory. We refer to figure \ref{fig:sqcdcov} for the quiver description of the $I_{N,M}$
interface and to figure \ref{fig:sqcdcov2} for the Seiberg-duality proof of duality-covariance. 
The composition of $I_{N,M}$ and $I_{M,S}$ can again be converted to a modification of $I_{N,S}$. 

\subsection{Duality walls for $SU(N)$ with $N_f = 2 N$}
\begin{figure}[h]
    \centering
 \includegraphics[width=.5\textwidth]{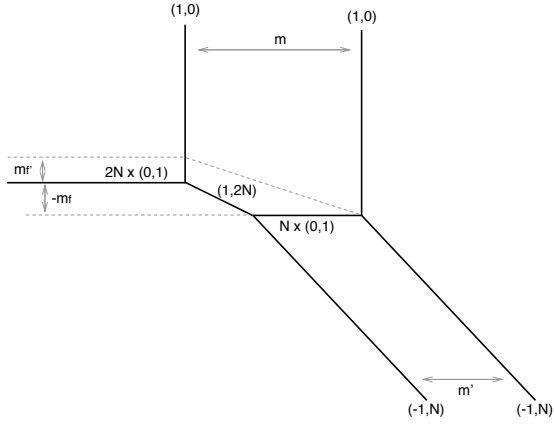}  
    \caption{The fivebrane web which engineers the UV completion of $SU(N)_{0}$, $N_f = 2N$ SQCD. 
    After removing the centre of mass, the non-normalizable deformation are the separation $m$ between the NS5 branes and the separation $\tilde m$ 
    between the $(-1,N)$ fivebranes. The vertical separation $m_f$ between the semi-infinite D5 branes and the intersection of one of the NS5 branes and the $(-1,N)$ fivebrane and instanton mass $m_i$ are related to $m$ and $m'$ as $m = m_i - N m_f$, $m' = m_i + N m_f$.}
    \label{fig:sqcd2n}
\end{figure}

The $SU(N)$ theory with $2N$ flavors is rather special: in the UV, two distinct Abelian global symmetries 
are expected to be promoted to an $SU(2)$. Essentially, they are the sum and difference of the instanton and baryonic 
$U(1)$ isometries. Correspondingly, we will find two commuting duality walls. In the fivebrane construction, the extra symmetry is due to two 
sets of parallel fivebranes. See figure \ref{fig:sqcd2n}

The first duality wall is defined precisely as before, 
i.e. set to zero at the boundary the fundamental half $X$ of the hypermultiplets on the right of the wall and 
anti-fundamental $Y'$ on the left of the wall, with a boundary superpotential 
\begin{equation}
W = b \det q + \Tr X' q Y \ .
\end{equation}

For the second wall, we replace $q$ with a set of bi-fundamental fields $\tilde q$ in the opposite direction, and 
set to zero at the boundary the anti-fundamental half $Y$ of the hypermultiplets on the right of the wall and 
fundamental $X'$ on the left of the wall, with a boundary superpotential 
\begin{equation}
W = \tilde b \det \tilde q + \Tr X \tilde q Y' \ .
\end{equation}

\begin{figure}[h]
    \centering
 \includegraphics[width=.75\textwidth]{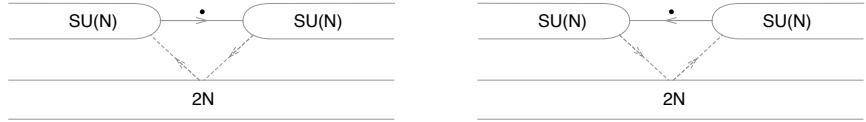}  
    \caption{The two duality walls for SQCD $N_f = 2N$. We include a superpotential coupling for the closed loop of three arrows.}
    \label{fig:sqcd2nwall}
\end{figure}

Both walls implement $Z_2$ symmetries: the composition of two walls of the same type flows to the identity, and 
they reflect one of the two fugacities $\lambda = i_r x^{N}$ or $\tilde \lambda = i_r x^{-N}$ while leaving the other one fixed. 

We can consider the concatenation of the two walls. That gives us a 4d $SU(N)$ gauge theory coupled to 
$q$, $\tilde q$ and the surviving half of the bulk hypermultiplet in the interval. If we pick one of the two possible orders of the composition, we find 
\begin{equation}
W = b \det q + \Tr X' q Y +\tilde b \det \tilde q + \Tr X' \tilde q Y''
\end{equation}
with $X'$ being a set of $2N$ fundamental chiral multiplets and $q$, $\tilde q$ anti-fundamentals. 

If we concatenate the walls in the opposite order, we find 
\begin{equation}
W = b \det q + \Tr X'' q Y' +\tilde b \det \tilde q + \Tr X \tilde q Y'
\end{equation}
with $Y'$ being a set of $2N$ anti-fundamentals and $q$ and $\tilde q$ fundamentals of the 4d gauge group. 

The two possibilities are precisely related by Seiberg duality! The mesons produced by the duality implement the 
switch in the boundary conditions for the hypermultiplets, and the baryons are re-mixed so that the $b$ and $\tilde b$ couplings 
match as well. Thus the two duality walls commute, as expected. 

\begin{figure}[h]
    \centering
 \includegraphics[width=.75\textwidth]{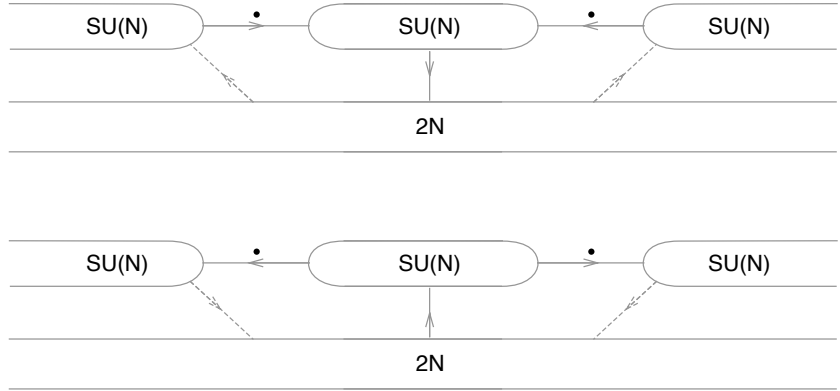}  
    \caption{The Seiberg duality demonstrating how the two duality walls for SQCD $N_f = 2N$ commute.}
    \label{fig:sqcd2ncomm}
\end{figure}

\subsection{Linear quivers}
\begin{figure}[h]
    \centering
 \includegraphics[width=.75\textwidth]{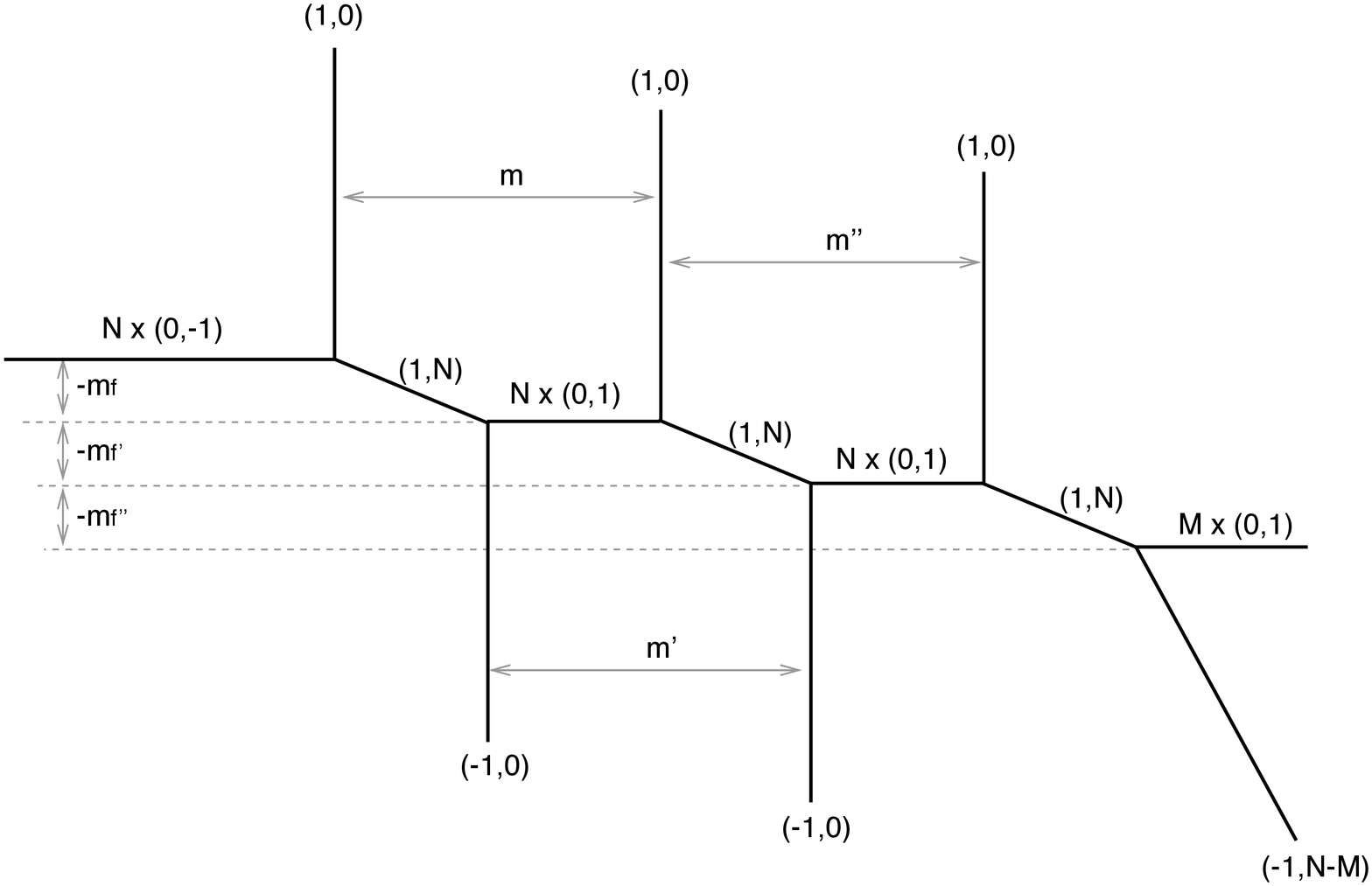}  
    \caption{The fivebrane web which engineers the UV completion of a $SU(N) \times SU(N)$ gauge theory with $N$ flavors at the left node and $M$ at the right node. 
    The five $U(1)$ global symmetries (two instanton symmetries and three hypermultiplet masses) are enhanced to $U(1)^2\times SU(2)\times SU(3)$
    because of the two sets of parallel fivebranes. The six mass deformations in the picture satisfy a relation: $m' = m + M m_f + N m_f'$}
    \label{fig:quiverweb}
\end{figure}

The duality walls we considered can be defined with minor changes in quiver gauge theories where 
one or more nodes satisfy a balancing condition $\pm \kappa = N_c - N_f/2$. In the language of fivebranes, 
if the quiver is engineered by a sequence of D5 brane stacks stretched between NS5 branes, 
the balancing condition insures that either the top pair of semi-infinite fivebranes associated to the gauge group are parallel, 
or the bottom. If $N_f = 2 N_c$ both pairs are parallel. See figure \ref{fig:quiverweb} for an example. 

A sequence of $k$ balanced nodes is expected to be associated in the UV to an $SU(k+1)$ global symmetry, enhancing a certain combination of 
the instanton and bi-fundamental hypermultiplet charges for these nodes. 

We want to understand the effect of a duality wall for a node of the quiver on the other nodes of the quiver, and figure out how 
the duality walls for different nodes match together. 

We can define the duality wall at a balanced node as we did for a single gauge group, 
leaving the other gauge groups and other hypermultiplets continuous at the interface.

\begin{figure}[h]
    \centering
 \includegraphics[width=.5\textwidth]{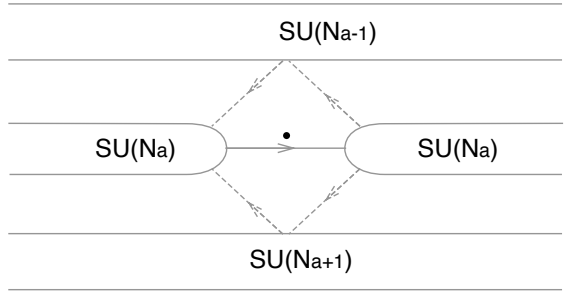}  
    \caption{Our schematic depiction of one of the duality walls for a quiver.}
    \label{fig:quiverwall}
\end{figure}

As the $X'$ and $Y$ fields for a given node are charged under the gauge groups at nearby nodes, but have different Abelian charges, 
in order for the corresponding symmetries to remain non-anomalous, we need to correct these Abelian charges by the instanton charge at the nearby nodes
on either sides of the interface. In terms of instanton fugacities, that means that the instanton fugacities at the nearby nodes will have to jump 
by the sum of the fugacities of $X'$ and $Y$, i.e. the fugacity $\lambda$ of $q$. That makes sense: the duality wall permutes 
two consecutive semi-infinite branes and the instanton symmetries at the other nodes are associated to the relative distance of nearby 
fivebranes. If we permute two fivebranes whose distance is associated to the fugacity $\lambda$, the distances from other fivebranes 
jump by plus or minus that distance and the fugacities jump by factors of $\lambda^{\pm 1}$. 

Let's denote the domain walls associated to nodes $a$ with positive balancing condition as $D^+_a$, and the ones associated to nodes
$a$ with negative balancing condition as $D^-_a$. If $N_f = 2 N_c$ at one node, both duality walls are available. 

It is easy to show that all $D^+_a$ commute with all the $D^-_a$. It is more interesting to show that each sequence of consecutive walls with the same sign 
satisfy the relations of a permutation group, i.e. $D^+_a D^+_{a+1} D^+_a = D^+_{a+1} D^+_a D^+_{a+1}$ and the same for $D^-_a$.

For the  $D^+_a D^+_{a+1} D^+_a = D^+_{a+1} D^+_a D^+_{a+1}$ relation, each side of the tentative equality gives rise to a 
four-dimensional $SU(N_a)$ gauge theory with $N_a+N_{a+1}$ flavors. For example, the left hand side gives 
\begin{equation}
W = b \det q + \Tr X' q Y + b' \det q' + \Tr X'' q' Y' + b'' \det q'' + \Tr X''' q'' Y'' \ .
\end{equation}
Seiberg duality appears to neatly exchange the interfaces corresponding to the two sides of the permutation group 
relation, up to a small mismatch concerning the $b' \det q'$ coupling for the intermediate interface in the composition:
$b'$ appears to couple on the two sides to two different operators with the same fugacities. The mismatch can likely be explained away by 
the possibility of operator mixing under Seiberg duality.  


\subsection{Exceptional symmetries in $SU(2)$ theories}
The UV completion of $SU(2)$ gauge theories with $N_f$ flavors is expected to have an enhanced $E_{N_f+1}$ global symmetry. 
This can be understood as a combination of the general UV enhancement for $SU(N)$ gauge theories and the 
enhancement of $U(N_f)$ to $SO(2N_f)$ due to the fact that the fundamental representation of $SU(2)$ is pseudo-real. 
Indeed, the $SU(2)$ enhancement involves a linear combination of $U(1)_{in}$ and the diagonal $U(1)$ subgroup of 
$U(N_f)$ and thus it combines non-trivially with the enhancement of $U(N_f)$ to $SO(2N_f)$. 

Correspondingly, we can find continuously many versions of our basic duality wall, each labelled by a choice of 
$U(N_f)$ subgroup in $SO(2N_f)$ and a splitting of the hypermultiplet scalars into $N$ ``X'' and $N$ ``Y'' complex scalar fields. 
It is most useful to look at domain walls which preserve a common Cartan sub-algebra of the global symmetry group,
implementing Weyl reflections in the UV.

If we denote the bulk quarks as $Q^i$, $i = 1, \cdots, 2 N_f$, we can consider duality walls for which the 
$X$ fields consist of $N_f - k$ quarks from the $i=1, \cdots, N_f$ range and $k$ quarks from the $i=N_f+1, \cdots, 2N_f$ range.
If we denote as $x_a$ the fugacities of the quarks, the overall fugacity of the $X$ fields will be defined as 
$x^{N_f} = \prod_{a \in X} x_a$. The domain walls invert $\lambda = i x^{N_f/2}$ and leave 
$i x^{N_f/2 - 4}$ and the ratios $x_a/x_{a'}$ for $a,a' \in X$ fixed. 

It is important to point out that not all splittings are simultaneously possible. 
There are two disconnected classes of choices of $X$ and $Y$ fields among the $Q^i$, 
distinguished by comparing the sign of their ``orientation'' $dX_1 dY_1 dX_2 dY_2 \cdots$. 
Intuitively, in order to interpolate between boundary conditions in different classes we need to add a single chiral doublet at the 
boundary, which contributes one unit to the discrete $Z_2$ gauge anomaly of $SU(2)$. Thus 
either boundary conditions with even $k$ are simultaneously non-anomalous, or boundary conditions with odd $k$
are simultaneously non-anomalous, but not both. 

Notice that $SU(2)$ gauge theories have no continuous theta angle, but have a discrete $Z_2$-valued theta angle. 
One unit of discrete $Z_2$ gauge anomaly at the boundary can be compensated by a shift of the bulk 
discrete theta angle. Thus we expect the two classes (even $k$ and odd $k$) of boundary conditions 
to be associated to the two different choices of bulk theta angle. 
Thus we have $2^{N_f-1}$ basic domain walls. 

In general, composing two such domain walls associated to splittings $(X,Y)$ and $(X',Y')$ 
will give an interface supporting an 4d $SU(2)$ gauge theory, 
with as many chiral quarks as the number of bulk flavors which belong to $X$ and $Y'$ (or equivalently $X'$ and $Y$). 
The relations in the Weyl group of $E_{N_f+1}$ must correspond to Seiberg-like dualities in the corresponding domain wall theories. 

For reasons of space, we will only verify these for the simplest non-trivial example, $N_f=2$.
In this case we have two basic duality walls, one involving $Q^1$ and $Q^2$, the other involving $Q^3$ and $Q^4$.
Both preserve the same $SU(2)$ subgroup of the $SO(4)$ global group, and mix the instanton symmetry with the other $SU(2)$ subgroup
to an $SU(3)$.  

At the level of fugacities, the first wall matches $i_r x =( i_\ell x')^{-1}$ and
$i_r x^{-3} = i_\ell (x')^{-3}$, while the second matches $i_r x^{-1} =i_\ell^{-1} x'$ and
$i_r x^{3} = i_\ell (x')^{3}$.

If we concatenate the two walls, the intermediate $SU(2)$ 4d gauge group will be coupled to three flavors, i.e. the six 
doublets $q$, $\tilde q$, $Q^1$, $Q^2$. In the IR, they will flow to a set of $15$ mesons. Two of them will be lifted by 
$b$ and $\tilde b$ and eight simply flip the boundary condition on the left and right hypermultiplets
so that we are left with $Q^1$ and $Q^2$ at both boundaries. The remaining ones give a set of bi-fundamental fields between the left and right gauge groups and a neutral singlet. The Pfaffian superpotential involving the $15$ mesons couples the singlet to the determinant of the bifundamental field and couples the bi-fundamental to the boundary values of the hypermultiplet. 

The final result is again a duality wall, combined with a permutation of the $Q^1$, $Q^2$ quarks with the 
$Q^3$, $Q^4$ quarks on one side of the wall. If we denote the two original duality walls as $D_1$ and $D_2$, 
and the trivial duality wall permuting the two sets of quarks as $D_3$, we find the relations
\begin{equation}
D_1 D_2 = D_2 D_3 = D_3 D_1  \qquad , \qquad D_2 D_1 = D_3 D_2 = D_1 D_3
\end{equation}
which agree well with the properties of the three permutations in $S_3$, the Weyl group of $SU(3)$. 

\section{Index calculations}
In this section, we consider the superconformal index (SCI) and the hemisphere index of a 5d SCFT at the UV fixed point. The superconformal index is a trace over the BPS operators in the CFT on $\mathbb{R}^D$, or over the BPS states on a sphere $S^{D-1}$ times $\mathbb{R}$ via the radial quantization~\cite{Kinney:2005ej}.  In $D=5$ dimensions, it is defined as~\cite{Kim:2012gu}
\begin{equation}
    I(w_a,\mathfrak{q};p,q) = {\rm Tr}(-1)^F p^{j_1+R}q^{j_2+R}\prod_{a}w_a^{F_a}\, \mathfrak{q}^k \ .
\end{equation}
$j_1, j_2$ and $R$ are the Cartan generators of the $SO(5)\times SU(2)_R$ bosonic algebra and $p,q$ are their fugacities. $F_a$ are the Cartans of the global symmetries visible in the classical Lagrangian and $w_a$ are the corresponding fugacities. $k$ is the instanton number and its fugacity is $\mathfrak{q}$.
This index can also be considered as a twisted partition function on $S^1\times S^4$, which was computed in~\cite{Kim:2012gu,Terashima:2012ra} using supersymmetric localization.

The hemisphere index is the supersymmetric partition function on an half of the sphere $D^4\subset S^4$ times $S^1$ with a specific boundary condition of the $D^4$. We can also interpret it as an index counting the BPS states on $S^1\times \mathbb{R}^4$ with Omega deformation, introduced in~\cite{Nekrasov:2002qd}. The deformation parameters $\epsilon_{1,2}$ are identified with the above fugacities as $p=e^{-\epsilon_1},q=e^{-\epsilon_2}$.
Roughly speaking, this index is an half of the superconformal index and thus the full sphere index (or SCI) can be reconstructed by gluing two hemisphere indices.
We will now use these indices to test our duality proposal.

\subsection{$SU(N)_N$ theories}
Let us begin by pure $SU(N)_N$ gauge theories. The hemisphere index with Dirichlet b.c. is given by
\begin{equation}
	II^N(z_i,\lambda;p,q) = (pq;p,q)^{N-1}_\infty \prod_{i\neq j}^N(pqz_i/z_j;p,q)_\infty Z_{\rm inst}^N(z_i,\lambda;p,q) \ .
\end{equation}
The ``gauge fugacity'' $z_i$ becomes here the fugacity of the boundary global symmetry.
$Z_{\rm inst}^N$ is the singular instanton contribution localized at the center of the hemisphere. 

The gauge theory on the full sphere can be recovered from two hemispheres with Dirichlet boundary conditions by gauging the diagonal $SU(N)$ boundary global symmetry. So the full sphere index can be written as
\begin{equation}
	I^N(\lambda;p,q) = \langle II^N| II^N \rangle \equiv  I_{V}^{N-1}\oint \frac{d\mu_{z'_i}}{\prod_{i\neq j}^N\Gamma(z_i/z_j)}  \overline{II^N(z_i,\lambda;p,q)}  II^N(z_i,\lambda;p,q)\ .
\end{equation}
The integrand includes the contribution of the 4d gauge multiplet, with $I_V \equiv (p;p)_\infty(q;q)_\infty$ being the contribution of the Cartan 
elements. The integration measure is simply $\frac{dz}{2 \pi i z}$. The overline indicates a certain  operation of ``complex conjugation'', which inverts all
gauge/flavor fugacities.

Other boundary conditions or interfaces can be obtained from Dirichlet boundary conditions by adding boundary/interface degrees of freedom and gauging the appropriate diagonal boundary global symmetries.
For example, if $I^{4d}_{N,M}(z_i, z_i';p,q)$ is the superconformal index of some interface degrees of freedom for an interface between $SU(N)$ and $SU(M)$ gauge theories, 
the sphere index in the presence of the interface becomes 
\begin{align}
&\langle I^{4d}_\CB| II^N \rangle \equiv  \langle II^N| \hat I^{4d}_{N,M}| II^M \rangle \equiv  I_{V}^{N+M-2} \cdot \\
&\cdot \oint \frac{d\mu_{z_i}}{\prod_{i\neq j}^N\Gamma(z_i/z_j)} \frac{d\mu_{z'_i}}{\prod_{i\neq j}^N\Gamma(z'_i/z'_j)}  \overline{II^N(z_i,\lambda;p,q)} I^{4d}_{N,M}(z_i, z_i';p,q) II^M(z'_i,\lambda;p,q)\ . \nonumber 
\end{align} \footnote{One can bring the 4d index under the conjugation. The inversion of fugacities can be understood as the difference in sign 
which appears when matching 5d and 4d fugacities for left or right boundary conditions}

Hemisphere indices, or sphere indices with an interface insertion, can be thought of as counting the number of boundary or interface local operators 
in protected representations of the superconformal group. 

Before going on, we should spend a few words on how to compute the correct instanton contribution $Z_{\rm inst}^N$ to the localization formula. 
The partition function is computed by equivariant localization on the moduli space of instantons. The instanton moduli spaces have singularities, 
whose regularization can be thought of as a choice of UV completion for the theory. The standard regularization for unitary gauge group 
is the resolution/deformation produced by a noncommutative background, or by turning on FI parameters in the ADHM quantum mechanics~\cite{Nakajima:1994nu,Nekrasov:1998ss}.

In principle, the standard regularization may not be the correct one to make contact with the partition function of a given UV SCFT. 
For SCFTs associated to $(p,q)$ fivebrane webs, the standard regularization is expected to be almost OK~\cite{Hayashi:2013qwa,Bao:2013pwa}: the correct instanton partition function is conjectured 
to be same as the standard instanton partition function up to some overall correction factor, independent of gauge fugacities and 
precisely associated to the global symmetry enhancement of the UV SCFT: each pair of parallel $(\pm 1,q)$ semi-infinite fivebranes  
contributes a factor of \footnote{PE$[f]$ denotes the plethystic exponent of single-letter index $f$.}
\begin{equation}
Z_{\rm extra}(\eta;p,q) = {\rm PE}\left[\frac{-\eta}{(1-p)(1-q)} \right] 
\end{equation}
to the correction factor, where $\eta$ is the fugacity for the global symmetry associated to the mass parameter corresponding to the separation between 
the parallel $(\pm 1,q)$ semi-infinite fivebranes. This correction factor has been extensively tested against the expected global symmetry enhancement 
of the superconformal indices. It appears to account for the decoupling of the massive W-bosons living on the six-dimensional world-volume of the semi-infinite fivebranes. 

The standard instanton partition function computed by using equivariant localization of~\cite{Nekrasov:2002qd,Nekrasov:2003rj} result takes the following contour integral form
\begin{align}
	\mathcal{Z}_{\rm QM}^N(z_i,\mathfrak{q};p,q) = \sum_{k=0}^\infty \mathfrak{q}^k \frac{(-1)^{k N}}{k!}\oint \prod_{I=1}^k\frac{d\phi_I}{2\pi i} e^{-\kappa\sum_{I=1}^k\phi_I} Z_{\rm vec}(\phi_I,z_i;p,q) \ ,
\end{align}
where $\kappa=N$ is the classical CS-level. The vector multiplet factor $Z_{\rm vec}$ is given in~(\ref{eq:SU-instanton-vector}).  It is known that the integral should be performed by using the Jeffrey-Kirwan (JK) method, which is first introduced in~\cite{1993alg.geom..7001J} and later derived in~\cite{Benini:2013xpa} for 2d elliptic genus calculations. See \cite{Hwang:2014uwa,Hori:2014tda,Cordova:2014oxa} for applications to 1d quantum mechanics and a detailed discussion of contour integrals.
See also appendix~\ref{sec:instanton-partition-function} for details on instanton partition functions.

The correction factor from the parallel semi-infinite NS5-branes is
\begin{equation}\label{eq:U(1)-factor}
  Z_{\rm extra}(\mathfrak{q};p,q) = {\rm PE}\left[\frac{-\,\mathfrak{q}}{(1-p)(1-q)} \right]  \ .
\end{equation}
Let us leave a few comment on this correction factor. This factor can also be read off from the residues $R_{\pm\infty}$ at infinity $\phi_I=\pm\infty$. $R_{\pm\infty}$ are associated to the noncompact Coulomb branch parametrized by vevs $\phi_I$ of the scalar fields in the vector multiplet. In fact, the above contour integral contains the contribution from the degrees of freedom along this Coulomb branch and it is somehow encoded in the $R_{\pm\infty}$.
The extra contribution is roughly an `half' of the $R_{\pm\infty}$. The residue at the infinity is in general given by a sum of several rational functions of $p,q$.
The `half' here means that we take only an half of them such that it satisfies two requirements: when we add it to the standard instanton partition function, 1) the full instanton partition function becomes invariant under inverting $x\equiv \sqrt{pq}$ to $x^{-1}$ and 2) it starts with positive powers of $x$ in $x$ expansion. 
The second requirement follows from the fact that the BPS states captured by the instanton partition function have positive charges under the $SU(2)$ associated to $x$.
This half then gives the extra contribution from the Coulomb branch and it also  coincides with the correction factor (\ref{eq:U(1)-factor}). We will see similar correction factors in the other examples below.

Since the Coulomb branch of the ADHM quantum mechanics dose not belong to the instanton physics of the 5d QFT, we should remove its contribution to obtain a genuine 5d partition function. So the correct instanton partition function of the 5d SCFT is expected to be
\begin{equation}
	Z^N_{\rm inst} (z_i,\lambda;p,q) = \mathcal{Z}_{\rm QM}^N(z_i,\lambda;p,q)/Z_{\rm extra}(\lambda;p,q) \ ,
\end{equation}
with $\mathfrak{q}=\lambda$ in this case.

At this point, we are ready to study the duality interface. The easiest way to do so is to look at the boundary condition obtained by acting 
with the duality interface on a Dirichlet boundary, i.e. the dual of Dirichlet boundary conditions. This consists of the duality interface degrees of freedom coupled to a single $SU(N)_N$ gauge theory, with the 
second $SU(N)$ global symmetry left ungauged. More general configurations can be obtained immediately by gauging that $SU(N)$ global symmetry. 

The 4d superconformal index of the duality interface degrees of freedom is simply
\begin{equation}
  \frac{\prod_{i,j=1}^N \Gamma(\lambda^{1/N} z_i/z'_j)}{\Gamma(\lambda)} \ ,
\end{equation}
where $z_i$ and $z'_i$ are the fugacities for the gauge group on the left and right of the wall.
The numerator factor comes from the bi-fundamental chiral multiplet $q$ and the denominator is from the singlet chiral multiplet $b$. 
The anomaly-free $U(1)_\lambda$ symmetry, which is a linear combination of $U(1)_{in}$ instanton symmetry and $U(1)_B$ baryonic symmetry, rotates the baryon operator $B={\rm det}\,q$ by charge $1$, so $B$ comes with the fugacity $\lambda$.
The contribution of $b$ is precisely the inverse of the contribution of a chiral multiplet with the same R-charge and fugacity as $B$. 

Thus the hemisphere index for dual Dirichlet boundary conditions is:
\begin{equation}\label{eq:duality-action}
  \hat{D}II^N \equiv I_{V}^{N-1}\oint \prod_{i=1}^{N-1}\frac{dz'_i}{2\pi iz'_i} \frac{ \prod_{i,j=1}^N \Gamma(\lambda^{1/N} z_i/z'_j) }{ \Gamma(\lambda) \prod_{i\neq j}^N\Gamma(z'_i/z'_j)} II^N(z'_i,\lambda;p,q) \ .
\end{equation}

If we have identified the correct duality interface, the hemisphere index for dual Dirichlet b.c. should actually coincide with the 
hemisphere index for Dirichlet b.c., up to a reflection of $U(1)_{in}$ instanton charges, i.e. an inversion of the instanton fugacity $\lambda\rightarrow \lambda^{-1}$. 
This motivates us to propose the following relation:
\begin{eqnarray} \label{eq:sunint}
  \hat{D}II^N(z_i,\lambda;p,q)=II^N(z_i,\lambda^{-1};p,q) \ .
\end{eqnarray}
This is a highly nontrivial relation. The instanton partition function in the hemisphere index on the right side of the wall has a natural expansion by positive powers of the instanton fugacity $\lambda$. On the other hand, the instanton partition function on the left side of the wall is expanded by the negative powers of $\lambda$. This relation is a very stringent test of our conjectural duality wall. 

We can test this conjectural relation for small $N$ and the first few orders in the power series expansion in $p$,$q$. 
We find that the relation holds with a particular choice of the integral contours. 
The contour should be chosen by the condition: $|p|,|q| \ll \lambda < 1$ while keeping the contour to be on a unit circle.
One can then check the duality relation order by order in the series expansion of $x \equiv \sqrt{pq}$.

For $SU(2)$ case, one finds
\begin{eqnarray}
  &&\hat{D}II^{N=2}(\lambda^{-1})= II^{N=2}(\lambda)  \\
   &=& 1 +(-\chi^{SU(2)}_{\bf 3}(z)+\lambda)x^2 + \chi^{SU(2)}_{\bf 2}(y)(-\chi^{SU(2)}_{\bf 3}(z)+\lambda)x^3  \nonumber \\
  && +\left(\left(1-\chi^{SU(2)}_{\bf 3}(y)\right)\chi^{SU(2)}_{\bf 3}(z)+\chi^{SU(2)}_{\bf 3}(y)\lambda+\lambda^2\right)x^4+\mathcal{O}(x^5) \ ,\nonumber
\end{eqnarray}
where $y\equiv \sqrt{p/q}$ and $\chi^{SU(N)}_{\bf r}(z)$ are the characters of dimension ${\bf r}$ representations of $SU(N)$ symmetry.
We have actually checked this relation up to $x^7$ order.

Similarly, for $SU(3)_3$ case, one finds
\begin{eqnarray}  
  &&\hat{D}II^{N=3}(\lambda^{-1}) = II^{N=3}(\lambda) \\
  &=& 1+(-\chi_{\bf8}^{SU(3)}(z) + \lambda)x^2+\chi^{SU(2)}_{\bf 2}(y)(-\chi_{\bf8}^{SU(3)}(x)+\lambda)x^3+\chi^{SU(2)}_{\bf 3}(y)\lambda x^4  \nonumber \\
  &&+\left(\chi_{\bf 10}^{SU(3)}(z) + \chi_{\bf \bar{10}}^{SU(3)}(z) + \chi_{\bf 8}^{SU(3)}(z)(1-\chi^{SU(2)}_{\bf 3}(y)-\lambda) +\lambda^2\right)x^4 + \mathcal{O}(x^5) \ , \nonumber
\end{eqnarray}
which is checked up to $x^5$ order.

The integral equation (\ref{eq:sunint}) is actually very constraining. We found experimentally that  
as long as we postulate 
\begin{equation}
II^N(z_i,\lambda;p,q) = 1 + \mathcal{O}(x) \ ,
\end{equation}
with positive powers of $\lambda$ only, we can use the integral equation order by order in $x$ to systematically 
reconstruct the full partition function. This is also the case for the hemisphere index with matters which we will now discuss.

\subsubsection{Example: $SU(N)_{N-N_f/2}$ theories with $N_f$ flavors}
The generalization to theories with flavors is straightforward. We first need to specify boundary conditions for the bulk hypermultiplets. We will use the boundary condition which sets the half of the hypermultiplet $Y$ to zero and couples the other half $X$ to the duality wall. The theory has the classical Chern-Simons coupling at level $\kappa\!=\!N\!-\!\frac{N_f}{2}$ that provides $N-\frac{N_f}{2}$ units of the cubic gauge anomaly. Given the boundary condition, the half of the hypermultiplet $X$ provides additional $\frac{N_f}{2}$ units of the cubic anomaly so that the total bulk cubic anomaly becomes $\kappa + \frac{N_f}{2} = N$. This will be exactly canceled by the boundary cubic anomaly when coupled to the duality wall.

The hemisphere index associated with this boundary condition is given by
\begin{equation}
  II^{N,N_f}(z_i,w_a,\mathfrak{q};p,q) = \frac{ (pq;p,q)^{N-1}_\infty\prod_{i\neq j}^N (pqz_i/z_j;p,q)_\infty }{ \prod_{i=1}^N\prod_{a=1}^{N_f} (\sqrt{pq}z_i/w_a;p,q)_\infty } Z^{N,N_f}_{\rm inst}(z_i,w_a,\mathfrak{q};p,q)  \ ,
\end{equation}
where $w_a$ are the $U(N_f)$ flavor fugacities.
The denominator factor in the 1-loop determinant is the contribution from the $X$.
The instanton partition function is the partition function of the ADHM quantum mechanics with additional degrees coming from the hypermultiplets. It is given by
\begin{align}
	\mathcal{Z}_{\rm QM}^{N,N_f}(z_i,\mathfrak{q};p,q) = \sum_{k=0}^\infty \mathfrak{q}^k \frac{(-1)^{k(N+N_f)}}{k!}\oint \prod_{I=1}^k\frac{d\phi_I}{2\pi i} e^{-\kappa\sum_{I=1}^k\phi_I} Z_{\rm vec}(\phi_I,z_i;p,q) \cdot \prod_{a=1}^{N_f}Z_{\rm fund}(\phi_I,m_a) \ ,
\end{align}
where $\kappa=N-N_f/2$ and $Z_{\rm fund}$ is the hypermultiplet contribution given in~(\ref{eq:SU-instanton-fund}). 
This partition function also contains correction factors associated to the Coulomb branch in the ADHM quantum mechanics. As explained in the previous subsection, the correction factor can be read off from the residues $R_{\pm\infty}$ at infinity $\phi_I=\pm \infty$, which is given by
\begin{align}\label{eq:U(1)-factor-SU(N)-flavors}
  Z_{\rm extra}^{N_f<2N}(w_a,\mathfrak{q};p,q) &= {\rm PE}\left[\frac{-\mathfrak{q}\prod_{a=1}^{N_f}w_a^{1/2}}{(1-p)(1-q)} \right] \ , \cr
  Z_{\rm extra}^{N_f=2N}(w_a,\mathfrak{q};p,q) &= {\rm PE}\left[\frac{-\mathfrak{q}\prod_{a=1}^{N_f}w_a^{1/2}}{(1-p)(1-q)} + \frac{-pq\,\mathfrak{q}\prod_{a=1}^{N_f}w_a^{-1/2}}{(1-p)(1-q)}\right] \ . 
\end{align}
When $N_f<2$, $R_{+\infty}$ is trivial and the single term in the letter index comes from the half of $R_{-\infty}$, whereas, when $N_f=2N$, each half of $R_{\pm\infty}$ gives each term in the letter index.

Let us couple it to the boundary theory. As explained above, we will multiply the contributions from the  4d vector and chiral multiplets living at the boundary, and integrate the gauge fugacities. Thus the dual of $Y=0$ Dirichlet boundary conditions is
\begin{equation}
  \hat{D} II^{N,N_f} \equiv \mathcal{I}_V^{N-1}\oint \prod_{i=1}^{N-1}\frac{dz'_i}{2\pi iz'_i}  \frac{ \prod_{i,j=1}^N \Gamma(\lambda^{1/N} z_i/z'_j) }{ \Gamma(\lambda) \prod_{i\neq j}^N\Gamma(z'_i/z'_j)} II^{N,N_f}(z'_i,w_a,\mathfrak{q};p,q) \ .
\end{equation}
In this notation the fugacity which is inverted by the duality operation is $\lambda$, defined by $\mathfrak{q}\equiv \lambda\prod_{a=1}^{N_f}w_a^{-1/2}$.

If we identified the correct duality wall, we should find 
\begin{equation}\label{eq:duality-hemisphare-flavor}
  \hat{D}II^{N,N_f}(z_i,w_a,\lambda;p,q) =   II^{N,N_f}(z_i,w'_a,\lambda^{-1};p,q) \ .
\end{equation}
The boundary superpotential $W=YqX'$ implies that the flavor fugacities in two sides of the wall should be identified as $w_a=\lambda^{1/N}w'_a$.

Again, this relation can be checked explicitly order by order in $x$ expansion: for example, we obtain 
\begin{align}
	&\hat{D}II^{2,2}(z,\lambda^{-1/2}w_a,\lambda^{-1}) = II^{2,2}(z,w_a,\lambda)\cr
	=& 1+ \chi^{SU(2)}_{\bf 2}(z)(w_1^{-1}+w_2^{-1})x + \chi^{SU(2)}_{\bf 2}(y)\chi^{SU(2)}_{\bf 2}(z)(w_1^{-1}+w_2^{-1})x^2 \\
	& + \left(\chi^{SU(2)}_{\bf 3}(z)(w_1^{-2}+(w_1w_2)^{-1}+w_2^{-2}-1) + (w_1w_2)^{-1}+\lambda+(w_1w_2)^{-1}\lambda\right)x^2 +\mathcal{O}(x^3) \nonumber \ ,
\end{align}
for $SU(2)$ with 2 flavors, and
\begin{align}
	&\hat{D}II^{3,1}(z,\lambda^{-1/3}w_1,\lambda^{-1}) = II^{3,1}(z,w_1,\lambda)\\
	=& 1+ \chi^{SU(3)}_{\bf 3}(z_i)w_1^{-1}x + \left(\chi^{SU(2)}_{\bf 2}(y)\chi^{SU(3)}_{\bf 3}(z_i)w_1^{-1}-\chi^{SU(3)}_{\bf 8}(z_i) + \chi^{SU(3)}_{\bf 6}(z_i)w_1^{-2} + \lambda \right)x^2+\mathcal{O}(x^3) \nonumber \ ,
\end{align}
for $SU(3)$ with 1 flavor. We have checked there relations up to $x^5$ order.

Again, the integral equation~(\ref{eq:duality-hemisphare-flavor}) is powerful enough so that we can reconstruct the full instanton partition function with fundamental matters order by order in $x$ expansion if we assume a natrual ``boundary condition'' as
\begin{equation}
    II^{N,N_f}(z_i,w_a,\mathfrak{q};p,q) = 1+ \mathcal{O}(x) \ .
\end{equation}

\section{Wilson loops}\label{sec:Wilson-Loops}
In this section we will use our duality walls in order to investigate the duality properties of line defects in 
the corresponding five-dimensional gauge theories. A BPS line defect intersecting (or ending on) a BPS boundary condition 
preserves the same supersymmetry as a chiral operator in a 4d gauge theory. 

A line defect which crosses our UV BPS Janus configuration will flow in the IR to a a pair of line defects in the two IR gauge theories on the two sides of the wall, meeting at a chiral local operator at the interface.  As the R-charge and global symmetries appear to be preserved under the RG flow,
that local operator should have zero global and R-charges. If the UV line defect preserves the enhanced UV global symmetry, then the IR line defects on the two sides of the wall will be identical. 

The natural BPS line defects in gauge theory are Wilson loop operators. A fundamental Wilson line can end on the duality wall 
on a $q$ local operator and then continue as a fundamental Wilson line on the other side of the wall. The intersection, though, 
would have R-charge $0$ but charge $1/N$ under the global symmetry $U(1)_\lambda$ inverted by the wall. There is an obvious way to ameliorate the problem: 
combine the gauge Wilson loop with a flavor Wilson loop for the $U(1)_\lambda$ symmetry, with flavor charge $-1/(2N)$. This completely cancels 
the global charges of $q$. 

At first sight, a dressing charge of $-1/(2N)$ may seem off-putting. It becomes natural though, if we imagine the Wilson loop to be the trajectory 
of a massive BPS particle in a theory with an extra massive flavor: we have seen in earlier sections that a duality-covariant charge assignment 
attributes $U(1)_\lambda$ charge $-1/(2N)$ to the hypermultiplets. 

It is also natural if we look at which BPS local operators may live at the end of 
the line defect. If the line defect has a $SU(2)_\lambda$-invariant UV completion, we expect the BPS local operators 
to fill up $SU(2)_\lambda$ representations. A bare fundamental Wilson loop can end on a hypermultiplet scalar in the anti-fundamental representation, 
but the resulting local operator has an inappropriate $U(1)_\lambda$ charge $1/(2N)$. 
If we dress the Wilson loop with the $U(1)_\lambda$ flavor Wilson loop, the hypermultiplet scalar at the end of the line defect becomes neutral under  
$U(1)_\lambda$.

If the theory has flavor, it is also possible to replace the $U(1)_\lambda$ flavor Wilson loop with an anti-fundamental $U(N_f)$ flavor Wilson loop. 
This is a better option for $N_f = 2N$, as the resulting loop has the correct properties to be invariant under both UV $SU(2)$ global symmetries. 

Next, we will test the hypothesis that the fundamental Wilson loop admits an $SU(2)_\lambda$-invariant UV completion 
with the help of the index. 

\subsection{Hemispheres with Wilson loop insertions}

We consider hemisphere indices with a Wilson loop insertion at the North pole. The BSP Wilson loops are inserted at the center of the hemisphere and wrap the $S^1$ circle. These enriched hemisphere indices count local operators sitting at the intersection of the Wilson loop and a BPS 
boundary. Similar statements apply for sphere partition functions, with or without the insertion of an interface, with Wilson loop insertions at the North pole, 
South pole or both. 

The main challenge in this calculation is to find the instanton partition function in the presence of a Wilson loop. 
Abstractly, a Wilson loop measures the gauge bundle at the origin and should be represented in the localization 
integral by the equivariant Chern character in the corresponding representation of the universal principal bundle over the instanton moduli space. 
Of course, as the instanton moduli space is singular, we face the usual regularization problem, with extra complications: even after we pick a regularization of the moduli space, we need to pick a regularization of the universal bundle over it.  
  
A canonical answer is well-known for fundamental Wilson loop insertions. It can be interpreted in terms of a modification of the ADHM construction, 
which adds extra fermionic matters which build the universal bundle in the fundamental representation (and antisymmetric powers of the fundamental representation as well) ~\cite{Tong:2014cha}. 

As general representations can be obtained as tensor powers of the fundamental representation, one can produce candidate 
equivariant Chern characters for these representations from the equivariant Chern character in the fundamental representation ~\cite{Shadchin:2004yx}. 
The equivariant Chern character of the universal bundle in the fundamental representation is given in~(\ref{eq:equivariant-Chern-E}). A list of equivariant Chern characters for the symmetric and antisymmetric, and adjoint representations are given in~(\ref{eq:equivariant-Chern-ex}).
One can compute those for other representations using the same method.

Given the Wilson loop and its equivariant Chern character, the equivariant localization states that the instanton partition function becomes
\begin{equation}\label{eq:Wilson-loop}
  \mathcal{W}_{QM,R}(z,\mathfrak{q};p,q) = Z_{\rm 1-loop}(z;p,q)\sum_{k=0}^\infty\mathfrak{q}^{k}\frac{1}{|W_k|} \oint [d\rho] \, Ch_{R}(z,\rho;p,q) \cdot Z_k(z,\rho;p,q) \ ,
\end{equation}
where $Ch_{R}$ is the equivariant Chern character of the universal bundle in representation $R$. $Z_{\rm 1-loop}$ is the 1-loop determinant and $Z_k$ is the $k$-instanton contribution without the Wilson loop. The contour integral needs to be evaluated using the JK-residue prescription. 

This answer may differ from the ``correct'' answer for a given UV completion of the theory 
 both by the usual overall correction factor (\ref{eq:U(1)-factor}) (or (\ref{eq:U(1)-factor-SU(N)-flavors})) and 
by extra corrections specific to the Wilson loop at hand.
In the following, we shall propose an integral relation satisfied by the hemisphere partition function 
with a properly defined Wilson loop insertion. This relation appears to uniquely fix the Wilson loop partition functions, order by order in $x$ expansion. 

We claim that the hemisphere partition function with a properly defined Wilson loop in representation $R$ in $SU(N)_{N-N_f/2}$ gauge theory with $N_f$ fundamental hypermultiplets satisfies the integral relation
\begin{equation}
  \hat{D} W_{R}^{N,N_f} (z_i,w_a,\lambda) = \lambda^{k(R)/N} W_{R}^{N,N_f}(z_i,w'_a,\lambda^{-1}) \ .
\end{equation}
The duality wall action $\hat{D}$ is the same as defined in (\ref{eq:duality-action}) and the flavor fugacities are also identified as $w_a=\lambda^{1/N}w'_a$. Here $k(R)$ is a positive integer number associated with the rank of the representation $R$. For example, the rank $n$ symmetric or anti-symmetric tensor representations have $k(R) = n$.

Therefore, duality wall maps the hemisphere index with a Wilson loop to itself dressed by a prefactor $\lambda^{k(R)/N}$, while inverting the instanton fugacity $\lambda \rightarrow \lambda^{-1}$. We will test this proposal with several examples momentarily.

We believe that this integral relation, combined with a 
\begin{equation}
W_{R}^{N,N_f} (z_i,w_a,\lambda) = \chi_R^{SU(N)}(z_i) + \mathcal{O}(x)
\end{equation}
``boundary condition'' fixes uniquely the Wilson loop index for all $R$. 

\subsection{Example: $SU(2)$ theories}
Let us first consider the fundamental Wilson loop in the $SU(2)$ gauge theory.
Practically, we will instead compute the partition function of $U(2)_2$ gauge theory and, after stripping off correction factors, we will regard it as the partition function of $SU(2)$ gauge theory.
For $U(2)_2$ theory, the hemisphere index from the formula (\ref{eq:Wilson-loop}) is given by
\begin{align}
  &\frac{\mathcal{W}^{2,N_f}_{QM,L+1}(a,w,\lambda)}{Z_{\rm 1-loop}(a,w)} \\
  =& \chi_{\bf L+1}^{SU(2)}(a) - \lambda \left[ \frac{\big\{\sqrt{a}\!+\!1\!/\!\sqrt{a}\!-\!(1\!-\!p)(1\!-\!q)\sqrt{a}\big\}_{\otimes^L\tiny\yng(1)}\prod_{a=1}^{N_f}(w_a-\sqrt{pqa})/\sqrt{w_a}}{(1-p)(1-q)(1-a)(1-pq a)}  +(a\rightarrow 1/a)\right]+\mathcal{O}(\lambda^2)\ , \nonumber
\end{align}
where $L+1$ denotes the $SU(2)$ representation of dimension $L+1$ and $\{Ch\}_{Y}$ stands for the tensor product of a character $Ch$ with product rule specified by a Young tableau $Y$. Here $Y=\otimes^L\tiny\yng(1)$, i.e. $L$-th symmetric product of a single box.

This index contains the usual correction factor $Z_{\rm extra}^{N_f}$ in (\ref{eq:U(1)-factor-SU(N)-flavors}).  We shall define a new Wilson loop index by removing the correction factor as
\begin{equation}
    W^{2,N_f}_{L+1}(a,w,\lambda) \equiv \mathcal{W}^{2,N_f}_{QM,L+1}(a,w,\lambda)/Z^{N_f}_{\rm extra}(w,\lambda) \ .
\end{equation}
In all cases with $N_f=0,L\le7$ and $N_f=1,L\le5$, and $N_f=2,L\le4$, we have confirmed that the Wilson loop index satisfies 
\begin{equation}
  \hat{D}W^{2,N_f}_{L+1}(a,\lambda^{1/2}w,\lambda) = \lambda^{L/2}\, W^{2,N_f}_{L+1}(a,w,\lambda^{-1}) \ .
\end{equation}
This relation has been checked for all cases at least up to $x^4$ order. It implies that the $SU(2)$ Wilson loops receive no additional corrections other than the usual correction factor (\ref{eq:U(1)-factor}).


\subsection{Example: $SU(3)$ theories}
Next, consider the pure $SU(3)_3$ theory with a Wilson loop in a representation labeled by a Young tableau $Y$. The Wilson loop index from the formula~(\ref{eq:Wilson-loop}) is
\begin{eqnarray}
	\frac{\mathcal{W}^3_{QM,Y}(z_i,\lambda)}{Z_{\rm 1-loop}(z_i)} &=& \chi_{Y}^{SU(3)}(z_i) - \lambda\bigg[\frac{\big\{(p+q-pq)z_1+z_2+z_3)\big\}_Y}{(1-p)(1-q)(1-z_1/z_2)(1-z_1/z_3)(1-pqz_1/z_2)(1-pqz_1/z_3)} \nonumber \\ 
	&&\qquad \qquad \qquad +(z_1,z_2,z_3\ {\rm permutations})\bigg] + \mathcal{O}(\lambda^2) \ .
\end{eqnarray}
Let us again define a new Wilson loop index (divided by the usual correction factor~(\ref{eq:U(1)-factor})):
\begin{equation}
    W^3_{Y}(z_i,\lambda) = \mathcal{W}^3_{QM,Y}(z_i,\lambda)/Z_{\rm extra}(z_i,\lambda) \ .
\end{equation}
For the rank $L$ symmetric representation denoted by $Y=\otimes^L\tiny\yng(1)$, we find the relation
\begin{equation}
	\hat{D}W^3_{\otimes^L\tiny{\yng(1)}}(z_i,\lambda;p,q) = \lambda^{L/3}\,W^3_{\otimes^L\tiny{\yng(1)}}(z_i,\lambda^{-1};p,q) \ ,
\end{equation}
till $L\le3$. This was confirmed at least up to $x^3$ order. 
Similarly, the Wilson loop index in the antisymmetric representation satisfies
\begin{equation}
    \hat{D}W^3_{\tiny{\yng(1,1)}}(z_i,\lambda;p,q) = \lambda^{2/3}\,W^3_{\tiny{\yng(1,1)}}(z_i,\lambda^{-1};p,q) \ ,
\end{equation}
which has been checked up to $x^3$ order. So for these representations, there would be no additional correction factors to the Wilson loop index.

As the last example, we consider the Wilson loop in the adjoint representation. We find that the index of this Wilson loop has an extra correction factor apart from the usual one (\ref{eq:U(1)-factor}) and it obeys
\begin{equation}
    \hat{D}\tilde{W}^3_{\tiny\yng(2,1)}(z_i,\lambda;p,q) = \lambda \tilde{W}^3_{\tiny\yng(2,1)}(z_i,\lambda^{-1};p,q) \ , 
\end{equation}
if we define
\begin{equation}\label{eq:Wilson-loop-SU(3)-adj}
    \tilde{W}^3_{\tiny\yng(2,1)}(z_i,\lambda;p,q) =  W^3_{\tiny\yng(2,1)}(z_i,\lambda;p,q) -\frac{\lambda}{2}II^3(z_i,\lambda;p,q) \ ,
\end{equation}
where $II^3$ is the bare hemisphere index. This relation has been confirmed up to $x^4$ order. The term proportional to the bare hemisphere index is the extra correction term which is not captured by the standard instanton partition function. We claim that the `correct' Wilson loop index is given by (\ref{eq:Wilson-loop-SU(3)-adj}) satisfying our duality relation.



\section{Duality and 't Hooft surfaces}\label{sec:thooft}
We can consider variant of the Higgsing procedure on $I_{N,N}$, which give a position-dependent vev to $v$ in order to produce 
a codimension two defect in the trivial interface, which is a surface defect in the 5d gauge theory. 
This is done by coupling the theory to a vortex configuration for $U(1)_e$ \cite{Gaiotto:2012xa,Gaiotto:2014ina}.

\subsection{Higgsing $I_{N,N}$}
It is useful to look at the index of the gauge theory in the presence of an $I_{N,N}$ domain wall as the expectation value of a certain operator 
between wavefunctions associated to the two hemispheres, as it is customarily done for $S^4_b$ partition functions in one lower dimension \cite{Drukker:2010jp}:
\begin{equation}
\langle I_L| \hat I |I_R \rangle = \oint d\mu_{z_i} d\mu_{\tilde z_i} I_L(z_i)  \frac{\prod_{i,j} \Gamma(\eta \tilde z_i/z_j)\prod_{i,a} \Gamma(\sqrt{pq/\eta} \lambda^{\frac{1}{2N}} z_i/w_a ) \Gamma(\sqrt{pq/\eta} \lambda^{-\frac{1}{2N}} w_a/\tilde z_i )}{\prod_{i\neq j} \Gamma(z_i/z_j)\Gamma(\tilde z_i/\tilde z_j)} I_R(\tilde z_i) \ .
\end{equation}
The standard Higgsing operation, associated to a constant vev for the bi-fundamental chiral multiplets, 
corresponds to looking for a pole at $\eta=1$, arising from the collision of $z_i = \eta \tilde z_i$ poles.  Everything cancels out and we are left with 
\begin{equation}
\oint d\mu_{z_i} I_L(z_i)  \frac{1}{\prod_{j,k} \Gamma(z_j/z_k)} I_R(z_i)
\end{equation}
i.e. the interface is gone.

Next, we can look for poles at $\eta= p^c$ for some power $c$, associated to a position-dependent vev for the bi-fundamental chiral multiplets with a zero of order $c$. 
These poles must arise from the collision of poles 
at $z_i = \eta \tilde z_i p^{n_i}$, which means that $N c= - \sum_i n_i \equiv -n$. 
The contributions from $v$ and from the $SU(N')$ vectormultiplets give  
\begin{equation}
\prod_{i\neq j} \frac{\Gamma(z_i/z_j p^{- n_i})}{\Gamma(z_i/z_j p^{n_j- n_i})}
\end{equation}
which give theta functions involving the gauge fugacities. 

The products $\prod_{i,a} \Gamma(\sqrt{pq/\eta} \lambda^{\frac{1}{2N}} z_i/w_a ) \Gamma(\sqrt{pq \eta} \lambda^{-\frac{1}{2N}} p^{n_i} w_a/z_i )$ 
give another set of theta functions. 
Thus we end up with a classic form for the action of a 't Hooft surface operator
\begin{align}
\langle I_L| O_n |I_R \rangle = \oint &\frac{d\mu_{z_i}}{\prod_{j,k} \Gamma(z_j/z_k)} I_L(z_i)  \sum_{n_i \geq 0}^{\sum n_j =n}  \prod_{i\neq j} \frac{\Gamma(z_i/z_j p^{- n_i})}{\Gamma(z_i/z_j p^{n_j- n_i})} \cdot \cr & \prod_{i,a}  \frac{\Gamma(\sqrt{pq} \lambda^{-\frac{1}{2N}} p^{n_i-n/(2N)} w_a/z_i )}{\Gamma(\sqrt{pq \eta} \lambda^{-\frac{1}{2N}} p^{-n/(2N)} w_a/z_i )} I_R(p^{n/N-n_i} z_i) \ .
\end{align}

As we obtained the operator as a half-BPS defect inside a half-BPS domain wall, this is actually a quarter-BPS object in the 5d gauge theory. The theta functions 
depending on the $w_a$ fugacities can be interpreted as contributions from $2N$ extra 2d $(0,2)$ Fermi multiplets added onto the bare 't Hooft surface in order to cancel 
a 2d gauge anomaly. In an half-BPS 't Hooft surface we would need to add whole $(0,4)$ Fermi multiplets. We should be able to restrict the $w_a$ fugacities 
in such a way to reproduce the contribution of  $N$ $(0,4)$ Fermi multiplets, but it does not seem urgent to do so. 

We can write down at first the $n=1$ defect. We can use the relation
\begin{equation}
\Gamma(p z) = \prod_{i\geq 0,j\geq 0} \frac{1-p^{i} q^{j+1} z^{-1}}{1-p^{i+1} q^j z} = (q z^{-1};q)_\infty (z;q)_\infty \prod_{i\geq 0,j\geq 0} \frac{1-p^{i+1} q^{j+1} z^{-1}}{1-p^{i} q^j z} = \theta(z;q) \Gamma(z)
\end{equation}
and denote as $\Delta_z$ a shift operator $z \to p z$ and 
\begin{equation} \Delta_{i} = \Delta_{z_i} \prod_{k\neq i} \Delta_{z_k}^{-1/N}
\end{equation}
to specialize  
\begin{equation}
O_n[w_a] =  \sum_{n_i \geq 0}^{\sum n_j =n}  \prod_{i\neq j} \frac{\Gamma(z_i/z_j p^{- n_i})}{\Gamma(z_i/z_j p^{n_j- n_i})} \prod_{i,a} \frac{\Gamma(\sqrt{pq} \lambda^{-\frac{1}{2N}} p^{n_i-n/(2N)} w_a/z_i )}{\Gamma(\sqrt{pq \eta} \lambda^{-\frac{1}{2N}} p^{-n/(2N)} w_a/z_i )} \prod_i \Delta_{z_i}^{n/N-n_i}
\end{equation}
to 
\begin{equation}
O_1[w_a] =  \sum_i  \prod_{k\neq i} \frac{1}{\theta(z_k/z_i)} \prod_{a} \theta(\sqrt{pq} (p \lambda)^{-\frac{1}{2N}} w_a/z_i)\Delta_i^{-1} \ .
\end{equation}
It is also useful to write the adjoint expressions
\begin{equation}
O_n  = \prod_i \overleftarrow\Delta^{n_i-n/N} _{z_i}  \left[\sum_{n_i \geq 0}^{\sum n_j =n}  \prod_{i\neq j} \frac{\Gamma(z_i/z_j p^{- n_j})}{\Gamma(p^{n_i-n_j} z_i/z_j)} \prod_{i,a} \frac{\Gamma(\sqrt{pq} \lambda^{-\frac{1}{2N}} p^{n/(2N)} w_a/z_i )}{\Gamma(\sqrt{pq \eta} \lambda^{-\frac{1}{2N}} p^{n/(2N)-n_i} w_a/z_i )} \right]
\end{equation}
and
\begin{equation}
O_1[w_a] = \sum_i \overleftarrow\Delta_i  \prod_{j\neq i} \frac{1}{\theta(z_i/z_j)} \prod_{a} \theta(\sqrt{pq} \lambda^{-\frac{1}{2N}} p^{1/(2N)-1} w_a/z_i ) \ .
\end{equation}

Although we know that $O_1[w_a]$ must commute with the duality wall, as it is obtained from Higgsing a duality-invariant interface, the check of this fact takes the form of a rather intricate-looking theta function identity. 
Let us denote the action of a duality wall on a boundary theory as 
\begin{equation}
\hat D |I_R \rangle  = \oint d\mu_{\tilde z_i} \frac{\prod_{i,j} \Gamma(\lambda^{1/N} z_i/\tilde z_j)}{\Gamma(\lambda) \prod_{i\neq j} \Gamma(\tilde z_i/\tilde z_j)} I_R(\tilde z_i, \lambda \to \lambda^{-1}) \ .
\end{equation}
We know that $\hat D$ and $\hat I$ commute: if we compose the interfaces 
\begin{equation}
\hat D \hat I = \oint d\mu_{\tilde z_i}  \frac{\prod_{i,j} \Gamma(\lambda^{1/N} z_i/\tilde z_j)}{\Gamma(\lambda) \prod_{i\neq j} \Gamma(\tilde z_i/\tilde z_j)} \prod_{i,j} \Gamma(\eta z'_i/\tilde z_j)\prod_{i,a} \Gamma(\sqrt{pq/\eta} \lambda^{-\frac{1}{2N}} \tilde z_i/w_a ) \Gamma(\sqrt{pq/\eta} \lambda^{\frac{1}{2N}} w_a/z'_i ) \ .
\end{equation}
and apply Seiberg duality, we get 
\begin{equation}
\oint d\mu_{\tilde z_i}  \frac{\prod_{i,j} \Gamma(\lambda^{1/N} \tilde z_j/z'_i)}{\Gamma(\lambda) \prod_{i\neq j} \Gamma(\tilde z_i/\tilde z_j)} \prod_{i,j} \Gamma(\eta \tilde z_j/z_i)\prod_{i,a} \frac{\Gamma(\sqrt{pq/\eta} \lambda^{-\frac{1}{2N}} w_a/\tilde z_i )}{\Gamma(\sqrt{pq \eta} \lambda^{-\frac{1}{2N}} w_a/z_i)}=\hat I \hat D \ .
\end{equation}

If we compare the residues of these expressions at $\eta = p^{-1/n}$, we find a theta function identity 
\begin{align}
\sum_i &\prod_{k} \frac{1}{ \theta(\lambda^{1/N} p^{1/N-1} z_k/\tilde z_i)} \prod_{j\neq i} \frac{1}{\theta(\tilde z_i/\tilde z_j)} \prod_{a} \theta(\sqrt{pq} (p\lambda)^{\frac{1}{2N}} p^{-1} w_a/\tilde z_i ) = \cr
\sum_i  &\prod_{k\neq i} \frac{1}{\theta(z_k/z_i)} \prod_{a} \theta(\sqrt{pq} (p \lambda)^{-\frac{1}{2N}} w_a/z_i) \prod_{k} \frac{1}{ \theta(\lambda^{1/N} p^{1/N-1} z_i/\tilde z_k)}
\end{align}
which would be challenging to prove directly. 

This suggests that Seiberg duality may be a useful trick to derive other properties of the 't Hooft surfaces. 

In particular, consider the composition of two interfaces $I_{N,N}$: 
\begin{align}
\hat I[w_a,\eta] \hat I[\tilde w_a,\tilde \eta] = \oint \frac{d\mu_{\tilde z_i}}{\prod_{i\neq j}\Gamma(\tilde z_i/\tilde z_j)} &\prod_{i,j} \Gamma(\eta \tilde z_i/z_j)\prod_{i,a} \Gamma(\sqrt{pq/\eta} \lambda^{\frac{1}{2N}} z_i/w_a ) \Gamma(\sqrt{pq/\eta} \lambda^{-\frac{1}{2N}} w_a/\tilde z_i ) \cr &\prod_{i,j} \Gamma(\tilde \eta z'_i/ \tilde z_j)\prod_{i,a} \Gamma(\sqrt{pq/\tilde \eta} \lambda^{\frac{1}{2N}} \tilde z_i/ \tilde w_a ) \Gamma(\sqrt{pq/\tilde \eta} \lambda^{-\frac{1}{2N}} \tilde w_a/z'_i ) \ .
\end{align}
Seiberg duality maps that to 
\begin{align}
\hat I[w_a,\eta] \hat I[\tilde w_a,\tilde \eta] =& \frac{\prod_{i,j} \Gamma(\eta \tilde \eta  z'_i/z_j)}{\prod_{a,b} \Gamma(\sqrt{\eta \tilde \eta}\tilde w_a/w_b)}\oint \frac{d\mu_{\hat z_a}}{\prod_{a\neq b}\Gamma(\hat z_a/\hat z_b)}\prod_{a,b} \Gamma(\sqrt{\eta}\tilde w_a/\hat z_b)\Gamma(\sqrt{\tilde \eta} \hat z_b/w_a)  \cr &\prod_{i,a} \Gamma(\sqrt{pq/(\eta \tilde \eta)} \lambda^{1/2N} z_i/\hat z_a) \Gamma(\sqrt{pq/(\eta \tilde \eta)} \lambda^{-1/2N} \hat z_a/z_i') 
\end{align}
i.e. 
\begin{align}
\hat I[w_a,\eta] \hat I[\tilde w_a,\tilde \eta] =& \frac{1}{\prod_{a,b} \Gamma(\sqrt{\eta \tilde \eta}\tilde w_a/w_b)}\oint \frac{d\mu_{\hat z_a}}{\prod_{a\neq b}\Gamma(\hat z_a/\hat z_b)}\prod_{a,b} \Gamma(\sqrt{\eta}\tilde w_a/\hat z_b)\Gamma(\sqrt{\tilde \eta} \hat z_b/w_a)  \hat I[\hat z_a, \eta \tilde \eta]  \ .
\end{align}

If we take a residue at $\eta = p^{-n/N}$, we are looking at a vev for the anti-baryon operator in the $SU(2N)$ gauge theory. 
We can look at contributions from $\hat z_a = \tilde w_a p^{m_a-n/(2N)}$ with $\sum_a m_a = n$: 
\begin{align}
\hat O_n[w_a] \hat I[\tilde w_a,\tilde \eta] =\sum_{m_a}^{\sum_a m_a = n} &\prod_{a\neq b} \frac{\Gamma(p^{-m_b} \tilde w_a/\tilde w_b)}{\Gamma(p^{m_a-m_b} \tilde w_a/\tilde w_b)} \cr &\prod_{a,b} \frac{\Gamma(\sqrt{p^{-n/N} \tilde \eta}p^{m_b} \tilde w_b/w_a) }{ \Gamma(\sqrt{p^{-n/N} \tilde \eta}\tilde w_b/w_a)}   \hat I[\tilde w_a p^{m_a-n/(2N)}, p^{-n/N} \tilde \eta] 
\end{align}
This is a striking formula which converts a 't Hooft surface acting on the interface into a ``flavor'' 't Hooft surface acting on the global symmetries of the interface. 

We can then take a second residue at $\tilde \eta = p^{- \tilde n/N}$, to get 
\begin{align}
\hat O_n[w_a] \hat O_{\tilde n}[\tilde w_a] =\sum_{m_a}^{\sum_a m_a = n} &\prod_{a\neq b} \frac{\Gamma(p^{-m_b} \tilde w_a/\tilde w_b)}{\Gamma(p^{m_a-m_b} \tilde w_a/\tilde w_b)} \cr &\prod_{a,b} \frac{\Gamma(\sqrt{p^{-(n+ \tilde n)/N}}p^{m_b} \tilde w_b/w_a) }{ \Gamma(\sqrt{p^{-(n+ \tilde n)/N}}\tilde w_b/w_a)}   \hat O_{n + \tilde n}[\tilde w_a p^{m_a-n/(2N)}] 
\end{align}

For example, setting $\tilde n =0$ we should have a recursion
\begin{align}
\hat O_1[w_a] =\sum_a \prod_{b\neq a} \frac{1}{\theta(\tilde w_a/\tilde w_b)} \prod_{b} \theta(p^{-1/{2N}} \tilde w_a/w_b)  \hat O_{1}[\tilde w_b p^{\delta_{a,b}-1/(2N)}] \ .
\end{align}
We can write this recursion term-by-term: 
\begin{align}
\prod_{b} \theta(\sqrt{pq} (p \lambda)^{-\frac{1}{2N}} w_b/z_i) =\sum_a  &\theta(\sqrt{pq} (p^2 \lambda)^{-\frac{1}{2N}} p \tilde w_a/z_i) \prod_{b\neq a} \frac{1}{\theta(\tilde w_a/\tilde w_b)}\cr & \prod_{b} \theta(p^{-1/{2N}} \tilde w_a/w_b)  
 \prod_{b \neq a} \theta(\sqrt{pq} (p^2 \lambda)^{-\frac{1}{2N}} \tilde w_b/z_i) \ . \quad
\end{align}

Setting $\tilde n =1$ we get
\begin{align}
\hat O_1[w_a] \hat O_1[\tilde w_a] =\sum_a \prod_{b\neq a} \frac{1}{\theta(\tilde w_a/\tilde w_b)} \prod_{b} \theta(p^{-1/N} \tilde w_a/w_b)  \hat O_{2}[\tilde w_b p^{\delta_{a,b}-1/(2N)}] \ .
\end{align}

\section{Codimension 2 defects}
A similar Higgsing procedure can also introduce codimension two (or three dimensional) defects in the 5d gauge theory,
possibly intersecting domain walls along two-dimensional defects. 
For example, starting from a ``UV'' gauge theory with a duality wall and turning on appropriate Higgs branch vevs, 
we are able to obtain an ``IR '' gauge theory modified both by a duality wall and a codimension two BPS defects. 
In other words, we obtain a duality domain wall for the combined system of a 5d gauge theory and a 3d defect in the gauge theory. 

For simplicity, we will focus on the simplest example: the RG flow from $SU(3)_2$ theory with $N_f=2$ fundamental hypermultiplets and pure $SU(2)_2$ 
gauge theory initiated by a vev of a mesonic operator. See~\cite{Gaiotto:2014ina} for more details. A position-dependent vev leaves behind a specific codimension two defect in 
the $SU(2)_2$ gauge theory, corresponding to a set of D3 branes ending on the five-brane web for pure $SU(2)_2$ gauge theory.
In this section, we aim to discuss the correction to the duality wall due to the presence of this defect.
We will first review the Higgsing procedure in the absence of the duality wall. 
\subsection{Higgsing in the absence of a duality wall}
We start with the hemisphere index of the UV theory, which is given by
\begin{equation}\label{eq:hemi-sphere-partition-ftn}
	II^{3,2}(z_i,w_a,\lambda;p,q) = \frac{(pq;p,q)_\infty^2\prod_{i\neq j}^3(pq z_i/z_j;p,q)_\infty}{\prod_{i=1}^3\prod_{a=1}^2(\sqrt{pq}z_i/w_a;p,q)_\infty} Z_{\rm inst}^{3,2}(z_i,w_a,\lambda;p,q) \ .
\end{equation}
We can Higgs this partition function by giving  nonzero vev to the mesonic operator, say $M^1_2 \equiv q^1\tilde{q}_2$. 
From the full sphere index point of view, the Higgsing procedure amounts to taking a residue at the pole corresponding to the meson operator. 
The full index has the poles of the form $w_1/w_2 = p^{r+1}q^{s+1}$ from the meson operator. 
Here, $r,s$ label the angular momentum of the meson operator along the four spatial directions. 
Thus $r=s=0$ means the meson operator has no position dependence, so it corresponds to giving a constant vev to the meson operator. 
Therefore, if we takes the residue at the pole $w_1/w_2=pq$, we end up with the superconformal index of the IR $SU(2)$ gauge theory without defect.

The residue at the pole with nonzero $r$ or $s$ gives rise to the full sphere index of the IR $SU(2)$ theory with a defect.
We will focus on the simplest defect with $r=1,\,s=0$.
In the contour integral expression, the pole at $w_1/w_2=p^2q$ appears when two sets of poles $(z_3 = w_1 (p\sqrt{pq})^{-1},z_3 = w_2 \sqrt{pq})$ and $(z_3 = w_1 \sqrt{pq}^{-1},z_3 = w_2 p\sqrt{pq})$ pinch the $z_3$ integral contour. These two sets corresponds to two different vacua of the defect. The full IR index with the defect can be obtained by the sum of residues from these two sets.

This Higgsing procedure can also be performed at the level of the hemisphere index.
This should give a certain extension of Dirichlet boundary conditions for the bulk theory 
to a boundary condition for the 3d defect. We will actually get two possible answers, which should correspond to 
two basic boundary conditions for the defect which constrain it in the IR to sit in either of the possible two vacua for the defect. 

The hemisphere index (\ref{eq:hemi-sphere-partition-ftn}) has poles at $z_3=w_1(p\sqrt{pq})^{-1}$ and $z_3=w_1\sqrt{pq}^{-1}$. One can Higgs the hemisphere index by taking residues at either of 
these poles and setting $w_1/w_2 = p^2q$. We first take the residue at the pole $z_3=w_1(p\sqrt{pq})^{-1}$. It gives the hemisphere index with a codimension two defect and a certain choice of boundary condition for the defect.
\begin{eqnarray}
  II^{(1)}(a,\mu,\lambda;p,q) &=& \lim_{\substack{w_1\rightarrow \mu_2p^2q,  \\ z_3\rightarrow w_1(p\sqrt{pq})^{-1} } }\frac{ \prod_{a=1}^2(\sqrt{pq}z_3/w_a;p,q)_\infty }{ (pq;p,q)_\infty\Gamma(p^{3/4}\mu^{-1}\sqrt{a}^\pm) } II^{3,2}(z_i,w_a,\lambda;p,q)  \\
  &=&   (pq;p,q)_\infty(pqa^\pm;p,q)_\infty  (p^{-1/4}\mu^{-1}\sqrt{a}^\pm ;q)_\infty^{-1} Z_{\rm inst}^{(1)}(a,\mu,\lambda;p,q) \ , \nonumber
\end{eqnarray}
where $a\equiv z_1/z_2,\, \mu\equiv(w_1w_2)^{3/4}$. Similarly, the residue at the second pole $z_3=w_1\sqrt{pq}^{-1}$ gives the hemisphere index with a codimension two defect 
and another choice of boundary condition for the defect.
\begin{eqnarray}
  II^{(2)}(a,\mu,\lambda;p,q) &=& \lim_{ \substack{w_1\rightarrow w_2p^2q, \\ z_3\rightarrow w_1\sqrt{pq}^{-1} } }\frac{ \prod_{a=1}^2(\sqrt{pq}z_3/w_a;p,q)_\infty }{ (pq;p,q)_\infty\Gamma(p^{-3/4}\mu^{-1}\sqrt{a}^\pm) } II^{3,2}(z_i,w_a,\lambda;p,q)  \\
  &=& (pq;p,q)_\infty (pqa^\pm;p,q)_\infty  (p^{1/4}q\mu^{-1}\sqrt{a}^\pm ;q)_\infty  Z_{\rm inst}^{(2)}(a,\mu,\lambda;p,q) \nonumber \ .
\end{eqnarray}
The functions $Z^{(1)}_{\rm inst}$ and $Z^{(2)}_{\rm inst}$ are the instanton partition functions with the fugacities tuned as required by the poles we picked.

These partition functions with defects are known to satisfy certain difference equations~\cite{Gaiotto:2014ina} (See also~\cite{Bullimore:2014awa}),
which encode the expansion of bulk line defects brought to the codimension two defects into a sum of line defects 
defined on the codimension two defects. 
The difference equation can be thought of as the quantization of the algebraic curve describing moduli space of supersymmetric parameter space of the 3d theory living on the defect. 
It also encodes the Seiberg-Witten curve for the 5d bulk gauge theory. 
The canonical coordinates on the moduli space are the parameter $\mu$ and its conjugate momentum $p_\mu$. When $q\neq 1$, they become non-commuting operators $p_\mu\mu = q\mu p_\mu$ and the algebraic curve written in terms of these coordinates is promoted to the difference equation.

The defect partition function $II^{(1)}$ with the first boundary condition satisfies experimentally the relation 
\begin{equation} \label{eq:diffrelation1}
p_\mu^{-1}-1-\sqrt{p}^{-1}(1+\lambda)\mu^{-2} + p^{-1}q^{-2}\lambda\, \mu^{-4} \, p_\mu = -p^{-1/4} \mu^{-1} \langle W^{(1)}_{\rm fund}\rangle \ .
\end{equation}
We denote by $\langle W^{(1)}_{\rm fund}\rangle$ the fundamental Wilson loop expectation value in the IR $SU(2)$ gauge theory in the presence of the codimension two defect. It can be obtained by Higgsing the fundamental Wilson loop of the UV $SU(3)$ gauge group as follows:
\begin{equation}
  \langle W^{(1)}_{\rm fund} \rangle = \lim_{\substack{w_1\rightarrow w_2p^2q,  \\ z_3\rightarrow w_1(p\sqrt{pq})^{-1} } } \left(\langle W^{3,2}_{\rm fund}\rangle - z_3\right)\sqrt{z_3} \ .
\end{equation}
In the Nekrasov-Shatashvili limit~\cite{Nekrasov:2009rc}, when $p\rightarrow 1$, this Wilson loop reduces to the fundamental Wilson loop of the pure $SU(2)$ theory.

On the other hand, the defect partition function $II^{(2)}$ with the second boundary condition satisfies experimentally the relation
\begin{equation}\label{eq:diffrelation2}
p_\mu-1-\sqrt{p}(1+\lambda)\mu^{-2} +pq^{2}\lambda\, \mu^{-4} \, p_\mu^{-1} = -p^{1/4}\mu^{-1} \langle W^{(2)}_{\rm fund}\rangle \ ,
\end{equation}
where 
\begin{equation}
  \langle W^{(2)}_{\rm fund} \rangle = \lim_{\substack{w_1\rightarrow w_2p^2q,  \\ z_3\rightarrow w_1\sqrt{pq}^{-1} } } \left(\langle W^{3,2}_{\rm fund}\rangle - z_3\right)\sqrt{z_3} \ .
\end{equation}

We have checked these difference equations numerically up to 3-instantons. We leave the analysis the physical meaning of these relations to future work.  

\subsection{Higgsing in the presence of a duality wall}
Let us now consider this Higgsing procedure when coupled to the duality domain wall. It leads to the duality wall action on the hemisphere index in the presence of codimension two defects. As we will see, the Hggsing can also introduce extra degrees of freedom localized at codimension two locus where the boundary intersects the codimension three defect. 

The hemisphere index of the UV $SU(3)$ theory coupled to the duality wall is given by
\begin{equation}
  \hat{D} II^{3,2} = I_V^2\oint \prod_{i=1}^{2}\frac{dz'_i}{2\pi iz'_i}  \frac{ \prod_{i,j=1}^3 \Gamma(\lambda^{1/3} z_i/z'_j) }{ \Gamma(\lambda) \prod_{i\neq j}^3\Gamma(z'_i/z'_j)} II^{3,2}(z'_i,w_a,\lambda;p,q) \ .
\end{equation}
This index satisfies the duality relation (\ref{eq:duality-hemisphare-flavor}). We shall Higgs the both sides of this relation by taking the residue at $w_1/w_2=p^2q$. 
The same Higgsing procedure as above leads to the following relations:
\begin{eqnarray}
   &&  II^{(1)}(a,\mu,\lambda^{-1})  \\
   &=& \frac{(p;p)_\infty(q;q)_\infty}{2!}\!\oint\! \frac{db}{2\pi ib}\! \left[ \frac{ \Gamma(\sqrt{\lambda}a^\pm b^\pm) }{ \Gamma(\lambda)\Gamma(b^\pm) } II^{(1)}(b,\sqrt{\lambda}\mu,\lambda)+ \frac{ \Gamma(\sqrt{p\lambda}a^\pm b^\pm) }{ \Gamma(\lambda)\Gamma(b^\pm) } Z_{2d}(a,b,\mu,\lambda) II^{(2)}(b,\sqrt{\lambda}\mu,\lambda) \right] \nonumber
\end{eqnarray}
and
\begin{equation}\label{eq:dualityop-defect2}
  II^{(2)}(a,\mu,\lambda^{-1}) = \frac{(p;p)_\infty(q;q)_\infty}{2!}\oint \frac{db}{2\pi ib} \frac{ \Gamma(\sqrt{\lambda}a^\pm b^\pm) }{ \Gamma(\lambda)\Gamma(b^\pm) }  II^{(2)}(b,\sqrt{\lambda}\mu,\lambda) \ .
\end{equation}

We can identify the collection of theta functions in the first relation as the 2d elliptic genus of some 2d degrees of freedom:
\begin{equation}
  Z_{2d}(a,b,\mu,\lambda;q) \equiv \theta(p^{-1/4}\mu^{-1}\sqrt{a}^\pm|q)^{-1}\theta(p^{-1/4}\sqrt{\lambda}\mu\sqrt{b}^\pm|q)^{-1} \ ,
\end{equation}
where $\theta(x;q) = (x;q)_\infty (qx^{-1};q)_\infty$. This appears to be the contribution of two $(2,0)$ fundamental chiral multiplets with appropriate global charges. 
Physically, the coefficients of these relations capture the 2d field content at the intersection of the duality wall and the codimension two defect, 
for a given choice of vacua on the two sides of the duality wall. 

We find it convenient to rewrite the relations as
\begin{equation}
  \left(\begin{array}{c} II^{(1)}(a,\mu,\lambda^{-1}) \\ II^{(2)}(a,\mu,\lambda^{-1}) \end{array}\right) = M_{2\times2}(a,b,\mu,\lambda) \left(\begin{array}{c} II^{(1)}(b,\mu,\lambda) \\ II^{(2)}(b,\mu,\lambda)\end{array}\right) \ .
\end{equation}
Here we have defined an integral operator
\begin{equation}
  M_{2\times2}(a,b,\mu,\lambda) \equiv \frac{(p;p)_\infty(q;q)_\infty}{2!}\oint \frac{db}{2\pi ib} \frac{ \Gamma(\sqrt{\lambda}a^\pm b^\pm) }{ \Gamma(\lambda)\Gamma(b^\pm) } 
  \left(\begin{array}{cc}1 \ \  & \frac{\Gamma(\sqrt{p\lambda}a^\pm b^\pm)}{\Gamma(\sqrt{\lambda}a^\pm b^\pm)}Z_{2d}(a,b,\mu,\lambda) \\  0 \ \ & 1\end{array}\right) \Delta_{\mu\rightarrow \sqrt{\lambda}\mu} \ ,
\end{equation}
with a shift operator $\Delta_{\mu\rightarrow \sqrt{\lambda}\mu}$ acting on $\mu$.

This integral operator can be though of acting on the partition functions of some boundary condition or interface for the 
bulk theory in the presence of the codimension two defect, computed in the IR with the defect sitting in either of its two vacua. 

We can derive the relation $M_{2\times 2}(a,b,\lambda^{-1})M_{2\times 2}(b,c,\lambda)=\delta_{ac}$ expected for a $Z_2$ duality wall by Higgsing the corresponding identity 
for the domain wall partition function in the $SU(3)_2$ gauge theory. It involves an interesting integral identity:
\begin{eqnarray}
  0&=&\oint \frac{db}{4\pi ib} \frac{ \Gamma(\sqrt{\lambda}^{-1}a^\pm b^\pm) }{ \Gamma(b^\pm)\theta(p^{-1/4}\sqrt{\lambda}\mu^{-1}\sqrt{b}^\pm|q) } \oint \frac{dc}{4\pi ic} \frac{ \Gamma(\sqrt{p\lambda}b^\pm c^\pm) }{\Gamma(c^\pm)\theta(p^{-1/4}\mu\sqrt{c}^\pm|q) }II^{(2)}(c,\mu,\lambda) \nonumber \\
  &&+ \oint \frac{db}{4\pi ib} \frac{ \Gamma(\sqrt{p\lambda^{-1}}a^\pm b^\pm) }{ \Gamma(b^\pm)\theta(p^{-1/4}\sqrt{\lambda}^{-1}\mu\sqrt{b}^\pm|q) } \oint \frac{dc}{4\pi ic} \frac{ \Gamma(\sqrt{\lambda}b^\pm c^\pm) }{ \Gamma(c^\pm)\theta(p^{-1/4}\mu^{-1}\sqrt{a}^\pm|q) }II^{(2)}(c,\mu,\lambda) \ . \nonumber 
\end{eqnarray}

\section{Duality walls between $Sp(N)$ and $SU(N+1)$ theories}
We have seen in the previous sections how the existence of the $Z_2$-duality interface of $SU(N)$ gauge theories 
is encoded at the level of the superconformal index in the properties of an ``elliptic Fourier transform''. In particular, the 
index of the domain wall degrees of freedom, combined with the $SU(N)$ vectormultiplet integration measure, 
provides the integral kernel for the elliptic Fourier transform. With a proper definition of the integration contours, 
the inverse of the elliptic Fourier transform is the elliptic Fourier transform itself, 
up to the inversion of the parameter $\lambda$ associated to the gauge theory instanton fugacity. 

There are other elliptic integral transformations with properties akin to the elliptic Fourier transform. In particular, there is a class of 
``A-C'' pairs of integral transformations introduced in ~\cite{2004math.....11044S} and reviewed below, such that the two members of each pair are 
inverse of each other. The two integral kernels can be decomposed into the product of vectormultiplet integration measures for $SU(N+1)$ 
and $Sp(N)$ respectively and a common residual kernel, again up to the inversion of a parameter $\lambda$.

It is natural to interpret the two integral transforms in each pair as describing the action of a single interface between 
some $Sp(N)$ and $SU(N+1)$ gauge theories onto the boundary conditions of either theory, with the property that the composition of two such interfaces 
(either from  $SU(N+1)$ to $Sp(N)$ and back to $SU(N+1)$ or vice versa) flows to the identity in the IR. 
More ambitiously, we may hope that such interface may actually be a duality interface, encoding a common UV 
completion for the two gauge theories. 

The matter content of the tentative duality interface appears to consist of a bi-fundamental chiral multiplet $q$ of $SU(N+1)\times Sp(N)$ together
with a chiral multiplet $M$ in the antisymmetric representation of $SU(N+1)$. The fugacity visible in the index are compatible with a 4d superpotential
\begin{equation}
  W = \mathrm{Tr}\, q \,M\,q^T\omega \ ,
\end{equation}
where $\omega$ is the symplectic form of $Sp(N)$.
These 4d matter fields have $(N+3)$ units of cubic anomaly for the $SU(N+1)$ global symmetry and various mixed 't Hooft anomalies. When we couple it to the 5d bulk theories, these anomalies should be canceled by anomalies arising from the 5d theories with certain boundary conditions.

Notice that for $N=1$ the matter content and couplings are precisely the same as for the $\IZ_2$ duality wall for $SU(2)$ gauge theories we 
defined in the first half of the paper. In this section we will thus set $N>1$.

The simplest possibility would be to couple such interface fields to pure 5d gauge theories on the two sides. If we assign charges $1/2$ and $-1$ 
to $q$ and $M$ under some global symmetry $U(1)_\lambda$, the cancellation of mixed anomalies will tie $U(1)_\lambda$ to (appropriate multiples) 
of the instanton symmetries on the two sides. An obvious obstruction to this idea is that the cancellation of the cubic anomaly for $SU(N+1)$
would require a Chern-Simons level $\kappa = N+3$, which should be excluded by the bound $|\kappa|\leq N+1$ 
proposed of \cite{Intriligator:1997pq} as a necessary condition for the existence of a UV fixed point. 

We can also add $N_f$ fundamental flavors on both sides of the wall, with the usual cubic superpotential coupling
\begin{equation}
  W = XqX' \ .
\end{equation}
involving the halves $X$ and $X'$ of bulk hypermultiplets for the for $SU(N+1)$ and $Sp(N)$ gauge theory respectively. 
This gives the constraint $\kappa = N+3-N_f/2$, which again violates the expected bound $|\kappa|\leq N+1-N_f/2$.
The interface glues together the $SU(N_f)$ flavor symmetry on the $SU(N+1)$ side to the $SU(N_f)$ subgroup of the 
$SO(2N_f)$ flavor symmetry on the $Sp(N)$ side and glues the instanton symmetries to appropriate combinations of $U(1)_\lambda$ and 
the $U(1)_f$ flavor symmetries on the two sides. 

Soldiering ahead and ignoring the apparent obstruction, we can compute the action of the conjectural duality wall onto $Sp(N)$
Dirichlet boundary conditions, i.e. the action of the C integral transform onto the appropriately dressed $Sp(N)$ instanton partition function
(with appropriate discrete theta angle). The result is very encouraging: for small $N$ we will find that the result of the integral transform admits 
a power series expansion in positive powers of the $SU(N+1)$ instanton fugacity, as it should be for an $SU(N+1)$ instanton partition function. 

Furthermore, the perturbative part of the answer is precisely right. We cannot compare the contribution with positive instanton number to 
a standard expression for the $SU(N+1)$ instanton partition function, as the usual ADHM localization integral itself becomes problematic if we violate the 
standard bound $|\kappa|\leq N+1$: the localization integral has poles at the origin or infinity of degree higher than $1$, which signal the presence of
spurious contributions from the Coulomb branch of the ADHM quantum mechanics, i.e. the singularity of the instanton moduli spaces. 
We have not been able to find a systematic way to deal with these poles and recover the desired answer. 

We are thus posed with two basic problem. The first question is to identify a UV completion of $SU(N+1)$ SQCD with $\kappa = N+3-N_f/2$,
endowed with a global symmetry enhancement $U(N_f) \to SO(2 N_f)$ and a second mass deformation to an $Sp(N)$ gauge theory with the same number of flavors. 
It would be nice to pinpoint a specific brane construction demonstrating the desired UV completion, but we will not do so. It should be straightforward to 
derive it from the proposal of \cite{Hayashi:2015fsa}. In the next sub-section we will sketch a field theory argument for the possibility of 
a UV completion with the appropriate enhanced global symmetry.

The second question is to find a prescription to compute the instanton partition function of that $SU(N+1)$ gauge theory which 
agrees with the C elliptic Fourier transform of the partition function of an appropriate $Sp(N)$ gauge theory. For $SU(3)$ gauge theories, we will propose a prescription of the instanton quantum mechanics in appendix~\ref{sec:partition-function-exotic-SU(3)} and show that the partition function of this quantum mechanics reproduces the result of the C elliptic Fourier transform.
Lacking such a prescription for the theories with higher rank gauge group, we will simply give some explicit 
calculations of the elliptic Fourier transform of $Sp(N)$ partition functions and extract from them the predicted form of the $SU(N+1)$ gauge theory instanton partition function. 

\subsection{Enhanced symmetry of $SU(N+1)$ theory}
Our first task is to test the possibility of an UV global symmetry enhancement $U(N_f) \to SO(2 N_f)$ for $SU(N)$ SQCD with $\kappa = N+2-N_f/2$.
We will follow and extend the analysis of conserved current multiplets arising from instanton operators 
proposed in~\cite{Tachikawa:2015mha} (See also~\cite{Zafrir:2015uaa}). 

Consider an instanton operator with instanton number `$1$' inserted at the origin of $\mathbb{R}^5$. 
It induces a nontrivial gauge configuration on a round $S^4$ surrounding the instanton operator. The quantum numbers of such an operator 
can be computed in analogy to monopole operators in 3d, by adding together classical contributions and the contributions which arise from the 
quantization of fermionic zero modes on this gauge field background.

Let us first consider the pure $SU(2)$ gauge theory. The $\mathcal{N}=1$ vector multiplet has a gaugino in the doublet of $SU(2)_R$ R-symmetry and  in the adjoint representation of the $SU(2)$ gauge symmetry. The gaugino provides 8 fermion zero modes $\lambda_{i\alpha}$ on the instanton moduli space, where $i=1,2$ labels a doublet of $SU(2)_R$ and $\alpha=1,2,3,4$ labels the spinor indices of $SO(5)$ isometry on $S^4$. The quantization of  these zero modes leads to 4 raising and 4 lowering operators and they construct sixteen states, i.e. $(\mu^+_{ij}, \psi^+_{i\alpha},J^+_\mu)$, forming a current multiplet which one identifies with a broken generator of the UV $SU(2)$ global symmetry. 
Here the superscript `$+$' denotes the instanton charge $+1$.

Next, we can consider an $SU(N)$ gauge theory. The one-instanton configuration can be embedded in the $SU(2)$ subsector while breaking the gauge symmetry to $SU(N-2)\times U(1)$. The generator of the $U(1)$ subgroup then takes the form
\begin{equation}
    {\rm diag}(N-2,N-2,-2,\cdots ,-2) \ .
\end{equation}
Thus the 1 instanton operator of the $SU(N)$ theory with a classical CS-level $\kappa$ has naive $U(1)$ gauge charge $(N-2)\kappa$. 
The gaugino can be decomposed into the adjoint of the $SU(2)\subset SU(N)$ and the adjoint of the $SU(N-2)$, and a bi-fundamental of the $SU(2)$ and $SU(N-2)$. 
The adjoint fermion of the $SU(2)$ provides the same fermionic zero modes as for $SU(2)$, generating the sixteen states. There are also additional fermionic zero modes from the bi-fundamentals. The quantization of these additional fermion modes leads to the raising operators $B_{ia}$ where $a$ denotes the fundamental of the $SU(N-2)$ subgroup. Imposing the $SU(N-2)$ gauge invariance one can construct the following states
\begin{equation}
    |0\rangle \ , \quad \epsilon^{a_1\cdots a_{N-2}}B_{i_1a_1}\cdots B_{i_{N-1}a_{N-2}} |0\rangle \ , \quad (B_{ia})^{2(N-2)}|0\rangle \ .
\end{equation}
where $|0\rangle$ is the ground state tensored by the broken current supermultiplet. These states carry $U(1)$ gauge charges $-(N-2)N,\,0,\,+(N-2)N$ respectively. 
Among these three states, the first and the third states carry appropriate $SU(2)_R$ charge for being a current multiplet. We also need to impose the $U(1)$ gauge invariance. 

Therefore, the instanton operator provides a broken current supermultiplet when the classical CS-level satisfies 
\begin{equation}
    \kappa \pm N = 0 \ .
\end{equation}
This supports the $U(1)_{in} \rightarrow SU(2)$ global symmetry enhancement of the $SU(N)_{\pm N}$ gauge theory at the UV fixed point.

We now consider $SU(N)$ gauge theory with fundamental hypermultiplets. The $N_f$ fundamental hypermultiplets induce on the instanton moduli space $N_f$ complex fermionic zero modes carrying the flavor charges and $U(1)$ gauge charge $N-2$. The quantization leads to $N_f$ raising operators $C_a,\, a=1,\cdots,N_f$ and they act on the states as
\begin{equation}\label{eq:fermion-zero-modes-states}
    C_{a_1}\cdots C_{a_r} |0\rangle \ ,
\end{equation}
where $0\le r \le N_f$. These states have $U(1)$ gauge charge $(N-2)(r-N_f/2)$ and flavor charges. We can construct the instanton operators by tensoring these states with the above gaugino contribution and imposing $U(1)$ gauge invariance. Then one can see that there exist candidate broken current supermultiplets having zero $U(1)$ gauge charge when
\begin{equation}\label{eq:instanton-U(1)-projection}
    \kappa \pm N + r-N_f/2 = 0 \ ,
\end{equation}
which may signal the symmetry enhancement of the UV CFT.

If we impose the standard bound $|\kappa|\le N - N_f/2$ as in ~\cite{Tachikawa:2015mha}, one finds that $r$ should be $0$ or $N_f$ 
and the broken current multiplet exists only when
\begin{equation}
   r = 0 : \ \kappa = -(N-N_f/2) \ , \qquad r = N_f : \ \kappa = N-N_f/2 \ .
\end{equation}
The surviving current multiplet with $r=0$ or $r=N_f$ is a singlet under $SU(N_f)$ flavor symmetry and carries the baryoninc $U(1)_B$ flavor charge $-N_f/2$ or $N_f/2$, respectively. 
Thus the $SU(N_f)\times U(1)_B\times U(1)_{in}$ global symmetry will be enhanced as expected to $SU(N_f)\times SU(2)_\pm \times U(1)_\mp$ at the UV fixed point by the instantonic conserved currents, where $\pm$ means linear combinations of the $U(1)_B$ and $U(1)_{in}$ current, namely $J^\mu_{in} \pm (2/N_f)J^\mu_B$. In particular, when $N_f=2N$, both $r=0$ and $r=2N$ states survive and the UV global symmetry is enhanced to $SU(N_f)\times SU(2)_+\times SU(2)_-$.

If we relax the bound on $\kappa$, though, other possibilities occur. Suppose we violate the bound by $n$: 
\begin{equation}\label{eq:constrain-SU-CFTs}
    |\kappa| \le N + n - N_f/2 \ ,
\end{equation}
We find that the broken current multiplets may exist if $r \le n$ or $r \ge N_f-n$.
The states with $r\le n$ can survive when
\begin{equation}
    \kappa = N+r-N_f/2 \ ,
\end{equation}
while the states with $r\ge N_f-n$ can survive when
\begin{equation}
 \kappa = -N+r-N_f/2  \ .
\end{equation}
These states provide candidate broken current multiplets in the rank $r$ antisymmetric representation of the $SU(N_f)$ flavor group.

There is no symmetry group whose adjoint representation is decomposed into irreps involving any rank $r>2$ antisymmetric representation of a subgroup. 
Thus we expect theories with $n>2$ to be truly incompatible with an UV completion. 
The constraint~(\ref{eq:constrain-SU-CFTs}) with $n=2$ agrees with the constraint conjectured from the $(p,q)$ 5-brane web realization of the 5d CFTs in~\cite{Bergman:2014kza,Kim:2015jba,Hayashi:2015fsa}. A similar analysis has been done in~\cite{Yonekura:2015ksa}.

For $r=1$ (or $r=N_f-1$) when $\kappa=N+1-N_f/2$ (or $\kappa=-N-1+N_f/2$) the candidate broken currents transform in the (anti-)fundamental representation of the $SU(N_f)$ flavor symmetry with the $U(1)_B$ charge $-N_f/2+1$ (or $N_f/2-1$). Therefore an UV CFT may exist with enhanced global symmetry $SU(N_f+1)\times U(1)$. The current multiplet of the $SU(N_f+1)$ is in the adjoint representation which is decomposed by current multiplets in the adjoint and a fundamental and an anti-fundamental representation of the subgroup $SU(N_f)$. The fundamental and anti-fundamental current multiplets are generated by following the above procedure in the instanton background. In particular, when $\kappa=\frac{1}{2}$ (or $\kappa=-\frac{1}{2}$) and $N_f = 2N+1$, an additional state with $r=N_f$ (or $r=0$) survives and it gives a current multiplet which is a singlet under the $SU(N_f)$ flavor symmetry. 
Thus in this case we have a bigger symmetry enhancement to $SU(2N+2)\times SU(2)$. Furthermore, when $\kappa=0$ and $N_f=2N+2$, both states $r=1$ and $r=N_f-1$ survive and provide two broken current multiplets in the fundamental and anti-fundamental representations. Therefore the symmetry of the UV CFT may be enhanced to $SU(2N+4)$.

Similarly, the instanton state with $r=2$ (or $r=N_f-2$) generates the broken current multiplet in the antisymmetric representation of the $SU(N_f)$ when $\kappa=N+2-N_f/2$ (or $\kappa=-N-2+N_f/2$). This suggests the global symmetry enhancement of $U(N_f)\times U(1)_{in} \rightarrow SO(2N_f)\times U(1)$ at the UV fixed point.
When $\kappa=1$ (or $\kappa=-1$) and $N_f=2N+2$, one more state with $r=N_f$ (or $r=0$) survive and it provides a current multiplet which is singlet under the $SU(N_f)$. So the enhanced symmetry of the UV fixed point becomes $SO(4N+4)\times SU(2)$.
When $\kappa=\frac{1}{2}$ (or $\kappa=-\frac{1}{2}$) and $N_f=2N+3$, two states with $r=2$ and $r=N_f-1$ (or $r=1$ and $r=N_f-2$) can provide current multiplets in the antisymmetric and the fundamental representations of the $SU(N_f)$ with different $U(1)_B$ charges, $-N+\frac{1}{2}$ and $N+\frac{1}{2}$ respectively. So the enhanced global symmetry of the UV CFT is $SO(4N+8)$. Lastly, when $\kappa=0$ and $N_f=2N+4$, two instanton states with $r=2$ and $r=N_f-2$ survive and they provide current multiplets in the rank $2$ and rank $N_f-2$ antisymmetric representation of the flavor symmetry. It has been conjectured in~\cite{Hayashi:2015fsa,Yonekura:2015ksa} that the $SU(N)_0$ gauge theory with $N_f=2N+4$ fundamental hypermultiplets is expected to be UV complete and has a 6d fixed point. The corresponding 6d theory is the $(D_{N+2},D_{N+2})$ minimal conformal matter theory~\cite{Heckman:2013pva,DelZotto:2014hpa}.

\begin{table}[!h]
\begin{center}
\begin{tabular}{|c|c||c|c|}
\hline
$N_f$ & $SU(N)_{\pm(N+1-N_f/2)}$  &  $N_f$ & $SU(N)_{\pm(N+2-N_f/2)}$\\
\hline
$\le 2N$ & $SU(N_f+1)\times U(1)$ & $\le 2N+1$ & $SO(2N_f)\times U(1)$\\
$2N+1$ & $SU(N_f+1)\times SU(2)$ & $2N+2$ & $SO(2N_f)\times SU(2)$ \\
$2N+2$ & $SU(N_f+2)$ & $2N+3$ & $SO(2N_f+2)$\\
\hline
\end{tabular}
\caption{Enhanced global symmetries of the 5d SCFTs. See also~\cite{Hayashi:2015fsa,Yonekura:2015ksa}.}
\end{center}
\end{table}

The discussion in this subsection strongly supports the duality proposed in this section. Following the fermion zero mode analysis above, the $SU(N+1)$ gauge theory with the CS-level $\kappa=N+3-N_f/2$ may admit a UV completion with a global symmetry $SO(2N_f)\times U(1)$ when $N_f\le 2N+2$ and $SO(2N_f)\times SU(2)$ when $N_f=2N+3$, 
which is the same as the expected UV global symmetry of the dual $Sp(N)$ gauge theory.

\subsection{From $Sp(N)$ to exotic $SU(N+1)$}\label{sec:duality-Sp-SU}
We first discuss the superconformal index and the instanton partition function of $Sp(N)$ gauge theory.
The superconformal index of the $Sp(N)$ gauge theory with $N_f$ fundamental flavors takes the form
\begin{align}\label{eq:superconformal-index-Sp}
    &I^{N,N_f}_{Sp}(w_a,\mathfrak{q}_{Sp};p,q) = \frac{(I_V)^N}{N!} \cdot \cr
     &\oint \prod_{i=1}^N\frac{dz_i}{2\pi iz_i} \left|\frac{\prod_{i>j}^N (z_i^\pm z_j^\pm;p,q)_\infty \prod_{i=1}^N(z_i^{\pm2};p,q)_\infty}{\prod_{i=1}^N\prod_{a=1}^{N_f}(\sqrt{pq}z_i^\pm/ w_a;p,q)_\infty} Z_{Sp,{\rm inst}}^{N,N_f} (z_i,w_a,\mathfrak{q}_{Sp};p,q)\right|^2 \ .
\end{align}

The function $Z_{Sp,{\rm inst}}^{N,N_f}$ is the instanton partition function of $Sp(N)$ gauge theory, which can be computed using localization of the path integral on the instanton moduli space given in~\cite{Nekrasov:2004vw,Shadchin:2004yx}. The 5d $Sp(N)$ instanton partition functions are studied in great detail in~\cite{Kim:2012gu,Hwang:2014uwa}. The results are summarized in appendix~\ref{sec:instanton-partition-function}.

The $Sp(N)$ gauge theory has $O(k)$ dual gauge group in the ADHM quantum mechanics. At each instanton sector we will compute two partition functions $Z_k^+$ and $Z_k^-$ for $O(k)_+$ and $O(k)_-$, respectively,
\begin{equation}
    Z^\pm_{k}(\alpha,m;\epsilon_{1,2}) = \frac{1}{|W|}\oint \prod_{I=1}^n\frac{d\phi_I}{2\pi i}\, Z_{\rm vec}^\pm(\phi,\alpha;\epsilon_{1,2})\prod_{a=1}^{N_f}Z^\pm_{\rm fund}(\phi,\alpha,m_a;\epsilon_{1,2}) \ ,
\end{equation}
with $k=2n+\chi$ and $\chi= 0$ or $1$. See appendix~\ref{sec:Sp-instanton-partition-function} for details.
In the following, we will assume that $\theta=0$ for odd $N+ N_f$ and $\theta=\pi$ for even $N+ N_f$ while choosing the same mass signs for all matter fields for notational convenience. 

The $k$ instanton partition function can be written as
\begin{align}
    Z^k_{Sp(odd)}(\alpha,m;\epsilon_{1,2}) &= \frac{1}{2}\left[Z_k^+(\alpha,m;\epsilon_{1,2})  + Z_k^-(\alpha,m;\epsilon_{1,2}) \right] \ ,\cr
    Z^k_{Sp(even)}(\alpha,m;\epsilon_{1,2}) &= \frac{(-1)^k}{2}\left[Z_k^+(\alpha,m;\epsilon_{1,2})  - Z_k^-(\alpha,m;\epsilon_{1,2}) \right] \ .
\end{align}
For instance, when $k=1$, there is no integral and the instanton partition function is simply given by sum of two partition functions
\begin{equation}
    Z^+_{k=1} = \frac{p^{3/2}q^{3/2}\prod_{a=1}^{N_f}w_a^{-1/2}(-1+w_a)}{(1-p)(1-q)\prod_{i=1}^N(1-\sqrt{pq}z_i^\pm)} \ , \quad Z^-_{k=1}=  \frac{p^{3/2}q^{3/2}\prod_{a=1}^{N_f}w_a^{-1/2}(1+w_a)}{(1-p)(1-q)\prod_{i=1}^N(1+\sqrt{pq}z_i^\pm)} \ ,
\end{equation}
for $O(1)_+$ and $O(1)_-$, respectively.

For higher instantons, we need to evaluate the contour integral over $O(k)$ Coulomb branch parameters using the JK-residue prescription.
For example, the 2-instanton partition function has a contour integral over one variable $\phi_1$ for $O(2)_+$ sector, whereas has no integral for $O(2)_-$ sector. The JK-prescription tells us that the poles we should pick up are
\begin{equation}
    \phi_1 \pm \alpha_i+\epsilon_+ = 0 \,, \quad 2\phi_1 +\epsilon_1 = 0 \,, \quad 2\phi_1 +\epsilon_2 = 0 \ ,  \quad (\text{`0'}\equiv 0 \ {\rm mod} \ 2\pi) \ .
\end{equation}
The sum over the JK-residues plus the $O(2)_-$ contribution gives the full 2-instanton partition function.

Furthermore, when $N_f=2N+4$, there exists a continuum in the instanton quantum mechanics associated to a classical noncompact Coulomb branch.
The partition function involves an extra contribution coming from this continuum which should be removed to obtain the correct QFT partition function.
We find that the extra contribution takes the form
\begin{equation}\label{eq:U1factor-Sp-Nf=8}
    Z_{Sp,{\rm extra}}^{N_f=8} = {\rm PE}\left[-\frac{1+pq}{2(1-p)(1-q)}\mathfrak{q}_{Sp}^2\right] \ .
\end{equation}
The half-integral coefficient in the letter index obviously shows that this is coming from a continuum. This correction factor can also be obtained by taking a half of the residue at infinity $\phi_I=\pm\infty$ in the integral formula. The QFT instanton partition function is therefore defined as
\begin{equation}
    Z_{Sp,{\rm inst}}^{N,N_f=8} = Z_{Sp,QM}^{N,N_f=8}/ Z_{Sp,{\rm extra}}^{N_f=8} \ ,
\end{equation}
where $Z_{Sp,QM}$ is the standard instanton partition function before removing the extra factor.

Next, we need to assemble the instanton partition function and 1-loop determinants into the hemisphere partition function for 
Dirichlet boundary conditions:
\begin{equation}
  II_{Sp}^{N,N_f}(z_i,w_a,\mathfrak{q}_{Sp};p,q) = \frac{\prod_{i>j}^N(pqz_i^\pm z_j^\pm)_\infty \prod_{i=1}^N(pq z_i^{\pm2};p,q)_\infty}{\prod_{i=1}^N\prod_{a=1}^{N_f}(\sqrt{pq}z_i^\pm/ w_a;p,q)_\infty} Z^{N,N_f}_{Sp,{\rm inst}}(z_i,w_a,\mathfrak{q}_{Sp};p,q) \ .
\end{equation}

The hemisphere index for the $SU(N+1)$ theory is similarly defined and takes the form
\begin{equation}
  II_{SU}^{N+1,N_f}(z_i,w_a,\mathfrak{q}_{SU};p,q) = \frac{\prod_{i\neq j}^{N+1}(pqz_i/z_j;p,q)_\infty}{\prod_{i=1}^{N+1}\prod_{a=1}^{N_f}(\sqrt{pq}z_i/w_a;p,q)_\infty} Z_{SU,{\rm inst}}^{N+1,N_f}(z_i,w_a,\mathfrak{q}_{SU};p,q) \ .
\end{equation}
with an a-priory unknown instanton contribution $Z_{SU,{\rm inst}}^{N+1,N_f}$.

The degrees of freedom on the duality wall have the 4d index contribution
\begin{equation}
  \frac{\prod_{i=1}^{N+1}\prod_{j=1}^N\Gamma(\sqrt{\lambda}z'_i z_j^{\pm})}{ \prod_{i>j}^{N+1} \Gamma(\lambda z'_iz'_j)} \ ,
\end{equation}
where $z'$ and $z$ are the fugacities for the bulk $SU(N+1)$ and $Sp(N)$ gauge groups.
To couple this to the 5d index, we need to multiply the 4d $Sp(N)$ vector multiplet contribution and integrate the $Sp(N)$ gauge fugacities $z$. The result is given by
\begin{equation}\label{eq:duality-action-Sp-SU}
  \hat{D}II_{Sp}^{N,N_f} = \oint d\mu_{z_i} \Delta^{(C)}(z,z',\lambda) II_{Sp}^{N,N_f}(z_i,\mathfrak{q}_{Sp}, w_a) \ ,
\end{equation}
where $w_a$ is the fugacity for $U(N_f)\subset SO(2N_f)$ flavor symmetry and
\begin{equation}
    \Delta^{(C)}(z,z',\lambda) =  \frac{I_V^N\prod_{i=1}^{N+1}\prod_{j=1}^N\Gamma(\sqrt{\lambda}z'_i z_j^{\pm})}{ \prod_{i>j}^{N+1} \Gamma(\lambda z'_iz'_j)\prod_{i>j}^N \Gamma(z_i^\pm z_j^\pm) \prod_{i=1}^N\Gamma(z_i^{\pm 2})} \ .
\end{equation}

Our conjecture is that the duality action $\hat{D}$ on the hemisphere index of the $Sp(N)$ gauge theory converts it into the hemisphere index of the $SU(N+1)$ gauge theory in the other side of the wall.
 So the following relation is expected to hold
\begin{equation}\label{eq:duality-relation-SU-Sp}
  \hat{D}II_{Sp}^{N,N_f}(z_i, w_a,\mathfrak{q}_{Sp};p,q) =II_{SU}^{N+1,N_f}(z'_i,w'_a,\mathfrak{q}_{SU};p,q)  \ .
\end{equation}
In this relation, the fugacities for the global symmetry in two sides of the wall should be identified as 
\begin{equation}\label{eq:parameter-identity-SU-Sp}
 w_a = \lambda^{1/2}w_a' \ , \quad \mathfrak{q}_{Sp} = \lambda^{(N+1)/2}\prod_{a=1}^{N_f}(w_a)^{-1/2} \ , \quad \mathfrak{q}_{SU} = \lambda^{-1}\prod_{a=1}^{N_f}(w_a')^{-1/2} \ .
\end{equation}
The first relation comes from the the constraint of the 4d superpotential. We determined the second and the third relations experimentally from the duality relations (\ref{eq:duality-relation-SU-Sp}) and (\ref{eq:duality-relatoin-Sp-SU}),
but they agree with the relations expected from cancellation of the mixed 't Hooft anomalies for the duality wall. 

The simplest example would be the duality action between $Sp(2)$ and $SU(3)$ gauge theories with $N_f$ flavors. To evaluate the integral in~(\ref{eq:duality-relation-SU-Sp}) and see the duality relation, we should choose a particular contour. We take the contour to be along a unit circle while assuming $x\ll \lambda <1$. 

 Acting with the duality wall, we find the following result for $N_f=0$:
\begin{align}
    & \hat{D}II^{2,0}_{Sp}(z_i,\mathfrak{q}_{Sp}) \equiv II^{3,0}_{SU}(z_i,\mathfrak{q}_{SU}) \\
    =& 1 + \left(-\chi^{SU(3)}_{\bf8}+\chi^{SU(3)}_{\bf3}\mathfrak{q}_{SU}\right)\left(x^2+\chi_{\bf2}^{SU(2)}(y)x^3+\chi_{\bf3}^{SU(2)}(y)x^4\right) \cr
    &+ \left(\chi_{\bf8}^{SU(3)}+\chi_{\bf10}^{SU(3)}+\chi_{\bf\overline{10}}^{SU(3)}-\left(\chi_{\bf3}^{SU(3)}+\chi_{\bf15}^{SU(3)}\right)\mathfrak{q}_{SU}+\chi_{\bf6}^{SU(3)}\mathfrak{q}_{SU}^2\right)x^4+\mathcal{O}(x^5) \nonumber
\end{align}
where $\chi_{\bf r}^{SU(3)}$ is the $SU(3)$ character of the dimension ${\bf r}$ irrep with fugacities $z_i$.
We checked that the right hand side agrees with the perturbative part of the $SU(3)$ hemisphere index and admits an expansion in non-negative powers of $\mathfrak{q}_{SU}$, up to the order 
$x^5$.

For general $N_f\le8$, we find
\begin{align}\label{eq:duality-hemisphere-Sp-SU}
    &\hat{D}II^{2,N_f}_{Sp}(z_i,w_a,\mathfrak{q}_{Sp}) \equiv II^{3,N_f}_{SU}(z_i,w_a',\mathfrak{q}_{SU}) \\
    & = 1 + \chi_{\bf 3}^{SU(3)}\chi_{\tiny\yng(1)}^{U(N_f)}\,x + \bigg[-\chi_{\bf 8}^{SU(3)}(z) + \chi_{\bf 6}^{SU(3)}\chi_{\tiny\yng(2)}^{U(N_f)}+\chi_{\bf\bar{3}}^{SU(3)}\chi_{\Lambda^2\tiny\yng(1)}^{U(N_f)} \cr
    & \quad +\chi_{\bf2}^{SU(2)}(y)\chi_{\bf 3}^{SU(3)}\chi_{\tiny\yng(1)}^{U(N_f)}+\left(\chi_{\bf3}^{SU(3)}+\chi_{\Lambda^2\tiny\yng(1)}^{U(N_f)}+\chi_{\Lambda^8\tiny\yng(1)}^{U(N_f)}\right)\prod_{a=1}^{N_f}\sqrt{w'_a}\mathfrak{q}_{SU} \bigg]x^2 + \mathcal{O}(x^3) \ ,\nonumber
\end{align}
where $\chi_{Y}^{U(N_f)}$ is the $U(N_f)$ character with fugacities $(w_a')^{-1}$ of a irrep labeled by a Young tableau $Y$. We have identified the parameters as~(\ref{eq:parameter-identity-SU-Sp}). The perturbative part on the right hand side agrees with that of the $SU(3)$ theory and the other parts are expanded by non-negative powers of $\mathfrak{q}_{SU}$. This relation has been checked at least up to $x^3$ order.

In appendix~\ref{sec:partition-function-exotic-SU(3)}, we shall suggest a UV prescription of the instanton moduli space of our exotic $SU(3)$ theory with matter fields, whose partition function precisely reproduces the right hand side. In addition, we will explicitly compute the superconformal index of this $SU(3)$ theory and show the desired global symmetry enhancement at the UV fixed point.

One can also consider the generalization to higher rank gauge theories. Acting with the duality wall on the hemisphere index of the $Sp(3)$ theories, we obtain
\begin{align}
    &\hat{D}II^{3,N_f\le3}_{Sp}(z_i,w_a,\mathfrak{q}_{Sp}) \equiv II^{4,N_f\le3}_{SU}(z_i,w'_a,\mathfrak{q}_{SU}) \\
    =& 1+\chi_{\bf4}^{SU(4)}\chi_{\tiny\yng(1)}^{U(N_f)}x + \bigg[-\chi_{\bf15}^{SU(4)}
    +(\chi_{\bf4}^{SU(4)})^2\chi_{\Lambda^2\tiny\yng(1)}^{U(N_f)} + \chi_{\bf10}^{SU(4)}\chi_{\tiny\yng(1)}^{U(N_f)}((w')^{-2}) \cr
    &+\chi_{\bf2}^{SU(2)}(y)\chi_{\bf4}^{SU(4)}\chi_{\tiny\yng(1)}^{U(N_f)} + \left(\chi_{\bf6}^{SU(4)}(z) + \chi_{\Lambda^2\tiny\yng(1)}^{U(N_f)}\right)\prod_{a=1}^{N_f}\sqrt{w_a}\mathfrak{q}_{SU}\bigg]x^2 + \mathcal{O}(x^3) \ . \nonumber
\end{align}
for $N_f\le3$, by identifying the parameters as (\ref{eq:parameter-identity-SU-Sp}).
We checked that the right hand side agrees with the perturbative part of the $SU(4)$ hemisphere index and admits an expansion in non-negative powers of $\mathfrak{q}_{SU}$, at least up to the order $x^4$.
 
Of course, the duality wall can also act in the opposite direction, from $SU(N+1)$ to $Sp(N)$. 
\begin{equation}\label{eq:duality-relatoin-Sp-SU}
    \hat{D}II^{N+1,N_f}_{SU}(z',w_a',\mathfrak{q}_{SU}) \equiv \oint d\mu_{z'} \Delta^{(A)}(z',z,\lambda) II^{N+1,N_f}_{SU}(z'_i,w_a',\mathfrak{q}_{SU}) = II_{Sp}^{N,N_f}(z_i,w_a,\mathfrak{q}_{Sp}) \ ,
\end{equation}
where the 4d index of the boundary degrees of freedom involving the 4d vector multiplet is given by
\begin{equation}
    \Delta^{(A)}(z',z,\lambda) =\frac{ I_V^N\prod_{i=1}^N\prod_{j=1}^{N+1}\Gamma(\sqrt{\lambda}^{-1}z_i^\pm/z'_j)}{\prod_{i\neq j}^{N+1}\Gamma(z_i'/z_j')\prod_{i>j}^{N+1}\Gamma(\lambda^{-1}(z'_iz'_j)^{-1})} \ .
\end{equation}
The contour is chosen along the unit circle with an assumption $x\ll \lambda^{-1}<1$ and the parameters are matched as (\ref{eq:parameter-identity-SU-Sp}).

Of course, this follows from the $CA$ and $AC$ inversion formula introduced in~\cite{2004math.....11044S}:
\begin{align}
    &\oint d\mu_{z'} \Delta^{(A)}(z',x,\lambda) \oint d\mu_{z} \Delta^{(C)}(z,z',\lambda) f(z) = f(x) \ , \cr
    &\oint d\mu_{z} \Delta^{(C)}(z,x,\lambda) \oint d\mu_{z'} \Delta^{(A)}(z',z,\lambda) f(z') = f(x) \ .
\end{align}
Note that the contours should be chosen along unit circles by assuming $x\ll \lambda^{-1}<1$ for the $A$-type integral, but by assuming $x\ll \lambda <1$ for the $C$-type integral as specified already.

\subsection{Wilson loops}
In this subsection, we will study the properties of BPS Wilson loops under the conjectural duality in the previous sections. We will focus on the simplest cases: fundamental Wilson loops of the $Sp(2)$ and $SU(3)$ gauge theories meeting at the interface. The Wilson loops on two sides of the wall are connected at the boundary by the bi-fundamental chiral multiplet $q$. 
The chiral multiplet $q$ has charge $\frac{1}{2}$ under the non-anomalous $U(1)_\lambda$ global symmetry. To cancel the global charge when it couples to the Wilson loops, we combine the gauge Wilson loops with a flavor Wilson loop for the $U(1)_\lambda$, with flavor charge $\frac{1}{2}$, which follows from the similar argument in section~\ref{sec:Wilson-Loops}.
We will compute hemisphere indices and test this duality property between two fundamental Wilson loops.

We first compute the hemisphere indices with fundamental Wilson loops inserted at the origin. We need to compute the instanton partition function in the presence of Wilson loops. As explained in section~\ref{sec:Wilson-Loops}, Wilson loops are represented by equivariant Chern characters in the localization, and that for the fundamental Wilson loop is given in~(\ref{eq:equivariant-Chern-E}). Then the localized partition function can be written in terms of the equivariant Chern characters as in~(\ref{eq:Wilson-loop}).

For the $Sp(N)$ gauge theory, the equivariant Chern character for the fundamental Wilson loop can be written, at $k$-instantons, as
\begin{align}
    Ch_{\rm fund}^+(e^{\alpha},e^\phi)&=\sum_{i=1}^N(e^{\alpha_i}+e^{-\alpha_i}) - (1-p)(1-q)(pq)^{-1/2}\sum_{I=1}^n(e^{\phi_I} + e^{-\phi_I} + \chi)\ , \\
    Ch_{\rm fund}^-(e^{\alpha},e^\phi)&=\sum_{i=1}^N(e^{\alpha_i}+e^{-\alpha_i}) - (1-p)(1-q)(pq)^{-1/2}\sum_{I=1}^n(e^{\phi_I} + e^{-\phi_I} + e^{i\pi}\chi)\ ,\nonumber
\end{align}
with $k=2n+\chi$ and $\chi=0$ or $1$. Here the superscripts $\pm$ means those for $O(k)_\pm$ sectors. Then the 1-instanton partition function can be written as
\begin{align}
    W_{k=1}^+ &= \frac{(pq)^{3/2}\left(\sum_{i=1}^N(e^{\alpha_i}+e^{-\alpha_i}) - (1-p)(1-q)(pq)^{-1/2}\right)\prod_{a=1}^{N_f}2\sinh\frac{m_a}{2}}{(1-p)(1-q)\prod_{i=1}^N(1-\sqrt{pq}e^{\pm\alpha_i})} \ , \cr
    W_{k=1}^- &= \frac{(pq)^{3/2}\left(\sum_{i=1}^N(e^{\alpha_i}+e^{-\alpha_i}) + (1-p)(1-q)(pq)^{-1/2}\right)\prod_{a=1}^{N_f}2\cos\frac{m_a}{2}}{(1-p)(1-q)\prod_{i=1}^N(1+\sqrt{pq}e^{\pm\alpha_i})} \ .
\end{align}
There could be extra instanton corrections to the Wilson loop index as we have seen in section~{\ref{sec:Wilson-Loops}.
For the cases in this section, however, we find that there is no such corrections up to certain order in $x$ expansion.

Now we consider the duality wall action on the hemisphere index of the $Sp(N)$ theory with the fundamental Wilson loop.
We propose that the fundamental Wilson loop partition function of the $Sp(N)$ theory is mapped to that of the $SU(N+1)$ theory after passing through the duality wall as follows:
\begin{equation}
    \hat{D}W_{\rm fund}^{Sp(N),N_f}(z_i,w_a,\lambda) = \lambda^{1/2}W_{\rm fund}^{SU(N+1),N_f}(z_i,w'_a,\lambda^{-1}) \ ,
\end{equation}
with the parameter identification in~(\ref{eq:parameter-identity-SU-Sp}). The duality action $\hat{D}$ is defined in the same way as in~(\ref{eq:duality-action-Sp-SU}), but the hemisphere indices $II^{N,N_f}_{Sp,SU}$ in both sides are replaced by the Wilson loop indices $W^{N,N_f}_{Sp,SU}$. The prefactor $\lambda^{1/2}$ is due to the $U(1)_\lambda$ flavor Wilson loop.

We compute the hemisphere indices of the $Sp(2)$ gauge theories and test this duality. 
We obtain
\begin{align}\label{eq:dual-action-Sp(2)-Nf=0}
    &\hat{D}W_{\rm fund}^{Sp(2),0}(z_i,\mathfrak{q}_{Sp}) \equiv \lambda^{1/2}W_{\rm fund}^{SU(3),1}(z_i,\mathfrak{q}_{SU}) \\
    =&\chi_{\bf3}^{SU(3)}(z)+\left(-\chi_{\bf15}^{SU(3)}(z)-\chi_{\bf\bar{3}}^{SU(3)}(z)^2+\chi_{\bf6}^{SU(3)}\mathfrak{q}_{SU}\right)x^2\cr
    & +\chi_{\bf2}^{SU(2)}(y)\chi_{\bf3}^{SU(3)}(z)x^3+\chi_{\bf2}^{SU(2)}(y)\chi_{\bf3}^{SU(3)}(z)^2\left(-\chi_{\bf\bar{3}}^{SU(3)}(z)+\mathfrak{q}_{SU}\right)x^3+\mathcal{O}(x^4)\nonumber \ ,
\end{align}
for $N_f=0$, and
\begin{align}\label{eq:dual-action-Sp(2)-Nf=1}
    &\hat{D}W_{\rm fund}^{Sp(2),1}(z_i,w,\mathfrak{q}_{Sp}) \equiv \lambda^{1/2}W_{\rm fund}^{SU(3),1}(z_i,w',\mathfrak{q}_{SU}) \cr
    =&\chi_{\bf3}^{SU(3)}(z)+\left((w_1')^{-1}\chi_{\bf3}^{SU(3)}(z)^2+(w_1')^{-1/2}\mathfrak{q}_{SU}\right)x + \left(-\chi_{\bf15}^{SU(3)}(z)-\chi_{\bf\bar{3}}^{SU(3)}(z)^2\right)x^2 \cr
    &+\left((w_1')^{-2}\chi_{\bf10}^{SU(3)}(z)+(w_1')^{-2}\chi_{\bf8}^{SU(3)}(z)+(w_1')^{-1}\chi_{\bf2}^{SU(2)}(y)\chi_{\bf3}^{SU(3)}(z)^2\right)x^2 \cr
    & + \left(\chi_{\bf6}^{SU(3)}(z)+(w_1')^{-2}\chi_{\bf3}^{SU(3)}(z)\right)(w_1')^{1/2}\mathfrak{q}_{SU}\,x^2+\mathcal{O}(x^3) \ ,
\end{align}
for $N_f=1$. 
We have checked that, for each $N_f=0,1$ case, the right hand side admits an expansion in non-negative powers of $\mathfrak{q}_{SU}$ and the perturbative part agrees with that of the $SU(3)$ gauge theory, up to $x^4$ order. It also turns out that the right hand sides agree up to $x^4$ order with the hemisphere indices of the $SU(3)$ theories with the fundamental Wilson loop whose instanton partition functions are computed using the UV prescription given in appendix~\ref{sec:partition-function-exotic-SU(3)}.

\section*{Acknowledgements}

The research of DG and HK was supported by the Perimeter Institute for Theoretical Physics. Research at Perimeter Institute is supported by the Government of Canada through Industry Canada and by the Province of Ontario through the Ministry of Economic Development and Innovation.

\appendix

\section{5d Nekrasov's instanton partition function}\label{sec:instanton-partition-function}

The moduli space of instantons has complicated singularities which are associated to one or more instantons shrinking to zero size. 
In the context of five-dimensional supersymmetric gauge theories, these field configurations are outside the obvious regime of validity 
of the gauge theory description of the theory. Correspondingly, the definition of the gauge theory instanton partition functions through equivariant localization 
on the instanton moduli spaces requires a prescription of how to deal with the singularities, which will depend on a choice of UV completion of the 
gauge theory. 

It is very challenging to work directly on the singular moduli spaces. Even in the absence of extra matter fields this was done only recently
\cite{2014arXiv1406.2381B} using the technology of equivariant intersection cohomology. Extra matter fields, in the form of hypermultiplets 
transforming in some representation of the gauge group, provide additional fermion zero modes in the instanton background which 
are encoded into some appropriate characteristic class inserted in the equivariant integral. The correct description of these characteristic classes 
over the singular instanton moduli space is poorly understood. 

The standard alternative to working with the singular moduli spaces, available for classical groups only, is to employ the 
ADHM technology to provide a resolution of the singularities in the monopole moduli space. The ADHM construction has a clear motivation 
in terms of a string theory UV completion. It realizes the instantons as D0 branes in presence of other brane systems which 
engineer the gauge theory itself. 

It is important to realize that this is not obviously the same as the quantum field theory UV completion 
we are after, which should involve some 5d SCFT or perhaps a 6d SCFT. Luckily, it appears that the answers computed 
by the ADHM construction can be easily corrected to sensible field theory answers, as long as the matter content of the gauge theory does admit 
a reasonable string theory lift. When that is not the case, it is not obvious that a construction of the correct bundle of fermion zero modes
will actually be available in the ADHM description of the moduli space. We will encounter some of these issues in the Sections 
\ref{sec:hypermultiplet-contribution} and \ref{sec:partition-function-exotic-SU(3)}. 
  
When the ADHM construction for a gauge group $G$ exists, it can be described as a one dimensional gauged linear sigma model of dual gauge group $\hat{G}$, called the ADHM quantum mechanic (ADHM QM). The Higgs branch of this theory coincides with the instanton moduli space. This theory has bosonic $SU(2)_1\times SU(2)_2\times SU(2)_R$ symmetry and 4 real supercharges $\bar{Q}_{\dot{\alpha}}^A$, where the $SO(4) =SU(2)_1\times SU(2)_2$ corresponds to the spatial $\mathbb{R}^4$ rotation and the $SU(2)_R$ is the R-symmetry in 5d. The indices $\alpha=1,2,\ \dot\alpha = 1,2, \ A=1,2$ are the doublets of $SU(2)_1,\ SU(2)_2,\ SU(2)_R$ symmetries respectively.
The ADHM QM consists of the (0,4) hypermultiplets
\begin{equation}
	(B_{\alpha\dot\alpha}, \lambda_{\alpha}^A) \ \ {\rm in \ adjoint \ rep}\,, \quad (q_{\dot\alpha},\psi^A) \ \ {\rm in \ fundamental \ rep}
\end{equation}
and the vector multiplet $(A_t,\phi,\bar\lambda_{\dot\alpha}^A)$. The bosonic fields in the hypermultiplets are called ADHM data.

In order to apply the ADHM construction to a five-dimensional gauge theory we need to find within the ADHM quantum mechanics a construction of the 
bundle of fermionic zero modes associated to the hypermultiplets. Concretely, that means adding extra fields to the quantum mechanics 
which add the appropriate fermionic bundle on top of the Higgs branch of the theory. If a string theory description of the 
gauge theory is available, one can usually read off from it the required extra degrees of freedom. 

If the instanton moduli space was not singular, it would be possible to derive simple relationships between the characteristic classes in the 
equivariant integral associated to hypermultiplets in different representations. If a string theory construction is not available
for some representation, one can try to guess an ADHM description for that representation by imposing the same relationship on the 
the corresponding characteristic classes/equivariant indices in the ADHM equivariant integral. 
Some equivariant indices for hypermultiplets in simple representations are given in~\cite{Shadchin:2004yx}. 
We will present below the equivariant indices and partition functions for the hypermultiplets used in the main text
and discuss the difficulties associated to this naive choice of UV completion. 

The instanton partition function takes the form of the instanton series expansion as
\begin{equation}
	Z_{\rm inst} = \sum_{k=0}^\infty \mathfrak{q}^k Z_k \ ,
\end{equation}
with an instanton counting parameter $\mathfrak{q}$.
The $Z_k$ is the $k$ instanton partition function. It is the supersymmetric Witten index of the 1d ADHM QM. It also admits a path integral representation. The supersymmetric localization was employed to evaluate this path integral of the ADHM quantum mechanics in~\cite{Nekrasov:2002qd,Nekrasov:2003rj}. See also~\cite{Hwang:2014uwa,Cordova:2014oxa,Hori:2014tda} for 1d localization calculations. We will now summarize some results.

\subsection{$SU(N)$ partition function}
The ADHM quantum mechanics has dual gauge group $\hat{G}=U(k)$ for $k$ instantons. In the bulk 5d theory, one can also turn on a classical CS coupling $\kappa$ when $N\ge3$. It induces a Chern-Simons coupling in the 1d quantum mechanics~\cite{Kim:2008kn,Collie:2008vc}.
\begin{equation}
	\kappa \int dt\, {\rm Tr}(A_t-\phi) \ .
\end{equation}
The $k$ instanton partition function takes the following integral expression
\begin{equation}
	Z_k(\alpha,m;\epsilon_{1,2}) = \frac{1}{k!}\oint \prod_{I=1}^k\frac{d\phi_I}{2\pi i}\, e^{-\kappa \sum_{I=1}^k\phi_I} Z_{\rm vec}(\phi,\alpha;\epsilon_{1,2}) \prod_{a}Z_{R_a}(\phi,\alpha,m_a;\epsilon_{1,2}) \ ,
\end{equation}
where $Z_{R_a}^k$ is the contribution from a hypermultiplet in $R_a$ representation and $m_a$ is the mass parameter. We will often use fugacities $z_i\equiv e^{\alpha_i}, w_a \equiv e^{m_a}$. The vector multiplet factor is
\begin{equation}\label{eq:SU-instanton-vector}
	Z_{\rm vec}(\phi,\alpha;\epsilon_{1,2}) = \frac{\prod_{I\neq J}^k2\sinh\frac{\phi_I-\phi_J}{2} \prod_{I,J}^k2\sinh\frac{\phi_I-\phi_J +2\epsilon_+}{2}}{\prod_{I,J}^k2\sinh\frac{\phi_I-\phi_J+\epsilon_1}{2}2\sinh\frac{\phi_I-\phi_J+\epsilon_2}{2} \prod_{i=1}^N\prod_{I=1}^k2\sinh\frac{\pm(\phi_I-\alpha_i)+\epsilon_+}{2}} \ .
\end{equation}
The hypermultiplet factor will be discussed later.

We still have the contour integral to be evaluated. The contour integral of the instanton partition function should be performed using the Jeffrey-Kirwan method~\cite{Hwang:2014uwa}. If the hypermultiplet factor has only fermionic contributions, as our naive expectation from the zero mode analysis in the 5d QFT, we need to take into account only the vector multiplet factor. The JK-prescription tells us that the residue sum of the following poles will give the final result.
\begin{equation}
	\phi_I -\alpha_i+\epsilon_+ = 0 \ , \quad \phi_I-\phi_J+\epsilon_1 = 0 \ , \quad \phi_I-\phi_J+\epsilon_2 = 0 \ ,
\end{equation}
with $I>J$. However, we will see that the hypermultiplets can introduce extra bosonic degrees for the UV completion of their zero modes. Thus they can also provide nontrivial JK-poles above the poles from the vector multiplet. We will discuss some examples below.

\subsection{$Sp(N)$ partition function}\label{sec:Sp-instanton-partition-function}
For $Sp(N)$ gauge theory, the ADHM quantum mechanics has $\hat{G}=O(k)$ dual gauge group. Since the $O(k)$ group has two disconnected components $O(k)_+$ and $O(k)_-$, we will get two partition functions $Z_k^+$ and $Z_k^-$ at each instanton sector. The $k$ instanton partition function is then given by a sum of these two functions. In addition, the $Sp(N)$ gauge theory has a $\mathbb{Z}_2$ valued $\theta$ angle associated with $\pi_4\left(Sp(N)\right)=\mathbb{Z}_2$~\cite{Douglas:1996xp}.
Two possible $\theta$ parameters lead to the following two different combinations~\cite{Bergman:2013ala,Bergman:2013aca}:
\begin{equation}
	Z_k^{Sp} = \left\{ \begin{array}{ll} \frac{1}{2}(Z_k^+ + Z_k^-) & \ , \  \theta = 0 \\   \frac{(-1)^k}{2}(Z_k^+ - Z_k^-) & \ , \ \theta = \pi \end{array} \right. \ .
\end{equation}
When the theory couples to more than one fundamental hypermultiplet, the $\theta$ angle becomes unphysical because it can be effectively absorbed by flipping the sign of a single mass of one fundamental matter.

The $k$ instanton partition function takes the form
\begin{equation}
	Z^\pm_k(\alpha,m;\epsilon_{1,2}) = \frac{1}{|W|}\oint \prod_{I=1}^n\frac{d\phi_I}{2\pi i}\, Z_{\rm vec}^\pm(\phi,\alpha;\epsilon_{1,2})\prod_{a}Z^\pm_{R_a}(\phi,\alpha,m_a;\epsilon_{1,2}) \ ,
\end{equation}
with $k=2n+\chi$ and $\chi= 0$ or $1$.
The Weyl factor is given by
\begin{equation}\label{eq:Ok-Weyl-factors}
	|W|^{\chi=0}_+ = \frac{1}{2^{n-1}n!} \ , \quad |W|^{\chi=1}_+ = \frac{1}{2^nn!} \ , \quad |W|^{\chi=0}_- = \frac{1}{2^{n-1}(n-1)!} \ , \quad |W|^{\chi=1}_- = \frac{1}{2^nn!} \ .
\end{equation}
The vector multiplet for $O(k)_+$ sector gives the contribution
\begin{eqnarray}
	Z^+_{\rm vec}= && 
	\left[ \frac{1}{2\sinh\frac{\pm\epsilon_-+\epsilon_+}{2}\prod_{i=1}^N2\sinh\frac{\pm\alpha_i+\epsilon_+}{2}}
	\prod_{I=1}^n \frac{2\sinh\frac{\pm\phi_I}{2}2\sinh\frac{\pm\phi_I+2\epsilon_+}{2}}{2\sinh\frac{\pm\phi_I\pm\epsilon_-+\epsilon_+}{2}}
	\right]^\chi \\
	&& \times \prod_{I=1}^n\frac{2\sinh\epsilon_+}{2\sinh\frac{\pm\epsilon_-+\epsilon_+}{2}\prod_{i=1}^N 2\sinh\frac{\pm\phi_I\pm\alpha_i+\epsilon_+}{2}}\cdot
	\frac{\prod_{I>J}^n2\sinh\frac{\pm\phi_I\pm\phi_J}{2}2\sinh\frac{\pm\phi_I\pm\phi_J+2\epsilon_+}{2}}{\prod_{I=1}^n2\sinh\frac{\pm2\phi_I\pm\epsilon_-+\epsilon_+}{2}\prod_{I>J}^n2\sinh\frac{\pm\phi_I\pm\phi_J\pm\epsilon_-+\epsilon_+}{2}} \ .\nonumber
\end{eqnarray}
For $O(k)_-$ sector, the vector multiplet contribution is
\begin{eqnarray}
	Z^-_{\rm vec} = &&
	\frac{1}{2\sinh\frac{\pm\epsilon_-+\epsilon_+}{2}\prod_{i=1}^N2\cosh\frac{\pm\alpha_i+\epsilon_+}{2}}
	\prod_{I=1}^n \frac{2\cosh\frac{\pm\phi_I}{2}2\cosh\frac{\pm\phi_I+2\epsilon_+}{2}}{2\cosh\frac{\pm\phi_I\pm\epsilon_-+\epsilon_+}{2}}
	 \\
	&& \times \prod_{I=1}^n\frac{2\sinh\epsilon_+}{2\sinh\frac{\pm\epsilon_-+\epsilon_+}{2}\prod_{i=1}^N 2\sinh\frac{\pm\phi_I\pm\alpha_i+\epsilon_+}{2}}\cdot
	\frac{\prod_{I>J}^n2\sinh\frac{\pm\phi_I\pm\phi_J}{2}2\sinh\frac{\pm\phi_I\pm\phi_J+2\epsilon_+}{2}}{\prod_{I=1}^n2\sinh\frac{\pm2\phi_I\pm\epsilon_-+\epsilon_+}{2}\prod_{I>J}^n2\sinh\frac{\pm\phi_I\pm\phi_J\pm\epsilon_-+\epsilon_+}{2}} \ , \nonumber 
\end{eqnarray}
with $k=2n+1$ and
\begin{eqnarray}
	Z^-_{\rm vec} = &&
	\frac{2\cosh\epsilon_+}{2\sinh\frac{\pm\epsilon_-+\epsilon_+}{2}2\sinh(\pm\epsilon_-+\epsilon_+)\prod_{i=1}^N2\sinh(\pm\alpha_i+\epsilon_+)}
	\prod_{I=1}^{n-1} \frac{2\sinh(\pm\phi_I)2\sinh(\pm\phi_I+2\epsilon_+)}{2\sinh(\pm\phi_I\pm\epsilon_-+\epsilon_+)}
	\\
	&& \times \prod_{I=1}^{n-1}\frac{2\sinh\epsilon_+}{2\sinh\frac{\pm\epsilon_-+\epsilon_+}{2}\prod_{i=1}^N 2\sinh\frac{\pm\phi_I\pm\alpha_i+\epsilon_+}{2}}\cdot
	\frac{\prod_{I>J}^{n-1}2\sinh\frac{\pm\phi_I\pm\phi_J}{2}2\sinh\frac{\pm\phi_I\pm\phi_J+2\epsilon_+}{2}}{\prod_{I=1}^{n-1}2\sinh\frac{\pm2\phi_I\pm\epsilon_-+\epsilon_+}{2}\prod_{I>J}^{n-1}2\sinh\frac{\pm\phi_I\pm\phi_J\pm\epsilon_-+\epsilon_+}{2}} \ , \nonumber
\end{eqnarray}
with $k=2n$.

\subsection{Hypermultiplets}\label{sec:hypermultiplet-contribution}
A hypermultiplet develops fermion zero modes in the instanton background. The presence of the fermion zero modes can be observed using an index theorem. Accordingly, it is expected that the bulk hypermultiplets induce  fermionic degrees on the instanton moduli space. When we attempt to engineer an ADHM quantum mechanics
description of these fermionic zero modes on the Higgs branch, however, extra bosonic degrees of freedom are in general required. 
Often these bosonic zero modes give rise to extra classical branches of vacua in the ADHM quantum mechanics, or extra continuum contributions to the spectrum,
which may be spurious from the point of view of the 5d gauge theory. 
In string theory constructions, they may describe D0 branes moving away from the brane system which engineers the 5d gauge theory. 
These spurious branches of vacua must be carefully subtracted from the final answer. 

We can give a few simple examples of this phenomenon. The instanton moduli space of a 5d gauge theory with an adjoint hypermultiplet has a string theory embedding. The instanton states can be interpreted as the D0/D4-brane bound states in this case. The 1d gauge theory living on the D0-branes is described by the ADHM quantum mechanics with additional matter fields corresponding to the bulk adjoint hypermultipet. This theory involves extra real 4 dimensional bosonic fields that parametrize the 4 transverse directions to the D4-branes in which the 5d gauge theory supports. The non-commutativity parameter (or FI parameter) in the 1d QM generally make these directions massive. However, when the commutativity is restored, these branches of vacua open up D0-branes (or instantons) can escape to infinity.

Similarly, the UV completion of instanton dynamics in $Sp(N)$ gauge theory with an antisymmetric and fundamental hypermultiplets has extra bosonic degrees of freedom from the hypermultiplets. Its string theory embedding is given by D0-D4-D8-O8 brane system~\cite{Aharony:1997pm}. The extra bosonic modes again parametrize the transverse directions to the D4-branes. In particular, the ADHM for this theory does not have noncommutative deformation of the space. Hence the observables computed using this UV completion in general involves extra contributions to be subtracted off. One can find examples in~\cite{Hwang:2014uwa}.

Next, we can describe our guess for the contribution of hypermultiplets in tensor powers of the fundamental representation, 
based on the prescription given in~\cite{Shadchin:2004yx}. If we could ignore the singularities, the hypermultiplets introduce vector bundles on the instanton moduli space, and the vector bundles are constructed by tensor products of an universal bundle $\mathcal{E}$. The tensor product structure of the vector bundle inherits that of the representation of the 5d hypermultiplet. We will now pretend that the same prescription can be applied to the ADHM-resolved moduli space of instantons. 
In~\cite{Shadchin:2004yx}, it was suggested that the equivariant index for the hypermultiplet can be computed by taking tensor product of the equivariant Chern character of the bundle $\mathcal{E}$, which is given by~\cite{Losev:2003py,Shadchin:2004yx}
\begin{equation}\label{eq:equivariant-Chern-E}
	Ch_\mathcal{E}(e^\alpha,e^\phi;p,q) = \chi_{\rm fund}(e^{\alpha_i}) - (1-p)(1-q)(pq)^{-1/2}\chi_{\rm fund}(e^{\phi_I}) \ ,
\end{equation}
where $\chi_{\rm fund}(e^{\alpha_i})$ and $\chi_{\rm fund}(e^{\phi_I})$ denote the character of the fundamental representations of the guage group $G$ and the dual gauge group $\hat{G}$, respectively.
For example, the equivariant indices for the hypermultiplets in the fundamental, symmetric, antisymmetric and adjoint representations are given by, respectively, 
\begin{align}\label{eq:equivariant-index}
	{\rm ind}_{\rm fund}(e^\alpha,e^\phi;p,q) &= \frac{\sqrt{pq}}{(1-p)(1-q)}Ch_\mathcal{E}(e^\alpha,e^\phi;p,q) \ ,\cr
	{\rm ind}_{\rm sym}(e^\alpha,e^\phi;p,q) &= \frac{\sqrt{pq}}{(1-p)(1-q)}Ch_{\mathcal{E}\otimes \mathcal{E}}(e^\alpha,e^\phi;p,q) \ , \cr
	{\rm ind}_{\rm anti}(e^\alpha,e^\phi;p,q) &= \frac{\sqrt{pq}}{(1-p)(1-q)}Ch_{\wedge^2 \mathcal{E}}(e^\alpha,e^\phi;p,q) \ , \cr
	{\rm ind}_{\rm adj}(e^\alpha,e^\phi;p,q) &= \frac{\sqrt{pq}}{(1-p)(1-q)}Ch_{\mathcal{E}\otimes \mathcal{E}^*}(e^\alpha,e^\phi;p,q) \ .
\end{align}
where the tensor product of the Chern character is defined using the usual tensor product rule as
\begin{align}\label{eq:equivariant-Chern-ex}
	Ch_{\mathcal{E}\otimes \mathcal{E}}(e^\alpha,e^\phi;p,q) &= \frac{1}{2}\left[Ch_{\mathcal{E}}(e^\alpha,e^\phi;p,q)^2 + Ch_{\mathcal{E}}(e^{2\alpha},e^{2\phi};p^2,q^2)\right] \ , \cr
	Ch_{\wedge^2\mathcal{E}}(e^\alpha,e^\phi;p,q) &= \frac{1}{2}\left[Ch_{\mathcal{E}}(e^\alpha,e^\phi;p,q)^2 - Ch_{\mathcal{E}}(e^{2\alpha},e^{2\phi};p^2,q^2)\right] \ , \cr
	Ch_{\mathcal{E}\otimes\mathcal{E}^*}(e^\alpha,e^\phi;p,q) &= Ch_{\mathcal{E}}(e^\alpha,e^\phi;p,q)\times Ch_{\mathcal{E}}(e^{-\alpha},e^{-\phi};p^{-1},q^{-1}) \ .
\end{align}
The equivariant indices in other representations can be obtained in the similar manner.
The resulting index computed in this way contains terms independent of the fugacity $e^{\phi_I}$ for $\hat{G}$. These terms amount to the perturbative contribution, so we will ignore them when we compute the instanton partition function .

The contribution to the instanton partition function of the hypermultiplets can be easily obtained using the relevant equivariant indices. There is a conversion rule for 5d calculation
\begin{equation}\label{eq:conversion-rule}
	{\rm ind}_R = \sum_i  n_i e^{z_i}  \ \rightarrow \ Z_R = \prod_{i}\left[2\sinh\frac{z_i}{2}\right]^{n_i}\ .
\end{equation}
Thus the plethystic exponential of the equivariant index yields the instanton partition function contribution of the hypermultiplet.
One can check that the contribution from an adjoint hypermultiplet computed using this prescription agrees with that from the localization of the ADHM quantum mechanics in~\cite{Kim:2011mv}.

Let us present explicit expressions for the hypermultiplets discussed in the main context.
For $SU(N)$ gauge theory, the fundamental hypermultiplet contribution is
\begin{equation}\label{eq:SU-instanton-fund}
	Z_{\rm fund}= \prod_{I=1}^k 2\sinh\frac{\phi_I-m}{2} \ ,
\end{equation}
with a mass parameter $m$. The antisymmetric hyper has the following contribution
\begin{equation}\label{eq:SU-instanton-antisymm}
	Z_{\rm asym} = \frac{\prod_{i=1}^{N}\prod_{I=1}^k2\sinh\frac{\phi_I+\alpha_i-m}{2} \prod_{I>J}^k2\sinh\frac{\phi_I+\phi_J-m-\epsilon_-}{2}2\sinh\frac{-\phi_I-\phi_J+m-\epsilon_-}{2} }
	{\prod_{I>J}^k2\sinh\frac{\phi_I+\phi_J-m-\epsilon_+}{2}2\sinh\frac{-\phi_I-\phi_J+m-\epsilon_+}{2} \prod_{I=1}^k2\sinh\frac{2\phi_I-m-\epsilon_+}{2}2\sinh\frac{-2\phi_I+m-\epsilon_+}{2}} \ .
\end{equation}
For $Sp(N)$ gauge theory, the fundamental representation has the contribution
\begin{equation}\label{eq:Sp-instanton-fund1}
	Z^+_{\rm fund} = \left(2\sinh\frac{m}{2}\right)^\chi \prod_{I=1}^n2\sinh\frac{\pm\phi_I+m}{2} \ ,
\end{equation}
for $O(k)_+$, and
\begin{equation}\label{eq:Sp-instanton-fund2}
	Z^-_{\rm fund} = 2\cosh\frac{m}{2}\prod_{I=1}^n2\sinh\frac{\pm\phi_I+m}{2} \ ,
\end{equation}
for $O(k)_-$ with $k=2n+1$, and
\begin{equation}\label{eq:Sp-instanton-fund3}
	Z^-_{\rm fund} = 2\sinh\frac{m}{2}\prod_{I=1}^{n-1}2\sinh\frac{\pm\phi_I+m}{2} \ ,
\end{equation}
for $O(k)_-$ with $k=2n$.
These are read off from the corresponding equivariant indices in~(\ref{eq:equivariant-index}). 

Next, we can assemble a modification of the bare ADHM quantum mechanics which would reproduce these modifications to the equivariant integrand. 
The contribution for the fundamental hypermultiplet implies that a fundamental matter induces a $(0,4)$ fermi multiplet in fundamental representation of $\hat{G}$ in the ADHM QM. This agrees with our expectation that the hypermultiplet develops fermion zero modes in the instanton background. On the other hand, the contribution from the antisymmetric hyper has factors in denominator as well as the factors in numerator. The numerator factors correspond to a fermi multiplet in the bifundamental representation of $G\times\hat{G}$ and a conjugate pair of fermi multiplets in the antisymmetric representation of $\hat{G}$. While, the denominator factors corresponds to a pair of $(0,4)$ hypermultiplets in the symmetric representation of $\hat{G}$. This means that the UV completion of the zero modes acquires nontrivial bosonic degrees which are not present in the zero mode analysis of the 5d QFT.

The computation of the 1d equivariant integral requires both an integrand and a choice of integration contour/prescription. 
The latter, in a sense, can be used to include or exclude the contribution of certain classical branches of vacua, by selecting which poles should 
be picked by the contour integral. The standard prescription in 1d localization computations is the JK-prescription. 
 To read the relevant poles from the JK-prescription, we should know the exact representations of the extra bosonic degrees under $\hat{G}$ rotation. However, although the recipe given in~\cite{Shadchin:2004yx} and in this section allows us to know the matter contents in the ADHM QM, it yet has an ambiguity in the exact representations of the multiplets. More precisely, it cannot distinguish a certain complex representation $R$ and its conjugation, i.e. `$\sinh\frac{R(\phi)+\cdots}{2}$' and `$-\sinh\frac{-R(\phi)-\cdots}{2}$'. Since we could not resolve this issue, we will give prescriptions for it case by case in the main context.

Further spurious contributions included by the standard JK-prescription have to be removed on a case-by-case basis. See~\cite{Hwang:2014uwa} for few examples.

\section{Partition functions of exotic $SU(3)$ theory}\label{sec:partition-function-exotic-SU(3)}
In this appendix, we propose a prescription to compute the instanton partition functions of the exotic $SU(3)$ theories with matters. With these results, we compute the hemisphere indices and then show that they agree with the hemisphere indices obtained in section~\ref{sec:duality-Sp-SU} using the duality wall action on the $Sp(2)$ hemisphere indices.

We are interested in the $SU(3)$ SQCD with $\kappa=5-N_f/2$, which obviously violates the bound $|\kappa| \le 3-N_f/2$ in~\cite{Intriligator:1997pq}. As mentioned before, when the theory violates this bound, the localization integral of the instanton partition function from the usual ADHM quantum mechanics encounters higher degree poles at the infinities $\phi_I=\pm \infty$. These poles are associated to the classical Coulomb branch of vacua in the ADHM quantum mechanics and not to the to the instanton moduli space which is described by the Higgs branch. Unfortunately, we do not know how to remove these spurious contributions when the degree of the pole is higher than 1.
In what follows, we will explain how to avoid having higher degree poles at infinity by introducing `pseudo' hypermultiplets in the instanton background.
We will add two (or more) `pseudo' hypermultiplets and integrate them out at the end. This will allow us to evaluate the instanton partition function without having the problem of the higher degree poles at infinity.

Let us first discuss the `pseudo' hypermultiplet and the ADHM quantum mechanics. The `pseudo' hypermultiplet is simply the hypermultiplet in the antisymmetric representation of the $SU(3)$. It should be equivalent to the fundamental hypermultiplet for the $SU(3)$ gauge theory. This is indeed the case for the perturbative analysis. However, the antisymmetric hypermultiplet affects the ADHM quantum mechanics in a different way from that of the fundamental hypermultiplet.
Strictly speaking, the ADHM quantum mechanics is designed for the $U(N)$ gauge theory since
it involves singular $U(1)$ instantons which is regularized by introducing extra UV degrees of freedom. Therefore, fermion zero modes from the antisymmetric hypermultiplet has a rather different UV completion than those from the fundamental hypermultiplet in the ADHM QM.

The fermionic zero modes from the antisymmetric hypermultiplet provide many non-trivial multiplets, not just fermi multiplets but possibly also hypermultiplets including extra bosonic zero modes, in the ADHM QM as depicted in figure~\ref{fig:adhm-SU(3)}.
\begin{figure}[h]
    \centering
 \includegraphics[width=.20\textwidth]{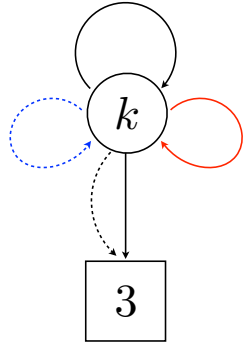}  
    \caption{Quiver diagram for $k$ instantons with an antisymmetric hypermultiplet.}
    \label{fig:adhm-SU(3)}
\end{figure}
The ADHM quantum mechanics is the $\mathcal{N}=(0,4)$ gauge theory of $U(k)$ gauge group with $SU(2)_1\times SU(2)_2\times SU(2)_R$ symmetry. See appendix~\ref{sec:instanton-partition-function} for details. We then add a bi-fundamental chiral fermion (black dashed arrow) of $U(k)$ and $SU(3)$ groups, and a $(0,4)$ fermi multiplet (blue dashed arrow), which is a doublet under the $SU(2)_1$ and in the antisymmetric representation of $U(k)$, and a hypermultiplet (red solid arrow) in the symmetric representation of $U(k)$.
This is equivalent to add to the instanton moduli space a vector bundle given by the antisymmetric product of the universal bundle in the fundamental representation.

We consider the $SU(3)$ gauge theory with two `pseudo' hypermultiplets and $N_f$ fundamental hypermultiplets. The $k$-instanton partition function from the ADHM QM can be written as
\begin{equation}
    \mathcal{Z}^{N_f}_{QM,k} = \frac{(-1)^{3+N_f}}{k!}\oint \prod_{I=1}^k\frac{d\phi_I}{2\pi i} e^{-\kappa\sum_{I=1}^k\phi_I} Z^k_{\rm vec}(\alpha,\phi;p,q) \prod_{a=1}^{N_f}Z^k_{\rm fund}(\phi,m_a;p,q)\prod_{a=1}^2Z^k_{\rm asym}(\phi,t_i;p,q)
\end{equation}
where $Z^k_{\rm vec},Z^k_{\rm fund},Z^k_{\rm asym}$ are given in~(\ref{eq:SU-instanton-vector}), (\ref{eq:SU-instanton-fund}), (\ref{eq:SU-instanton-antisymm}), respectively.
We will set the classical CS-level $\kappa=4-N_f/2$. One can easily see that the integral then has a simple pole at infinity $\phi_I=-\infty$, which is now controllable.

We are essentially interested in the theory with $\kappa=5-N_f/2$ and without the `pseudo' matters. This theory can be obtained by integrating out two `pseudo' hypermultiplets. We will send their mass parameters $t_i$ to infinity. Then it will effectively shift the bare CS-level by $+1$ and the low energy theory will have the  CS-level $\kappa=5-N_f/2$ as desired for our exotic theory. To avoid the higher degree poles at infinity, we shall integrate out the `pseudo' matters after evaluating the contour integrals. It thus allows us to compute the instanton partition function of the exotic $SU(3)$ theory without facing higher degree poles at the infinity. This procedure can be interpreted as a UV prescription of the $SU(3)$ instanton moduli space at the exotic CS-level.  Here the `pseudo' hypermultiplets are used as a UV regulator.
We will restrict ourselves to the cases with $N_f\le8$\footnote{
One may notice that the integral has higher degree poles at infinity when $N_f>8$. We may be able to resolve this by introducing one more `pseudo' hypermultiplet, but we will not discuss these cases.} for which we can consider the dual SCFT with $Sp(2)$ gauge group.

The contour integral will be evaluated using the JK-residue prescription.
One then notices that the `pseudo' matter contributions provide additional nonzero JK-residues. For example, at one instanton, the JK-residues at the following poles are nonzero:
\begin{equation}
    2\phi_1-t_a - \epsilon_+ = 0 \ , \ \ (`0' \equiv 0 \ {\rm mod} \ 2\pi) \ .
\end{equation}
Summing over all JK-residues including both from the vector multiplet and from the `pseudo' hypermultiplets, we can compute the partition function with `pseudo' matters. 

This is not quite our final answer. To obtain the QFT partition function, we need to strip off some overall factor associated to 
the extra bosonic flat directions introduced by the `pseudo' hypermultiplets. We conjecture that the extra factor is given by
\begin{align}\label{eq:extra-factor-exotic-SU(3)}
    &Z_{\rm extra,pseudo}^{N_f} = {\rm PE}\bigg[\mathfrak{q}_{SU}\,f^{N_f}(w_a,\tau_a;p,q)\bigg] \ , \cr
    &f^{N_f}= \frac{-\sqrt{\tau_1\tau_2}\prod_{a=1}^{N_f}\sqrt{w_a}}{(1-p)(1-q)(1-pq\tau_1/\tau_2)(1-pq\tau_2/\tau_1)} \bigg[
    pq(1+pq)\left(\chi_{\Lambda^2\tiny\yng(1)}^{U(N_f)}(1/w)+(\tau_1\tau_2)^{-1}\chi_{\Lambda^6\tiny\yng(1)}^{U(N_f)}(1/w)\right) \cr
    & \qquad \qquad  + (pq)^{3/2}(\tau_1+\tau_2)\left(1+(\tau_1\tau_2)^{-1}\chi_{\Lambda^4\tiny\yng(1)}^{U(N_f)}(1/w)+(\tau_1\tau_2)^{-2}\chi_{\Lambda^8\tiny\yng(1)}^{U(N_f)}(1/w)\right)\bigg] \ ,
\end{align}
where $\tau_a\equiv e^{-t_a}$ and $\chi_{\Lambda^L\tiny\yng(1)}^{U(N_f)}$ is the character of the rank $L$ antisymmetric irrep of the $U(N_f)$ flavor group with fugacities $1/w_a$. For example, $\chi_{\Lambda^2\tiny\yng(1)}^{U(N_f)}(1/w)=\sum_{a>b}^{N_f}(w_aw_b)^{-1}$. Note that this extra factor is independent of the $SU(3)$ gauge fugacities and thus it indeed corresponds to the degrees of freedom decoupled from the 5d QFT. We have checked that, after subtracting off this factor, the  instanton partition function has no poles for $\tau_a$ and is a finite polynomial in $\tau_1$ and $\tau_2$, as expected, at 1-instanton for all $N_f$ and up to 2-instantons for $N_f<6$. 

There is the usual correction factor coming from the continuum along the noncompact Coulomb branch. It is associated to the residues at infinity $\phi_I=\pm\infty$. We obtain
\begin{align}\label{eq:continuum-factor-exotic-SU(3)}
    Z_{\rm extra,cont}^{N_f} &= {\rm PE}\left[-\frac{\mathfrak{q}_{SU}\prod_{a=1}^{N_f}\sqrt{w_a}}{(1-p)(1-q)\sqrt{\tau_1\tau_2}}\right] \ , \cr
    Z_{\rm extra,cont}^{N_f=8} &= {\rm PE}\left[-\frac{\mathfrak{q}_{SU}}{(1-p)(1-q)}\left(\sqrt{\tau_1\tau_2}^{-1}\prod_{a=1}^8\sqrt{w_a} + pq\sqrt{\tau_1\tau_2}\prod_{a=1}^8\sqrt{w_a}^{-1} \right)\right] \ .
\end{align}
The `correct' partition function can then be written as
\begin{equation}
    \mathcal{Z}_{\rm inst}^{N_f}(z_i,w_a,\tau_a,\mathfrak{q}_{SU};p,q) = \mathcal{Z}_{\rm QM}^{N_f}/Z_{\rm extra}^{N_f} \ , \quad (Z^{N_f}_{\rm extra} \equiv Z_{\rm extra,cont}^{N_f}\cdot Z^{N_f}_{\rm extra,pesudo}) \ .
\end{equation}
where $\mathcal{Z}_{QM}$ is the partition function of the ADHM QM evaluated with the JK-prescription.

We now integrate out the `pseudo' hypers. We will send their masses to infinity $t_a\rightarrow -\infty$\footnote{We can also take the limit $t_a\rightarrow \infty$. Then we will get the theory with CS-level $\kappa=3-N_f/2$.} and take the leading contribution. By rescaling the instanton fugacity as $\mathfrak{q}_{SU}\sqrt{\tau_1\tau_2} \rightarrow \mathfrak{q}_{SU}$, we will end up with the instanton partition function of the $SU(3)$ theory with $N_f$ flavors and the CS-level $\kappa=5-N_f/2$:
\begin{equation}
    Z_{\rm inst}^{3,N_f}(z_i,w_a,\mathfrak{q}_{SU};p,q) \equiv \lim_{\tau_1,\tau_2\rightarrow\infty}\mathcal{Z}^{N_f}_{\rm inst}(z_i,w_a,\tau_a,\mathfrak{q}_{SU}/\sqrt{\tau_1\tau_2};p,q) \ .
\end{equation}

Taking into account the extra factors carefully, we compute 1-instanton partition functions for $N_f\le8$ and obtain
\begin{align}
    Z_{{\rm inst},k=1}^{3,N_f} 
    =& \left(\chi_{\bf3}^{SU(3)}(z) + \chi_{\Lambda^2\tiny\yng(1)}^{U(N_f)}+ \chi_{\Lambda^8\tiny\yng(1)}^{U(N_f)}
    \right)\left(x^2+\chi_{\bf2}^{SU(2)}(y)x^3+\chi_{\bf3}^{SU(2)}(y)x^4\right) \cr
    & +\left(\chi_{\Lambda^5\tiny\yng(1)}^{U(N_f)}-\chi_{\bf\bar{3}}^{SU(3)}(z)\chi_{\tiny\yng(1)}^{U(N_f)}\right)\left(x^3+\chi_{\bf2}^{SU(2)}(y)x^4\right) \cr
    &+ \left(\chi_{\bf\bar{6}}^{SU(3)}(z) - \chi_{\bf\bar{3}}^{SU(3)}(z)\chi_{\Lambda^4\tiny\yng(1)}^{U(N_f)}-\chi_{\bf3}^{SU(3)}(z)\chi_{\Lambda^6\tiny\yng(1)}^{U(N_f)}\right)x^4 +\mathcal{O}(x^5) \ .
\end{align}
Combining the 1-loop determinant, we have checked that the hemisphere index of our exotic $SU(3)$ theory yields exactly the right hand side of the duality relation~(\ref{eq:duality-hemisphere-Sp-SU}) between the $Sp(2)$ and $SU(3)$ theories, in all examples at least up to $x^3$ order. This result supports the UV prescription of the exotic $SU(3)$ theory in this section.

Similarly, we can compute the Wilson loop index of the exotic $SU(3)$ theories using the above UV prescription. An Wilson loop in a representation $R$ inserts the corresponding equivariant Chern character into contour integral of the instanton partition function. At $k$-instantons, the Wilson loop index before integrating out the pseudo hypers can be written as
\begin{equation}
    \mathcal{W}^{N_f}_{QM,k} = \frac{(-1)^{3+N_f}}{k!}\!\oint \prod_{I=1}^k\frac{d\phi_I}{2\pi i}  Ch_{R}(\alpha,\phi)\cdot e^{-\kappa\sum_{I=1}^k\phi_I}Z^k_{\rm vec}(\alpha,\phi) \prod_{a=1}^{N_f}Z^k_{\rm fund}(\phi,m_a)\prod_{a=1}^2Z^k_{\rm asym}(\phi,t_i) \ ,
\end{equation}
where $Ch_{R}(\alpha,\phi)$ is the equivariant Chern character of the vector bundle in the representation $R$. We will focus only on the Wilson loop in the fundamental representation whose equivariant Chern character is given in~(\ref{eq:equivariant-Chern-E}).
The contour integral is again evaluated using the JK-prescription.
Since the Wilson loop insertion does not change the pole structure of the integrand, we can pick up the same poles as before.

As we have seen above, the partition function involves the correction factors from the Coulomb branch and the extra bosonic degrees of the `pseudo' matters given in~(\ref{eq:continuum-factor-exotic-SU(3)}) and (\ref{eq:extra-factor-exotic-SU(3)}), which we should subtract off.
Due to the same reason as without Wilson loops, we expect the correct Wilson loop index has no poles for the mass parameter $\tau_a$ of the `pseudo' matters. However, even after subtracting the correction factors in~(\ref{eq:continuum-factor-exotic-SU(3)}) and (\ref{eq:extra-factor-exotic-SU(3)}), we notice that the Wilson loop index still has poles for $\tau_a$. We find that the Wilson loop receives an additional correction when $N_f>0$. For example, if we define new Wilson loop indices taking the form
\begin{align}
    \mathcal{W}_{\rm fund}^{3,0}(z,w,\tau,\mathfrak{q}_{SU}) &= \mathcal{W}_{QM,{\rm fund}}^{3,0}/Z_{\rm extra}^{N_f=0} \ , \\
    \mathcal{W}_{\rm fund}^{3,1}(z,w,\tau,\mathfrak{q}_{SU}) &= \mathcal{W}_{QM,{\rm fund}}^{3,1}/Z_{\rm extra}^{N_f=1} +\mathfrak{q}_{SU}  \frac{\sqrt{pq\tau_1\tau_2/w_1}(1+pq)}{(1-pq\tau_1/\tau_2)(1-pq\tau_2/\tau_1)}II^{3,1} \ , \cr
     \mathcal{W}_{\rm fund}^{3,2}(z,w,\tau,\mathfrak{q}_{SU}) &= \mathcal{W}_{QM,{\rm fund}}^{3,2}/Z_{\rm extra}^{N_f=2} + \mathfrak{q}_{SU} \frac{\sqrt{pq\tau_1\tau_2/(w_1w_2)}(1+pq)(w_1+w_2)}{(1-pq\tau_1/\tau_2)(1-pq\tau_2/\tau_1)}II^{3,2}\ , \nonumber
\end{align}
where $II^{3,N_f}$ is the bare hemisphere index without Wilson loops, these new indices have no poles for $\tau_a$. We have checked this till 2-instantons.
Thus we suggest that the `correct' Wilson loop index with `pseudo' matters should be this new index.

Let us integrate out the `pseudo' hypers by rescaling the instanton fugacity as $\mathfrak{q}_{SU}\sqrt{\tau_1\tau_2}\rightarrow \mathfrak{q}_{SU}$ and taking the limit $t_a\rightarrow -\infty$.
It leads to the Wilson loop index of the exotic $SU(3)$ theory, given by
\begin{equation}
    W_{\rm fund}^{3,N_f}(z,w,\mathfrak{q}_{SU};p,q) \equiv \lim_{\tau_1,\tau_2\rightarrow\infty}\mathcal{W}^{3,N_f}_{\rm fund}(z,w,\tau,\mathfrak{q}_{SU}/\sqrt{\tau_1\tau_2};p,q) \ .
\end{equation}
 We have also checked that this Wilson loop index yields the results in~(\ref{eq:dual-action-Sp(2)-Nf=0}) and (\ref{eq:dual-action-Sp(2)-Nf=1}) obtained from the duality wall action on the dual $Sp(2)$ hemisphere indices, up to $x^4$ order.

\subsection{Superconformal indices}
Now, we compute the superconformal indices for the $Sp(2)$ and $SU(3)$ theories and check the duality conjecture. Let us first discuss the $Sp(2)$ theories. The superconformal index is defined in~(\ref{eq:superconformal-index-Sp}). For $N_f<8$, we find
\begin{eqnarray}\label{eq:superconformal-index-Sp(2)}
    I_{Sp}^{2,N_f} &=& 1 + \left(1+\chi^{SO(2N_f)}_{\bf adj}\right)x^2 + \left(\chi_{\bf 2}^{SU(2)}(y)\left(2+\chi_{\bf adj}^{SO(2N_f)}\right)+\mathfrak{q}_{Sp}\chi_{\bf \bar{S}}^{SO(2N_f)}+\mathfrak{q}_{Sp}^{-1}\chi_{\bf \bar{S}^*}^{SO(2N_f)}\right)x^3\cr
    &&+\left(\chi_{\bf3}^{SU(2)}(y)\left(2+\chi_{\bf adj}^{SO(2N_f)}\right) + \chi_{\bf 2}^{SU(2)}(y) \left(\mathfrak{q}_{Sp}\chi_{\bf \bar{S}}^{SO(2N_f)}+\mathfrak{q}_{Sp}^{-1}\chi_{\bf \bar{S}^*}^{SO(2N_f)}\right)\right)x^4 \cr
    &&+\left(2+\chi^{SO(2N_f)}_{\bf adj\otimes adj} -\chi^{SO(2N_f)}_{\bf fund}(w_a^2)\right)x^4 + \mathcal{O}(x^5) \ ,
\end{eqnarray}
where $\chi^{SO(2N_f)}_{\bf r}$ is the character of the  ${\bf r}$ irrep  of $SO(2N_f)$ symmetry with fugacities $w_a$ and $\bar{S}$ denotes the conjugate spinor representation and $\bar{S}^*$ is the complex conjugation of $\bar{S}$. $\chi^{SO(2N_f)}_{\bf fund}(w_a^2)$ denotes the fundamental character with fugacities $w_a^2$.
For $N_f=8$, we compute
\begin{eqnarray}\label{eq:superconformal-index-Sp(2)-8flavors}
    I_{Sp}^{2,N_f=8} &=& 1 + \left(\chi^{SU(2)}_{\bf 3}(\mathfrak{q}_{Sp})+\chi^{SO(16)}_{\bf adj}\right)x^2 \cr
    &&+ \left(\chi_{\bf 2}^{SU(2)}(y)\left(1+\chi^{SU(2)}_{\bf 3}(\mathfrak{q}_{Sp})+\chi_{\bf adj}^{SO(16)}\right)+\chi^{SU(2)}_{\bf 2}(\mathfrak{q}_{Sp})\cdot\chi_{\bf \bar{S}}^{SO(16)}\right)x^3 \cr
    &&+\left(\chi_{\bf3}^{SU(2)}(y)\left(1+\chi^{SU(2)}_{\bf 3}(\mathfrak{q}_{Sp})+\chi_{\bf adj}^{SO(16)}\right) + \chi_{\bf 2}^{SU(2)}(y) \cdot \chi^{SU(2)}_{\bf 2}(\mathfrak{q}_{Sp}) \cdot \chi_{\bf \bar{S}}^{SO(16)}\right)x^4 \cr
    &&+\left(2+\chi^{SO(16)}_{\bf adj\otimes adj} 
    +\chi^{SU(2)}_{\bf 3}(\mathfrak{q}_{Sp})\left(1+\chi^{SO(16)}_{\bf adj}\right)-\chi^{SO(16)}_{\bf 136}\right)x^4 + \mathcal{O}(x^5) \ .
\end{eqnarray}
Here $\chi^{SO(16)}_{\bf 136}$ is the character of the rank $2$ symmetric representation of $SO(16)$. This theory has an enhanced $SU(2)\times SO(16)$ global symmetry  at the UV fixed point.
There are additional BPS states at $x^2$ order corresponding to the conserved currents with instanton fugacity $\mathfrak{q}_{Sp}$ and all BPS states properly arrange themselves to form representations of the enhanced symmetry. Thus the result is consistent with the symmetry enhancement.

We now turn to the $SU(3)$ theories. The superconformal index of the general $SU(N)$ SQCD can be written as
\begin{align}
    & I^{N,N_f}_{SU}(w_a,\mathfrak{q}_{SU};p,q) \\
    & = \frac{(I_V)^{N-1}}{N!}  \oint \prod_{i=1}^{N-1}\frac{dz_i}{2\pi i z_i} \left|\frac{\prod_{i\neq j}^{N} (z_i/z_j;p,q)_\infty  }{ \prod_{i=1}^{N}\prod_{a=1}^{N_f}(\sqrt{pq}z_i/w_a;p,q)_\infty } Z_{SU,{\rm inst}}^{N,N_f} (z_i,w_a,\mathfrak{q}_{SU};p,q)\right|^2 \ , \nonumber
\end{align}
with $\prod_{i=1}^{N}z_i = 1$. Our exotic theory has the classical CS-level $\kappa=N+2-N_f/2$ which only enters in the instanton partition function.

For our $SU(3)$ theories, the instanton partition functions are given in the previous section, so the superconformal index computation is straightforward. We find that the results perfectly agree with the indices of the dual $Sp(2)$ theories computed in~(\ref{eq:superconformal-index-Sp(2)}) and (\ref{eq:superconformal-index-Sp(2)-8flavors}), once we identify the fugacities of two dual theories as (\ref{eq:parameter-identity-SU-Sp}).
This has been checked at least up to $x^4$ orders.
This result provides a strong evidence for the duality conjecture of the $Sp(2)$ and $SU(3)$ theories and also the symmetry enhancements of the $SU(3)$ theories at UV fixed points.

\bibliographystyle{JHEP}

\bibliography{5d-paper}

\end{document}